\documentclass[longauth,traditabstract]{aa}  
\usepackage{txfonts}
\usepackage{wrapfig}
\usepackage{graphicx}
\usepackage{amsmath, amssymb } 
\usepackage{enumitem}
\usepackage[toc, title]{appendix}
\usepackage{hyperref}
\hypersetup{
    colorlinks=true,
    linkcolor=blue,
    filecolor=magenta,      
    urlcolor=cyan,
    citecolor=blue
}
\usepackage[nameinlink,noabbrev]{cleveref}
\Crefname{equation}{Eq.}{Eqs.}
\Crefname{eqnarray}{Eq.}{Eqs.}
\Crefname{section}{Sect.}{Sects.}
\Crefname{figure}{Fig.}{Figs.}
\crefname{equation}{Equation}{Equations}
\crefname{section}{Section}{Sections}
\crefname{figure}{Figure}{Figures}
\creflabelformat{equation}{#2#1#3}

\usepackage{nameref}
\usepackage{aalongtable}
\usepackage{xspace}
\usepackage{caption}
	\newcommand{\nustar}{\emph{NuSTAR}\xspace}		
    \newcommand{\xmm}{\emph{XMM-Newton}\xspace}
	\newcommand{\Msun}{\mbox{$M_{\odot}$}\xspace}
	\newcommand{\lbol}{L$_{\rm bol}$\xspace} 
    \newcommand{\ecut}{E$_{\rm cut}$\xspace}
    \newcommand{\ledd}{$\lambda_{\rm Edd}$\xspace}
    \newcommand{\mbh}{M$_{\rm BH}$\xspace}

\defcitealias{Matzeu23}{M23}

\begin{document}

    \title{SUBWAYS: Supermassive Black Hole Winds in X-rays}

   \subtitle{V. Properties of hot coronae in quasars at intermediate redshift}

   \author{S. Peluso
          \inst{1,2,3} \thanks{\email{sara.peluso@eso.org}}
        \and G. Lanzuisi\inst{1}
          \and
        A. Comastri\inst{1}
        \and
        M. Brusa\inst{2,1}
        \and
        M. Giustini\inst{4}
        \and
        G. Miniutti\inst{4}
        \and
         S. Bianchi\inst{5}
        \and
         V. E. Gianolli\inst{6}
        \and
        R. Middei\inst{7,8}
        \and
        P-O. Petrucci\inst{9}
        \and
        L. Borrelli\inst{2,1}
        \and
        E. Amenta\inst{2,10}
        \and
        E. Bertola\inst{11} 
        \and
        B. De Marco\inst{12}
        \and
        A. De Rosa\inst{13}
        \and
        S. Kraemer\inst{14}
        \and
        G. Kriss\inst{15}
        \and
        Y. Krongold\inst{16} 
        \and
        S. Mathur\inst{17}
        \and 
        A. Merloni\inst{18}
        \and
        E. Nardini\inst{11}
        \and
        F. Panessa\inst{13}
        \and
        E. Piconcelli\inst{19}
        \and
        G. Ponti\inst{20}
        \and
        F. Ricci\inst{5}
        \and
        A. Tortosa\inst{19}
        \and 
        L. Zappacosta\inst{19}
        \and
        R. Serafinelli \inst{21, 7 }
        \fnmsep
          }

   \institute{INAF -- Osservatorio di Astrofisica e Scienza dello Spazio di Bologna (OAS), Via Gobetti 93/3, I-40129 Bologna, Italy
  \and
Dipartimento di Fisica e Astronomia (DIFA), Università di Bologna, via Gobetti 93/2, I-40129 Bologna, Italy 
\and 
European Southern Observatory (ESO), Karl Schwarzschild Strasse 2, D-85748 Garching bei Muenchen, Germany
         \and
         Centro de Astrobiología (CAB), CSIC-INTA, Camino Bajo del Castillo s/n, 28692 Villanueva de la Cañada, Madrid, Spain 
         \and
              Dipartimento di Matematica e Fisica, Università degli Studi Roma Tre, Via della Vasca Navale 84, 00146 Roma, Italy
            \and
            Dep. of Physics and Astronomy, Clemson University, Kinard Lab of Physics, 140 Delta Epsilon Ct, Clemson, SC 29634, USA
            \and
            INAF Osservatorio Astronomico di Roma, 10 Via Frascati 33, 00078 Monte Porzio Catone, RM, Italy
            \and 
            Space Science Data Center, 12 Agenzia Spaziale Italiana, Via del Politecnico snc, 00133 Roma, Italy
            \and
            Univ. Grenoble Alpes, CNRS, IPAG, 38000 Grenoble, France
            \and
            INAF – Istituto di Radioastronomia, Via P. Gobetti 101, 40129 Bologna, Italy
        \and
            INAF – Osservatorio Astrofisco di Arcetri, Largo E. Fermi 5, 50127 Firenze, Italy
            \and
             Departament de Fis\'{i}ca, EEBE, Universitat Polit\`ecnica de Catalunya, Av. Eduard Maristany 16, S-08019 Barcelona, Spain
             \and
               INAF – Istituto di Astrofisica e Planetologia Spaziali, Via Fosso del Cavaliere, 00133 Roma, Italy
             \and
                Department of Physics, Institute for Astrophysics and Computational Sciences, The Catholic University of America, Washington, DC 20064, USA 
            \and
                Space Telescope Science Institute, 3700 San Martin Drive, Baltimore, MD 21218, USA
            \and 
                Instituto de Astronom\'ia, Universidad Nacional Aut\'onoma de M\'exico,  Circuito Exterior, Ciudad Universitaria, Ciudad de M\'exico 04510, M\'exico 
             \and   
             Department of Astronomy, University of Michigan, 1085 South University Avenue, Ann Arbor, MI 48109, USA
             \and
             Max-Planck-Institut für extraterrestrische Physik (MPE), Gießenbachstraße 1, D-85748 Garching bei München, Germany
             \and
            INAF – Astronomical Observatory of Rome, Via Frascati 33, 00040 Monte Porzio Catone, Italy
            \and 
            INAF – Osservatorio Astronomico di Brera, Via Bianchi 46, 23807 Merate (LC), Italy
            \and
            Instituto de Estudios Astrofísicos, Facultad de Ingeniería y Ciencias, Universidad Diego Portales, Avenida Ejército Libertador 441, Santiago, Chile
}
   \date{}
\abstract
{
We present the X-ray analysis of coronal properties in a statistically representative sample of 23 mostly radio-quiet AGN from the SUBWAYS campaign (SUpermassive Black holes Winds in XrAYs), focusing on quasars at redshifts $0.1<z<0.4$ and bolometric luminosities $2\times10^{44}<L_{bol} (\rm erg/s)<2\times10^{46}$.
The main aim of this work is to investigate the properties of the hot corona through the study of the hard X-ray band emission, including a proper treatment of the soft X-ray band.
High-quality X-ray spectra from \xmm, complemented by \nustar data extending up to 30–40 keV in the rest frame, are available for this sample.
The soft X-ray band (0.3–2 keV) spectrum is best fitted by a warm corona model with a median temperature of $\sim0.40$ keV, and an optical depth in the range $\tau\sim20-40$, consistent with previous results on lower luminosity sources.
The hard X-ray band is well described using a hot corona model, with a median high-energy cut-off of $\sim87$ keV, at the lower end of the distribution of typical values found in Seyfert galaxies ($\sim100-200$ keV). The derived median value of the optical depth ($\tau\sim 1-5$) suggests the presence of a moderately optically thick corona.
Combining the SUBWAYS results with literature samples at low and high redshift, we assemble the largest sample to date of AGN with $E_{cut}$ and accretion parameter measurements, finding a significant anticorrelation of $E_{cut}$ with both $\lambda_{\rm Edd}$ and \lbol\, with the median $E_{cut}$ decreasing from $\sim250-300$ keV at low accretion rates and luminosities to $\sim90-100$ keV at high accretion rates and luminosities — consistent with enhanced coronal cooling, possibly driven by pair-production.
These results favor cooler, optically thicker coronae in luminous AGN compared to those in lower-luminosity Seyfert galaxies. 

}

   \maketitle

\section{Introduction}
\label{Introduction}

Active Galactic Nuclei (AGN) are galaxies with a supermassive black hole (SMBH) at their center, typically having masses of $10^5$–$10^{10} \ M_{\odot}$, which actively accretes surrounding matter \citep{Salpeter64}. In most cases, it is expected that the accretion process generates a radiatively efficient, optically thick, and geometrically thin thermal disk that emits across the electromagnetic spectrum, peaking in the ultraviolet (UV) and extreme ultraviolet (EUV) bands \citep{Shakura73}.
 
A fraction of the available gravitational energy is dissipated in a hot plasma near the SMBH, generally referred to as the hot corona, with temperatures up to a few hundred keV, comparable to the virial temperature of the electrons at a few gravitational radii \citep{Gilfanov2014}. The hot coronal plasma upscatters the soft UV photons from the accretion disk to X-ray energies \citep{Sunyaev1980}. The resulting X-ray spectrum is approximated by a power law with a high-energy cut-off:  $N(E) \propto E^{-\Gamma} \exp\left(-\frac{E}{E_{cut}}\right)$, where $\Gamma$ is the photon index of the power law, and $E_{cut}$ denotes the e-folding energy for an exponential decline. Values for the high-energy cut-off between $E_{cut}=50$ keV and $E_{cut}=300$ keV were first obtained in relatively small samples of Seyfert galaxies thanks to the broad-band capabilities of  BeppoSAX  \citep[e.g.,][]{Matt00}  and INTEGRAL \citep[e.g.,][]{Panessa2008,Molina2009} satellites.  

The up-scattered photons emitted towards the accretion disk are partially reflected, with a characteristic spectrum dominated by a hump at 10-30 keV and strong emission lines, most notably the Iron K$\alpha$ line (e.g. 6.4 keV). The disk reflection spectrum from the innermost accretion disk regions is subject to special and general relativistic effects, which can be used to infer the spacetime geometry in the immediate vicinity of the central SMBH (e.g., \citealt{Fabian2005}). 
Reflection from material at larger radii in the disk also gives rise to an X-ray reflection spectrum, which is, however, unaffected by relativistic effects due to the large distance from the central SMBH and slower orbital motion.

The current and widely accepted physical interpretation of these observed properties is based on the thermal Comptonization mechanism. In this framework, the temperature $kT_e$ and optical depth $\tau$ of the hot coronal plasma are closely linked to the measured spectral parameters, the high-energy cut-off and the photon index, offering a way to extract information about the coronal conditions from broadband X-ray observations  \citep{Beloborodov99, Petrucci01}. 
In the early model description (e.g. \citealt{HaardtMaraschi91}), a substantial fraction of the gravitational power is dissipated in a layer of hot plasma (hot phase) on top of a cold accretion disk (cold phase). By requiring an energy balance between the two phases, a self-consistent solution for the hot-phase temperature and optical depth is derived. A good agreement with the observations is obtained for $\tau \leq 1$ and electron temperatures in the $\sim 50-300$ keV interval \citep{Fabian2015}.

Comptonization models have been expanded in several ways in the literature, and some have been adapted for spectral fitting of X-ray spectra  \citep{Fabian_2017,Tortosa18a,Middei19Moca,Pal2024}. While the physical mechanism remains the same, a wider parameter space is explored, including plasma in the moderately optically thick regime ($\tau > 1$), various geometrical configurations (i.e., spherical versus slab), and covering factors (i.e., patchy corona, see \citealt{HaardtMaraschi93}).    

The temperature, optical depth, and geometrical configuration of the hot coronal plasma are not straightforward to obtain from observations, due to approximations in Comptonization models, the degeneracy between $kT_e$ and $\tau$, and the statistical quality of the X-ray spectra. In this context, recent X-ray polarimetric studies have provided new insights into the geometry of the corona, suggesting that the commonly adopted lamp-post geometry may be disfavored relative to a slab-like one\footnote{Nevertheless, this geometry remains widely used in theoretical models, as it offers a mathematically convenient representation.} \citep{Tagliacozzo2023, Gianolli23, Gianolli24, 2024Gianolli, Friederic2024}. 
Given the relevance of the coronal geometry in the interpretation of the emerging spectrum, it is useful to recall that, assuming a slab geometry, an 
empirical relation of E$_{cut} = 2-3 \  kT_e$\label{Empirical_Relation_Ecut_kTe} is often adopted. The factor 2 applies to an optically thin plasma ($\tau < 1$), while a factor of 3 is appropriate in the optically thick ($\tau \gg 1$) limit \citep{Petrucci00,Petrucci2001}.
The determination of the hard X-ray spectrum, however, becomes increasingly challenging when the coronal plasma is pushed to extreme physical conditions. At very high energies $ h\nu\sim$1 MeV, interactions can dominate over inverse Compton scattering, leading to the production of electron–positron pairs. Pair production acts as a thermostat mechanism, ensuring that the coronal temperature remains within a stable range by continuously balancing the energy distribution \citep{Fabian2015, Fabian_2017, Ricci2018}.
The onset of this regime depends on the electron temperature, the radiative compactness ($\ell$), proportional to the ratio between source luminosity and size of the emitting region, and the optical depth of the hot corona \citep{Cavaliere1980}. 

Below 2-3 keV, the X-ray emission of AGN is instead often characterized by a soft excess: extrapolating the hard X-ray continuum to lower energies largely underestimates the observed soft flux, highlighting a different origin. The spectral shape of the soft excess emission is usually well described by one or more blackbody components \citep{Holt1980, Pravdo81, Singh85, Arnaud1996}, with remarkably similar temperatures of the order of 0.1-0.2 keV, at odds with what one would expect if the soft excess is due to thermal emission from the disk, where the temperature should scale with the BH mass (i.e., $T \propto M^{-1/4}$).

The true physical origin of this component is, therefore, still a matter of debate. Two main different interpretations have been proposed: relativistically blurred, ionized disk reflection \citep{Ballantyne2001} and Comptonization by a warm, optically thick corona distinct from the hotter one that produces the hard X-ray continuum \citep{Petrucci13,Petrucci18}.
In the first hypothesis, the soft excess originates from the innermost regions of an ionized disk around a rapidly rotating SMBH illuminated by the hot corona.
Conversely, the second hypothesis assumes a layer of electrons with a temperature $kT_e\sim 0.1-1$ keV and an optical depth $\tau\sim 10-20$, called the warm corona, placed above the surface of the accretion disk \citep{Petrucci13}. In this case, the UV photons from the inner disk interact with the warm corona electrons, producing the soft X-ray excess observed in the spectrum through a thermal Comptonization process, similar to that of the hot corona.

Several observational campaigns have aimed at constraining the physical parameters of the hot corona in local AGN, primarily through the application of phenomenological models (e.g., \citealt{Wardzinski2000, Fabian2015, Ursini15, Tortosa18a, Akylas21, Pal2024}).
Furthermore, \nustar also allowed this analysis to be extended to a few high-z galaxies \citep{Lanzuisi19, Bertola22} thanks to its higher sensitivity and imaging capabilities \citep{Harrison13}. Recent studies of the hot corona have found median high-energy cut-off values in the range $E_{cut} \sim 100 - 200\ \text{keV}$, with a possible anticorrelation between accretion rate and high-energy cut-off found in some samples \citep{Ricci2018} but not in others (e.g., \citealt{Tortosa18a, Serafinelli2024}). However, significant uncertainty remains in these measurements and their dependence on the physical properties of the corona, motivating further investigation \citep{Ricci2017, Kamraj2022}.

This study aims to fill the gap in redshift ($z=0.1-0.4$) and luminosity ($L_{bol}=2\times 10^{44}-2\times10^{46}$ erg/s) by analyzing high-quality X-ray data, from the combination of \xmm and \nustar facilities, of 23 QSOs at intermediate redshift within the SUBWAYS project\footnote{\url{https://sites.google.com/inaf.it/subways-project/}} \citep{Brusa2022}, providing more accurate estimates of the electron temperature and optical depth of the warm and hot coronae in an underexplored parameter space. Combining SUBWAYS results with literature studies at lower and higher redshifts, the dependence of the corona properties on the accretion parameter is investigated. 

We present the SUBWAYS project and data reduction in \Cref{sec:Sample and Data reduction}. We describe the spectral analysis in \Cref{sec:Spectral Analysis} and the results in \Cref{sec:Results}. The discussion and final remarks of this study are presented in  \Cref{sec:Discussion} and  \Cref{sec:Conclusions}. 
Unless otherwise stated, uncertainties are quoted at the 90$\%$ confidence level.
Throughout the paper we adopt standard cosmological parameters ($H_0=70 $  km s$^{-1}$ Mpc$^{-1}$, $\Omega_m=0.3$, $\Omega_\Lambda=0.7$).

\section{Sample and Data reduction}
\label{sec:Sample and Data reduction}
\subsection{SUBWAYS sample}

SUBWAYS is an \xmm large programme 
awarded in AO18 with a total exposure time of 1.45 Ms. 
The primary objective of the SUBWAYS project was to conduct a systematic search for sub-relativistic accretion disk winds, known as ultra-fast outflows (UFOs), which are among the most powerful AGN winds and a viable driver of large-scale galactic outflows \citep{KingPounds15,Costa2020,WARD2024}.
Research on UFOs has been largely restricted to AGN in the local Universe, primarily because high count rates are required to detect their weak absorption features in X-ray spectra. 
Through deep \xmm pointings, SUBWAYS has expanded the systematic search for UFOs to quasars with bolometric luminosities in the range $L_{bol}\sim2\times10^{44}-2\times10^{46}$ erg/s and redshifts z$\sim0.1-0.4$ (\citealt{Brusa2022,Matzeu23}, M23 hereafter). The target selection was performed by identifying sources within the cross-match of the 3XMM-DR7\footnote{\url{http://xmmssc.irap.omp.eu/Catalogue/3XMM-DR7/3XMM_DR7.html}}, SDSSDR14\footnote{\url{https://skyserver.sdss.org/dr14/en/home.aspx}},
and PG QSO catalogs \citep{SchmidtGreen83}, ensuring they have a high enough count rate to collect $\sim10^4$ source counts in the 4-10 keV band in one \xmm orbit (see \citetalias{Matzeu23} for details). 
The sample was restricted to radio-quiet AGN in order to exclude jet-dominated sources and ensure a clearer view of UFO features. However, a parallel radio analysis has shown that some of the sources are classified as intermediate between radio-quiet and radio-loud (see Amenta et al. 2026). Despite this, the expected contribution of radio jet in the X-ray band is subdominant with respect to the X-ray emission and is therefore not expected to significantly affect this analysis.
Initially,  
The sample consists of 23 mostly radio-quiet AGN, 17 of which were observed with \xmm between May 2019 and June 2020 (\Cref{SUBWAYS_info_obs}), at sufficient depth to allow for a significant detection of typical $EW\sim 50-150$ eV UFO absorption features. The remaining 6 objects were retrieved from archival data. UFOs have been detected in $\sim$30\% of the SUBWAYS sample, on the basis of a careful X-ray analysis of the deep \xmm observations \citepalias{Matzeu23}. 

Additional observational campaigns have complemented the \xmm one: co-eval optical-UV data were obtained through the Hubble Space Telescope COS spectrograph, to further investigate lower ionization outflows in the optical-UV energy band \citep{Mehdipour23}, and radio data from the VLA (B configuration, L- and C-band) were awarded to study the origin of the radio emission in this sample of AGN (Amenta et al. 2026 sub.).

Most importantly for this work, coordinated observations were obtained with \nustar (PI: S. Bianchi) for 19 targets, totaling 600 ks of exposure time. The extensive energy coverage of \nustar facilitates an in-depth analysis of the coronal properties, as the hot corona emission dominates this high-energy range.

\nustar observations were performed between June 2020 and November 2021 (\Cref{SUBWAYS_ID}), and are complemented by archival data in six cases.  
For the sources included in the \xmm SUBWAYS program and \nustar follow-up, the observations of the two telescopes are spaced by 5 to 17 months, except for three cases of quasi-simultaneity (less than one month apart). For the archival sources, the separation can be even larger (several years).
Given the BH mass range of the sample  (Log(M$_{\rm BH})=7-9$, we may expect X-ray variability on timescales shorter (days) than the months-long separation between our epochs \citep{McHardy06,Ponti12}. However, a comparison of the overlapping 3-10 keV band reveals flux differences within $\sim20\%$ for most sources (accounted for in the fits using a cross-calibration, see \autoref{sec:Spectral Analysis}), suggesting they did not undergo drastic state changes.
Two sources from the archival sample (namely HB891257+286 and PG1114+445) have been observed by \xmm multiple times (six to eleven, \citetalias{Matzeu23}). For this work, we focused on the longest pointings within a single epoch (2004 for HB891257+286 and 2010 for PG1114+445) to minimize spectral variability within the \xmm spectra.
Further details are provided in \Cref{SUBWAYS_info_obs}, which summarizes the basic properties of the SUBWAYS high energy observations, comprising a total of 42 \xmm and 26 \nustar observations.

\subsection{Data reduction}
We used \xmm spectra for PN, MOS1 and MOS2 with data reduced as described in \citetalias{Matzeu23}. In that work, the data reduction was performed using the Science Analysis System (SAS) v21.0 and adopting an optimized filtering method (following \citealt{Piconcelli04}) to clean background flared time intervals, designed to maximize the signal-to-noise ratio in a given band, 4-10 keV in this case. Considering the flux of the SUBWAYS sources (\Cref{SUBWAYS_info_obs}), this approach allowed us to retain most of the \xmm exposure time (see \citetalias{Matzeu23} for details).
PN, MOS1 and MOS2 were grouped using \texttt{ftgrouppha} tool within Heasoft v.6.30\footnote{\url{https://heasarc.gsfc.nasa.gov/docs/software/lheasoft/}}, applying the optimal binning scheme proposed by \cite{KaastraBleeker16}.

The \nustar data were processed using the \nustar Data Analysis Software package (NuSTARDAS v2.1.1) within Heasoft v.6.30.
Calibrated and cleaned event files were generated using the calibration files from the \nustar CALDB v.20220331\footnote{\url{https://heasarc.gsfc.nasa.gov/docs/heasarc/caldb/}} and applying standard filtering criteria with the \texttt{Nupipeline} task.
For each source, the background spectra for FPMA and FPMB were extracted from a source-free region with a radius of $75''$. Different source extraction radii were tested to optimize the source SNR across the full NuSTAR energy band (3–79 keV).
The optimal source radius typically lies between $50''$ and $80''$, corresponding to $\sim60-80\%$ of the encircled energy fraction for the \nustar point spread function \citealt{Hongjun2014}.
The \nustar spectra were grouped using the \texttt{ftgrouppha} following the optimal binning scheme as used for \xmm data.
For most sources, significant detections (SNR $>3$) are achieved only up to $\sim25$ keV. This is due to their relatively low flux levels and the presence of strong emission lines dominating the \nustar background spectra at 25-30 keV \citep{Wik2014}. 
\section{Spectral Analysis}
\label{sec:Spectral Analysis}
We performed a joint spectral analysis of \xmm and \nustar data for each of the 23 sources included in the SUBWAYS sample. The broadband fit was performed from 0.3 keV to $\sim20-50$ keV, depending on the SNR of the \nustar spectra of each source (with most sources reaching 40 keV), aiming to obtain the most accurate modeling of both the soft excess and the hot corona\footnote{The complete spectral atlas of the sample is available at \url{https://doi.org/10.5281/zenodo.18860380}.}. 
The inclusion of \nustar data is crucial to better constrain the high-energy cut-off of the power-law continuum, leading to a more accurate characterization of the coronal properties. A cross-calibration constant is included in the fit to take into account flux variability between \xmm and \nustar observations (typically within $20\%$), while the spectral shape is assumed to be the same between the two data sets.
The spectral analysis extends the analysis presented in \citetalias{Matzeu23} by introducing a more advanced modeling of the different components. All emission and absorption features included in the \citetalias{Matzeu23} analysis are preserved in the present analysis.

For the soft excess, we test the warm-corona model, rather than the competing blurred reflection model. This choice is primarily driven by the general lack of evidence for relativistically broadened Fe K emission lines across the sample. 
Referring to the detailed spectral analysis of the Fe K region presented in \citetalias{Matzeu23} (see their Table 2), there are only a few sources (specifically PG\,1352$+$183, PG\,0804$+$761, and LBQS\,1338$-$0038, and marginally  PG\,1114$+$445) that exhibit emission features with total line widths $\sigma > 300$\,eV and equivalent widths in the range $EW \sim 100-300$\,eV, which could be interpreted as broadened emission lines. 
However, as discussed in \citetalias{Matzeu23}, these features are likely arising from a blend of the neutral Fe\,K$\alpha$ and ionized species (e.g., Fe\,XXV) lines. Inspection of the line scans (e.g., Fig.~E.1 and Fig.~4 in \citetalias{Matzeu23}) supports this interpretation, revealing distinct narrow features at $\sim6.4$\,keV and $\sim7$\,keV. When these ionized components are self-consistently modeled, the features appear narrow rather than relativistically broadened. Even if the relativistic reflection model could be successfully applied to these specific sources, the majority of the sample (16 out of 23 sources) shows no clear evidence of a broad iron line component. This means the available data quality is insufficient to efficiently constrain its many free parameters. While relativistic reflection has been applied to sources with less prominent iron lines in the literature \citep[e.g., Ton S180,][]{Nardini12}, the broad components in those cases ($\sigma \sim 400$\,eV, $EW \sim 200$\,eV) are still relatively stronger than what is observed in the bulk of our sample.
Modeling the soft excess with relativistic reflection in the absence of a broad iron line often requires adopting extreme physical conditions, such as high disk densities ($n \sim 10^{19}$\,cm$^{-3}$), which can lead to parameter degeneracies and less robust constraints \citep[see e.g.,][]{Madathil24}. On the other hand, the warm corona model only depends on two main parameters (temperature and optical depth) and works well for featureless spectra. We therefore follow this approach for its simplicity and because it allows us to compare with similar samples, especially in the local Universe.

To best characterize the soft and hard X-ray emission of our sources, we test two different models, including the following components, using XSPEC notation, as:  

\begin{itemize}[label=\textbullet]
    \item Model-1: \texttt{const$\cdot$TBabs$\cdot$XABS$\cdot$(compTT+xillver)}
    \item Model-2: \texttt{const$\cdot$TBabs$\cdot$XABS$\cdot$(nthcomp+xillverCp)}
\end{itemize}

The choice of adopting two different model components for the soft and hard X-ray bands is primarily motivated by their distinct, yet comparable, descriptions of the continuum, which result in different free parameters.


We model the soft excess in the warm corona scenario using either the \texttt{compTT} (Model-1) or \texttt{nthcomp} (Model-2) components, while the primary continuum and its associated reflection are accounted for by either \texttt{xillver} (Model-1), or \texttt{xillverCp} (Model-2) from the \texttt{relxill} package\footnote{\url{https://www.sternwarte.uni-erlangen.de/~dauser/research/relxill/}}.
The combination of model components follows a logical rationale. In particular, \texttt{nthcomp} is paired with \texttt{xillverCp} because both describe the continuum using the same Comptonization framework, ensuring consistency between the soft and hard X-ray bands.
This combination is often used \citep[e.g.,][]{Middei18,Porquet18,Petrucci18}, allowing us to directly compare our results with the literature. However, \texttt{nthcomp} is not the only Comptonization model used in the literature; testing \texttt{compTT} provided an alternative set of spectral parameters and also enabled direct comparison with results from other studies. We note that the \texttt{nthcomp} model component has been superseded by the newer convolution model \texttt{thcomp}\footnote{\url{https://heasarc.gsfc.nasa.gov/docs/software/xspec/manual/node322.html}}. However, to fully exploit this model, a thorough modeling of the optical–UV seed photons spectrum must be carried out, including e.g., \xmm Optical Monitor data.
A dedicated and comprehensive study of the broad-band SED of SUBWAYS sources will be presented in a forthcoming paper (Gianolli et al., in preparation).

The constant (\texttt{const}) accounts for the cross-calibration of the detectors on board \xmm and \nustar. 
This constant is fixed to 1 for the PN data, while it is left free to vary for the MOS detectors.
For the \nustar modules (FPMA and FPMB), the \texttt{const} values are determined by fitting \nustar and PN data jointly in the overlapping 3–10 keV band using a simple power-law model, with tied photon index. The best-fit value of \texttt{const} found in the 3-10 keV energy band is then fixed in the broadband spectral fitting to account for flux variability between the \xmm and \nustar pointings and to avoid adding another free parameter strongly correlated with reflection intensity and the high-energy cut-off.
\begin{figure}[!t]
    \centering
    \includegraphics[width=0.9\linewidth]{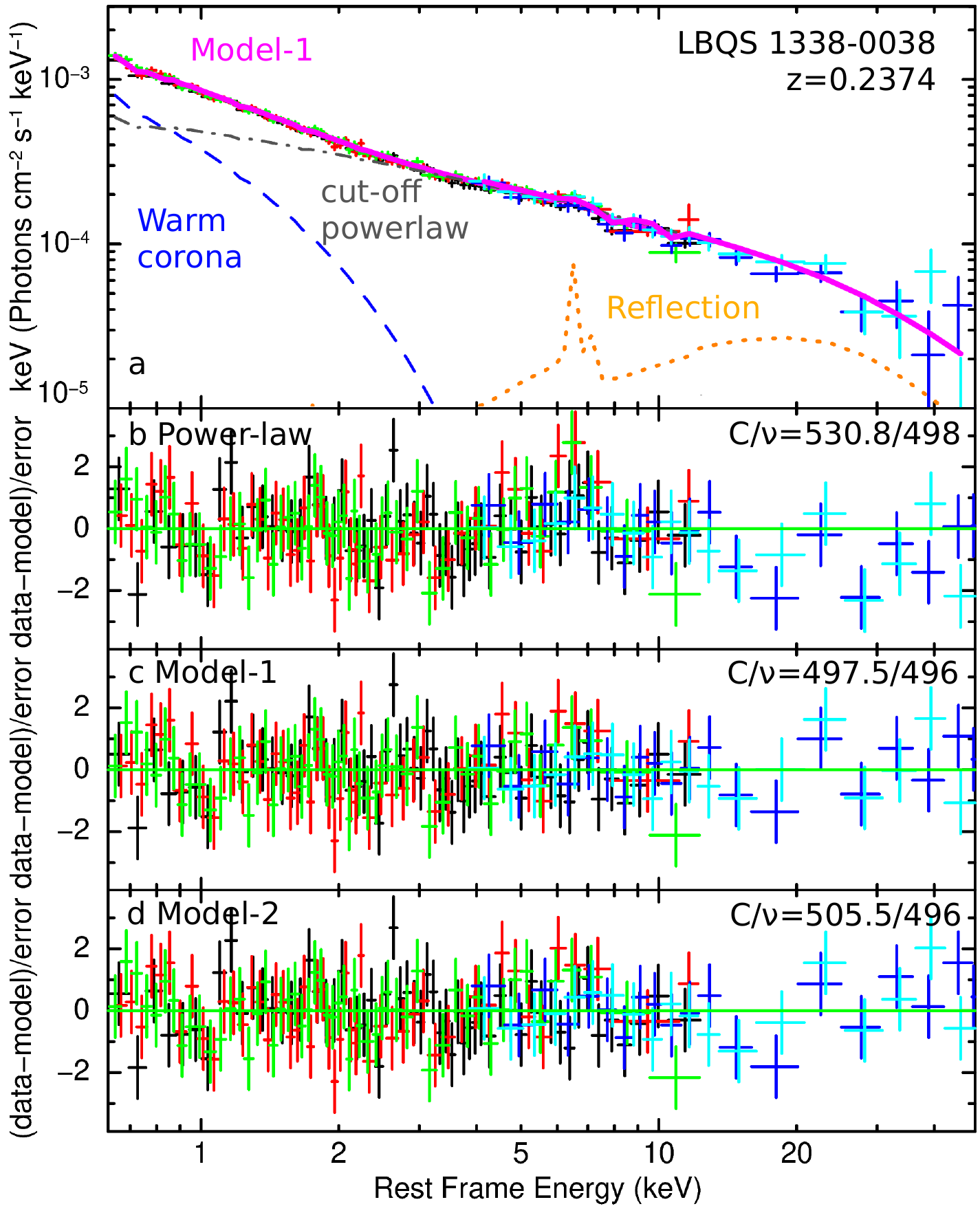}
    \caption{ \footnotesize {\bf a:} Spectrum of LBQS1338-0038 fitted with Model-1. Data points are PN (black), MOS1 (red), MOS2 (green), FMPA (blue), and FMPB (cyan). 
    The warm corona component (\texttt{compTT}) is represented by the dashed blue curve, while the reflection-only component from \texttt{xillver} is shown by the dotted orange curve. The primary cut-off power-law is shown in gray, and the total best-fit model for Model-1 is in magenta.  {\bf b:} Residuals with respect to a simple black body+power-law model from \citetalias{Matzeu23}, without any reflection or high-energy cut-off.
    {\bf c:} Residuals with respect to the fit with Model-1. 
   {\bf d:} Residuals with respect to the fit with Model-2. 
   The improvement in the fit is significant when reflection and high-energy cut-off are included ($\Delta Cstat\sim30$ for two more parameters), while both Model-1 and Model-2 provide statistically acceptable fits.
    }
    \label{Spectra}
\end{figure}
The Galactic absorption is modeled with \texttt{TBabs} \citep{Wilms00} adopting the hydrogen column densities specific for each source line of sight, derived from the HI4PI Collaboration maps \citep{HI4PI16}. 
A fully covering, mildly ionised (warm) absorber is needed to reproduce the observed spectra of 7 out of 23 sources, and we include it using \texttt{XABS} tables \citep{Kaastra96spex, Steenbrugge03xabs}  with column density ($N_H$) and ionization parameter ($\log \xi$\footnote{$\xi=\frac{L}{n_Hr^2}$, where L is the ionizing luminosity, $n_H$ is the Hydrogen column density and r is the distance from the ionizing source}) free to vary. The resulting values are consistent with previous findings reported in \citetalias{Matzeu23}.

The specific models properties are discussed in \Cref{sec:Model-1} and  \Cref{sec:Model-2}. The spectral features originating in the reflection of the primary radiation on the accretion disk (Compton hump and Iron K$\alpha$ emission line) are self-consistently included in both models.
In the following, we describe the two models and their free and fixed parameters in detail.
\begin{figure*}[t!]
     \centering
     \includegraphics[width=0.48\linewidth]{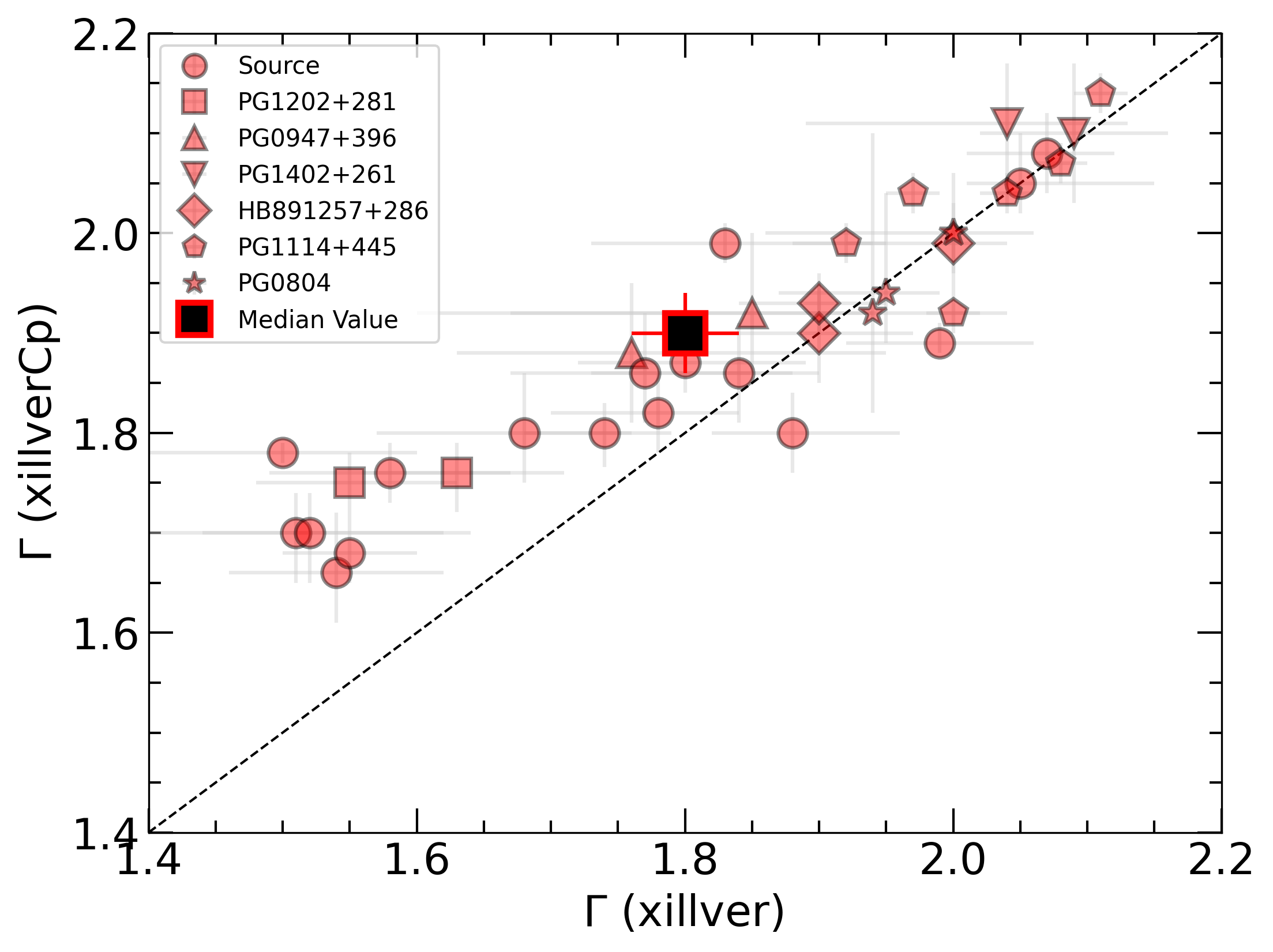}
     \hspace{0.5cm}
     \includegraphics[width=0.48\linewidth]{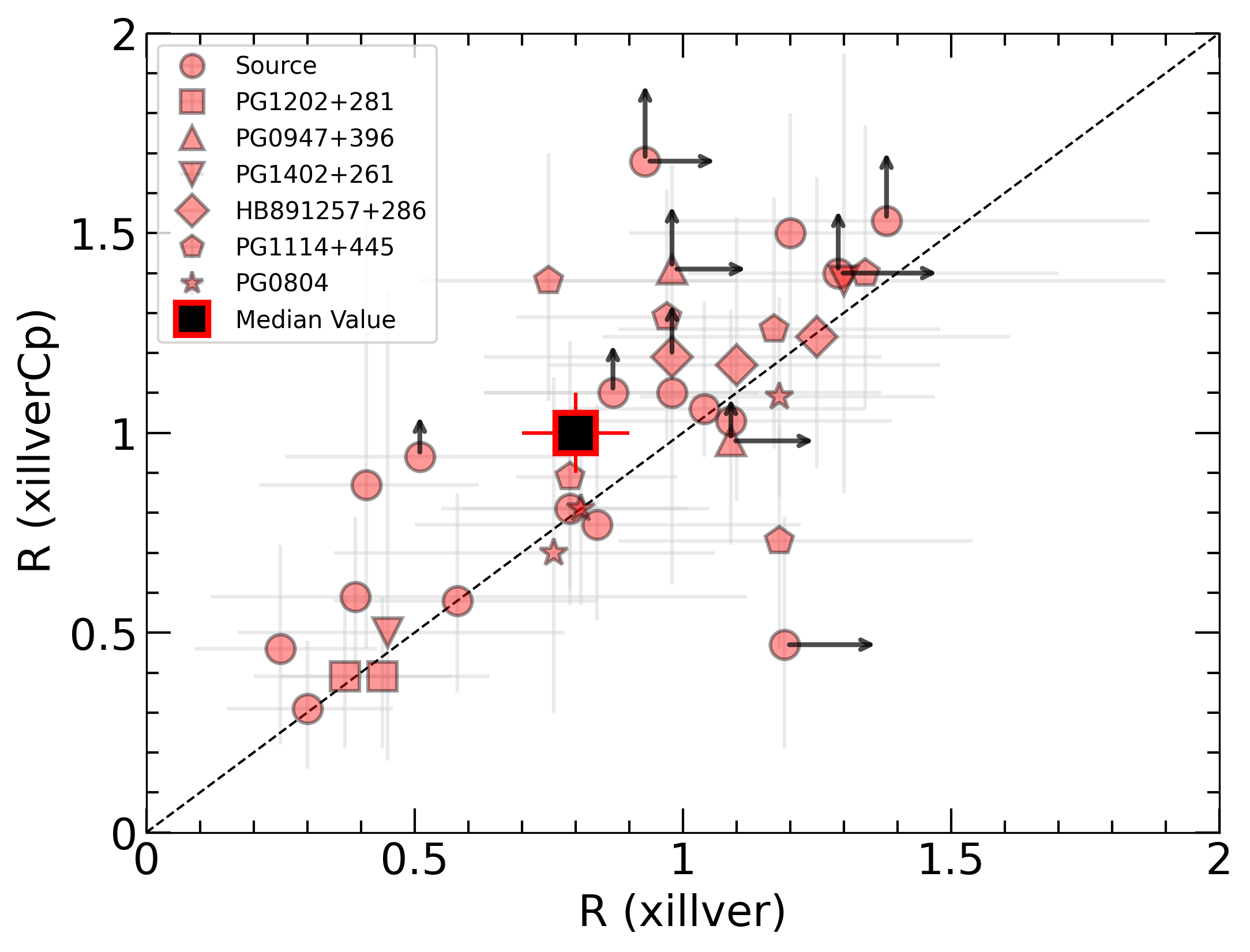}
     \caption{\footnotesize 
      Comparison of the power-law photon index (left side) and of the reflection parameter (right side) among the two models, \texttt{xillver} (x-axis) and \texttt{xillverCp} (y-axis). The single observations are marked as Source with a circle, while multiple observations for a single source are distinguished by the markers and they are explicit in the legend. In black the median value with errors at $90 \%$ confidence level computed with parametric fit using a log-normal function (see \Cref{sec:hotcoronaproperties} for further details).    
     }
     \label{Gamma_R_comparison}
 \end{figure*}

\subsection{Model-1}
\label{sec:Model-1}
In Model-1, the soft excess is parameterized by the \texttt{compTT} model \citep{Titarchuk94} by solving the Kompaneets equation \citep{Kompaneets57}.
The \texttt{compTT} free parameters are the electron temperature ($kT_e^{\text{warm}}$) and optical depth $\tau$.
The spectral distribution of the seed photons accounts for the UV thermal emission of the accretion disk and follows a Wien law. The seed photon temperature is fixed at $kT_{bb}=0.01$ keV. The dependence of the results on this assumption is negligible: the change in the slope of the Comptonized spectrum is less than $\sim0.1$ for seed photons temperatures in the range 5-500 eV \citep{HaardtMaraschi91}. 


The primary emission and its associated reflection features are modeled using the \texttt{xillver} model, which consists of a phenomenological power-law with an exponential cut-off and incorporates non-relativistic reflection off an optically thick, ionized constant density slab \citep{Garcia10,Garcia13}. The reflection parameter $R$ is defined as the ratio of the intensity of the reflected emission to that of the directly observed primary continuum, in the source’s rest frame. For a semi-infinite, plane-parallel disk, a value of $R = 1$ implies equal intensity between the reflected and direct components. The assumed geometry for the hot corona is point-like within the lamp-post framework. 
Several parameters in \texttt{xillver} are fixed to standard values to reduce the number of free parameters: the iron abundance is set to Solar, and the disk inclination angle is fixed at to $60^\circ$.  
Although Type-1 AGN are typically observed at lower inclinations, we adopt $60^\circ$ as a representative value for an isotropic distribution, having no information on the actual inclination of the systems in our sample; we note that assuming a more face-on geometry would enhance the observed reflected emission, therefore requiring systematically lower best-fit R values (typically by 50\% going from $60^\circ$ to $30^\circ$) to reproduce the same spectral features, without significantly impacting the overall spectral shape or the high-energy cut-off measurement.

The disk density is fixed to the default value of $n_H \sim 10^{15}$ cm$^{-3}$. The ionization parameter ($\xi$) can dramatically alter the shape of the reflected spectrum. However, in the case of SUBWAYS spectra, there is no clear evidence of strong ionized Fe emission lines above the $6.4$ keV Fe k$\alpha$ emission\footnote{With the possible exception mentioned at the beginning \Cref{sec:Spectral Analysis}, see \citetalias{Matzeu23} for details.}.
In fact, when left as a free parameter, the best fit values are always consistent with neutral gas. Therefore, for the sake of simplicity we fixed it at $\rm log\xi=0$.

\subsection{Model-2}
\label{sec:Model-2}
In Model-2, the soft excess is modeled with \texttt{nthcomp}  \citep{Zdziarski96,Zycky99} which describes the thermal Comptonization of UV photons from the accretion disk by warm electrons, while the continuum is described by the \texttt{xillverCp} component \citep{Garcia13,Garcia14}. \texttt{XillverCp} employs a Comptonization continuum described by the same \texttt{nthcomp} model used for the warm corona. The fixed parameters remain the same as in \texttt{xillver}, but it returns the plasma electron temperature  $kT_e^{hot}$, instead of the high energy cut-off $E_{cut}$.\\

The \texttt{nthcomp} free parameters are the electron temperature ($kT_e^{warm}$) of the scattering plasma and the photon index ($\Gamma$) of its power law emission. The slope $\Gamma$ serves as a proxy for $\tau$, indeed the latter is not an explicit parameter , but it can be derived
using  relations from \citet{Longair11} such as 
 \begin{equation}
\Gamma = -\frac{1}{2}+\sqrt{\frac{9}{4}+\gamma},
\label{Longair}
\end{equation}
 where $\gamma$ depends on the Compton $y$--parameter, and it can be expressed for spherical geometry as:
\begin{equation}
\gamma_{sphere} = \frac{\pi^2}{3} \ \frac{1}{\theta \cdot(\tau + \frac{2}{3})^2},
\label{Longair_sphere}
\end{equation}
and for slab geometry as:
\begin{equation}
\gamma_{slab} = \frac{\pi^2}{12} \ \frac{1}{\theta\cdot (\tau + \frac{2}{3})^2}.
\label{longair_slab}
\end{equation}

The dimensionless electron temperature $\theta$ is defined as:
\begin{equation}
\theta = \frac{kT_e}{m_e c^2}
\label{theta}
\end{equation}
where $m_e$ is the electron mass and $c$ the speed of light.
It is also possible to use the \citealt{Beloborodov99} approximations to recover $\tau$ (\Cref{gamma_tau} and \Cref{Compton_parameter}). In this model component, it is possible to specify the seed photon spectral distribution (blackbody or disk-blackbody), and we choose the disk-black body consistently with \texttt{compTT} choice.


\section{Results}
\label{sec:Results}

Broadband spectral fitting is performed for all sources in our sample; as a representative case, we show in \Cref{Spectra} the data and the unfolded best-fit model for LBQS1338. The residuals are shown relative to the baseline model from \citetalias{Matzeu23} consisting of a black body component for the soft-excess, and a power law for the primary continuum (panel b), to model-1 (panel c), and model-2  (panel d; see caption for details). 
From model-1 and 2 best fits, we derive hot and warm corona parameters. In particular, we refer to the best fits of Model-1 to derive $\Gamma$, $E_{\text{cut}}$ and $R$, and to those of Model-2 to derive $\Gamma$, $kT_e^{hot}$ and $R$.
For the warm corona, the physical properties obtained are the electron temperature ($kT^{\text{warm}}_e$) for both models, while $\tau$ is obtained from Model-1 and $\Gamma$ from Model-2.  Parameters that are set to a fixed value are presented in \Cref{sec:Model-1}, \Cref{sec:Model-2} for Model-1 and Model-2, respectively.
\begin{figure*}[t!]
    \centering
        \includegraphics[width=0.48\linewidth, height=6cm]{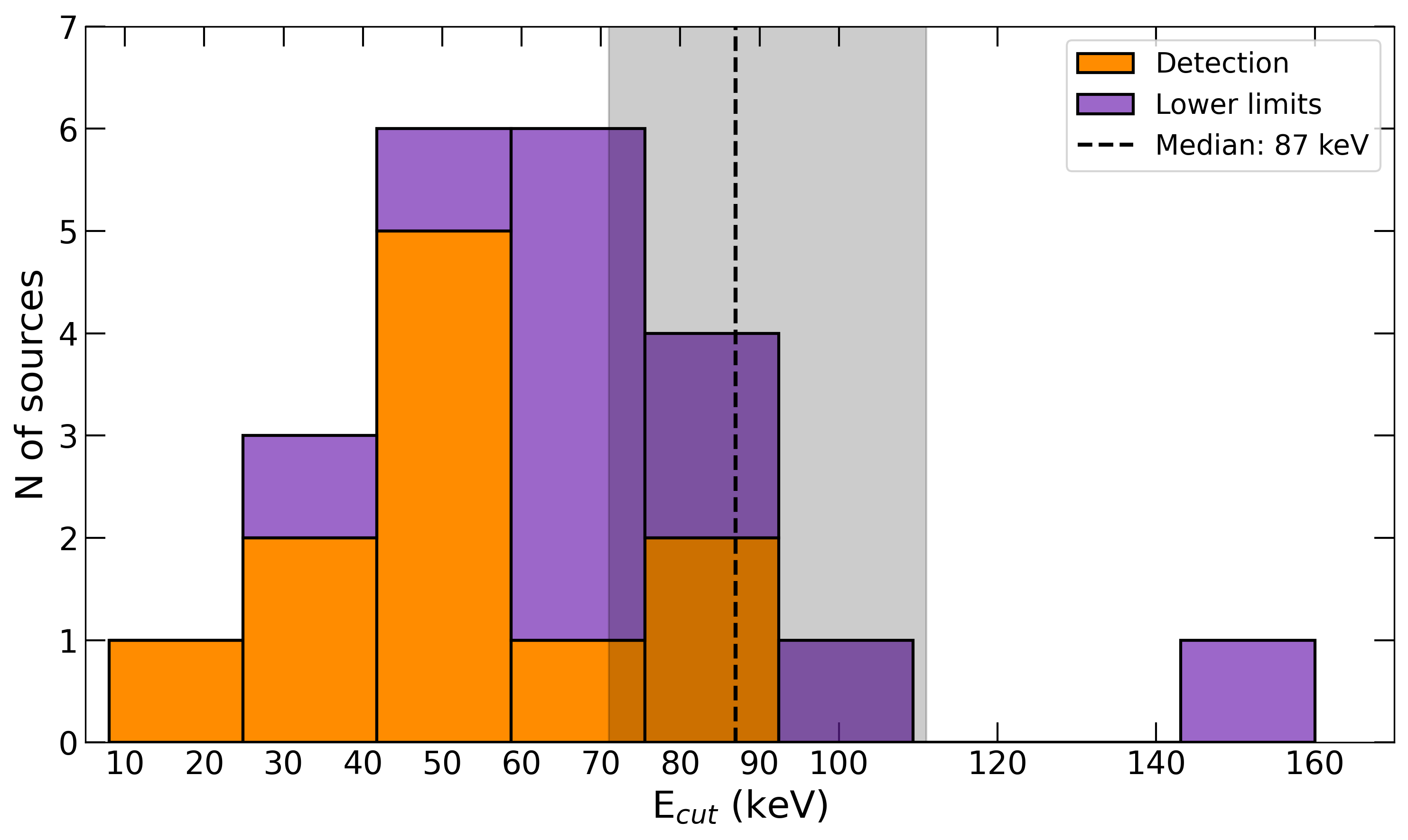}
        \hspace{0.5cm}
        \includegraphics[width=0.48\linewidth, height=6cm]{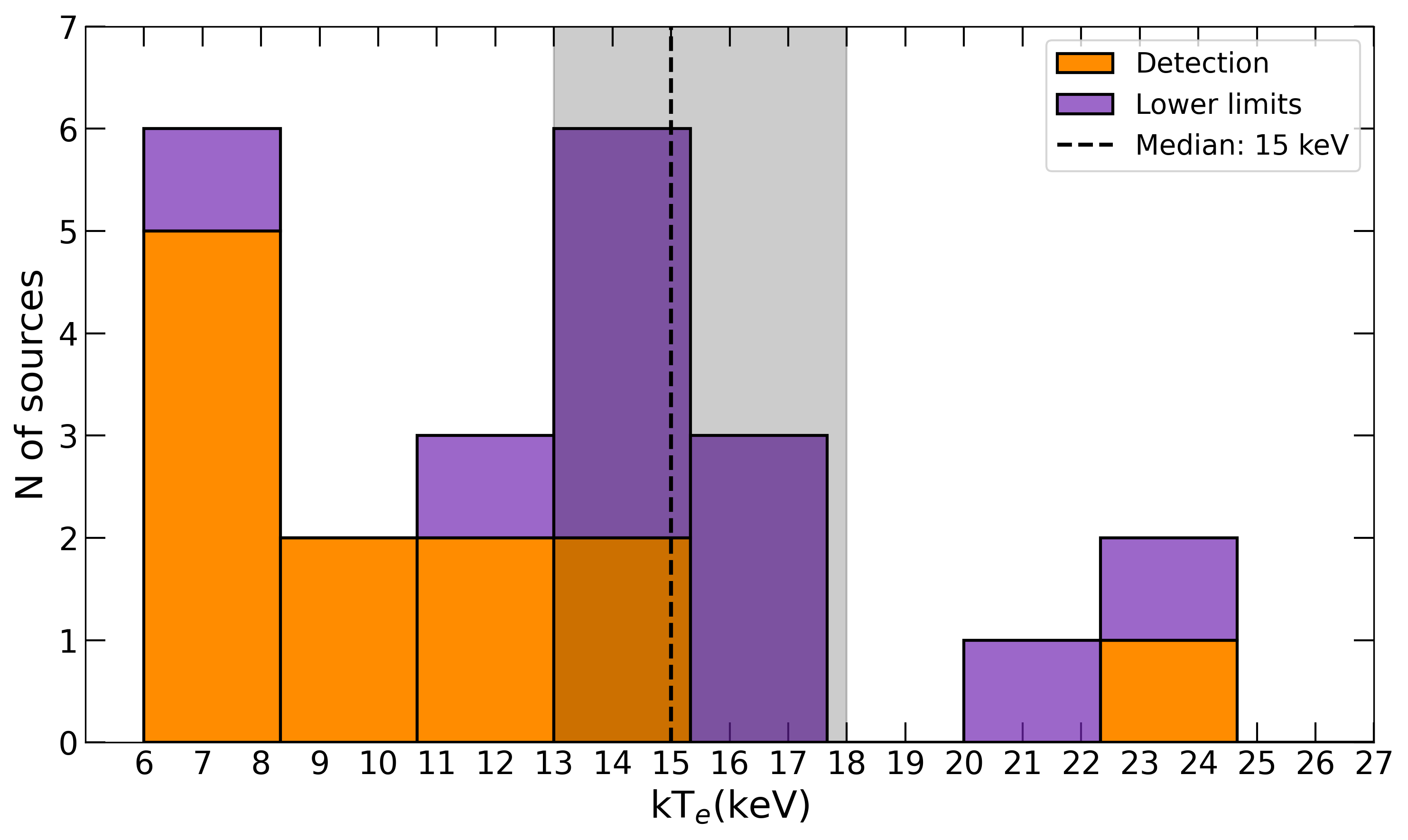}
    \caption{\footnotesize \textbf{Left}: High-energy cut-off distribution for the hot corona using \texttt{xillver}.   
    \textbf{Right}: Electron temperature distribution for hot corona from \texttt{xillverCp}. Detections are shown in orange, while lower limits are shown in violet, stacked over the detections. The median values, taking into account lower limits, are marked by the dashed line with $68\%$ confidence level (light grey areas) and are computed by using the parametric fit using a log-normal function(see \cref{sec:hotcoronaproperties} for more details).}
    \label{Ecut_kt}
\end{figure*}

The parameters $\Gamma,~\tau$ and kT$_e$ are related through standard comptonization models formulae. 
For this work we adopt the equations of Section 3.2. In the following, we present the best-fit values obtained from the spectral analysis for each source of the sample.

\subsection{Continuum properties: \texorpdfstring{$\Gamma$, R}{Gamma, R}}
\label{sec:Results_gamma}
The best-fit values for the continuum parameters $\Gamma$ and $R$ are reported in \Cref{Model2} for Model-1 and in \Cref{Model3} for Model-2.
The median value of the photon index peaks at approximately $\Gamma\sim1.8$, in agreement with the typical values observed in AGN (e.g., \citealt{NandraPounds94,Piconcelli04}).
The two models provide consistent estimates within the 90$\%$ confidence error for the $\Gamma$ parameter for most of the sources, although \texttt{xillverCp} tends to find steeper photon indices, as shown in \Cref{Gamma_R_comparison} (left), especially for flat spectra ($\Gamma\lesssim1.8$). This is a known effect (see e.g., \citealt{Middei18,Diaz20,Middei20_Mrk359}) due to the sharper high-energy curvature of a full thermal Comptonization spectrum like \texttt{nthcomp} used in \texttt{xillverCp}, with respect to the more gradual rollover of the cut-off power-law. 
In the 10–30 keV band, this gives a stronger curvature for the same $\Gamma$ and $E_{\text{cut}}/kT_e$ pairs, so fits with \texttt{xillverCp} tend to compensate with a steeper $\Gamma$, especially for hard spectra.

Then, we analyze the distribution of the reflection parameter $R$ and compute its median for each model. 
In \Cref{Gamma_R_comparison} (right), the R values for the two models are compared. The results are broadly consistent between the two models, with a distribution centered around $R\sim1$, with a significant spread (R$\sim0.5-1.5$).
For Model-1, the reflection intensity is constrained in 27 out of 33 observations ($81\%$), with the remaining measurements providing lower limits. Similarly, for Model-2, the reflection component is constrained in 24 out of 33 observations ($73\%$), with the remaining measurements providing lower limits. 
In both models, values of $R>1$ are found in some source.
Although values of R could in principle exceed unity, reflection fractions much larger than 1 would require ad hoc explanations, including source variability and deviations from the assumption of an infinite planar reflector. We therefore restrict the explored range to $R\leq 2$.

We note that the reflection fraction R obtained with Model-2 (xillverCp) is significantly higher than that inferred with Model-1 in several sources. One possible explanation is that the Comptonized illuminating continuum in xillverCp is intrinsically more curved than a phenomenological cutoff power law, resulting in a weaker Compton hump per unit incident flux. The higher R, therefore, compensates for differences in the assumed continuum shape.
We also note that the median R value derived here for SUBWAYS
(median R$\sim0.8-1.0$ depending on the model adopted)
is higher than what is found in comparably luminous QSOs detected in deep \nustar surveys (median R$=0.43$) \citep{Zappacosta18} (derived with the same inclination angle of $60^\circ$), although with a similar large spread 
($0.5-1.5$ vs. $0.06-1.5$ inter-quartile ranges). SUBWAYS spectra, however, have much higher quality, which allows us to simultaneously constrain R and the high-energy cut-off (fixed to 200 keV in \citealt{Zappacosta18}), therefore obtaining much narrower constraints on R.

\subsection{Hot corona properties: \texorpdfstring{$E_{cut}$, $kT_e^{hot}$}{Ecut, kTehot}}
\label{sec:hotcoronaproperties}
The results for $E_{cut}$ and $kT_e^{hot}$ are listed in \Cref{Model2} and \Cref{Model3}, respectively. We first analyze the $E_{\text{cut}}$ distributions obtained from Model-1 in \Cref{Ecut_kt} (left). 
The high-energy cut-off $E_{\text{cut}}$ is constrained in 13 out of 33 observations ($39\%$), while for the remaining sources, we measured lower limits in the range 25-150 keV. 

To properly account for lower limits in the results, we employ the survival analysis tool Asurv \citep{Asurv1990} and its non-parametric Kaplan–Meier estimator (KM; \citealt{Kaplan1958}) to handle censored data. This computes the median and the 25th and 75th percentiles of the $E_{cut}$ distribution. Detections are treated as events, and lower limits are treated as right-censored entries. For sources with multiple observations, we used a single representative value per source. When multiple measurements yielded well-constrained values, we adopted the median; when only lower limits were available, we conservatively adopted the lowest limit.

However, the Kaplan–Meier estimator is limited in its ability to constrain the median when the sample is heavily censored (censoring rates $>50\%$). The SUBWAYS sample falls into this regime, with $60\%$ of the $E_{cut}$ values being lower limits. To robustly recover the median in this high-censoring regime, we also tested a parametric approach: we modeled the $E_{cut}$ distribution with four different parametric forms: Exponential, Weibull, Log-Normal, and Log-Logistic. These models were fitted to the data distribution using Maximum Likelihood Estimation via the {\sf lifelines} Python package \citep{DavidsonPilon2019}. Model selection was performed using the Akaike Information Criterion (AIC; \citealt{Akaike74}) and the Bayesian Information Criterion (BIC; \citealt{Schwarz78}). The Log-Normal model consistently provided the lowest AIC/BIC scores and was therefore adopted as our fiducial parametric description of $E_{cut}$.
The median of the distribution was then estimated directly from the best-fit Log-Normal function. To estimate the uncertainty at $1\sigma$ around the median, we employed a bootstrap technique with $N=1000$ resamples. In each iteration, the data were resampled with replacement (preserving the censoring flags), the model was refit, and the median was recorded. 

We report the central 68\% confidence interval of the median value, derived from this bootstrap distribution: $E_{cut}\sim87\ (71-111)$ keV, which is marked with a dashed line in \Cref{Ecut_kt} (left). For comparison, Asurv provides a median of 71 keV, with the 25th-75th percentiles ranging from 50 to 115 keV.
In both cases, this value is lower than the typical values ($\sim 150-200$ keV) measured in the literature for local AGN \citep{Fabian2015, Ricci2018, Tortosa18a, Balokovic20}. We investigate this difference further in \Cref{sec:Discussion}.

\begin{figure*}[!t]
\centering
\includegraphics[width=0.48\linewidth,height=6.5cm]{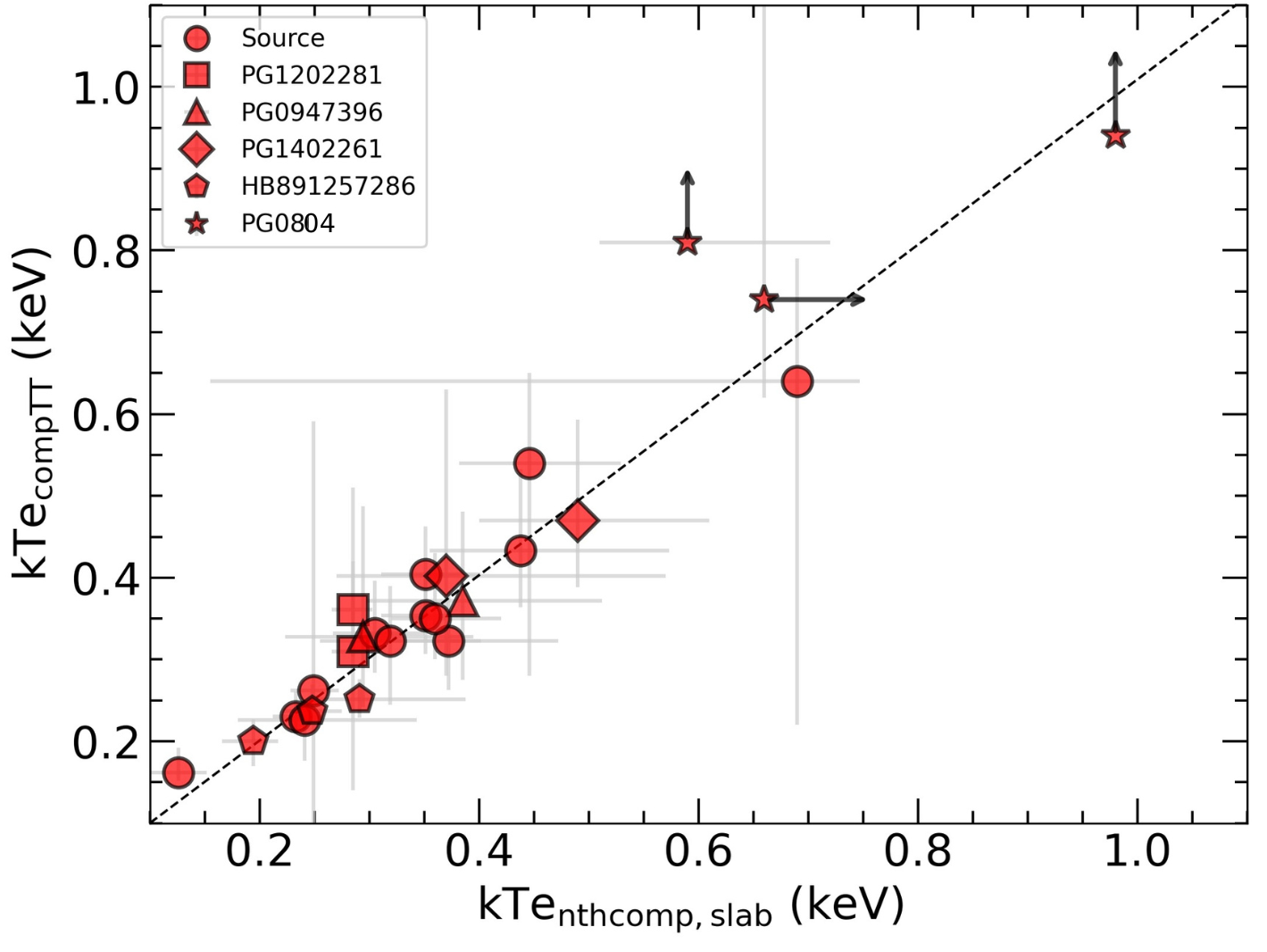}
\hspace{0.5cm}
\includegraphics[width=0.48\linewidth,height=6.5cm]{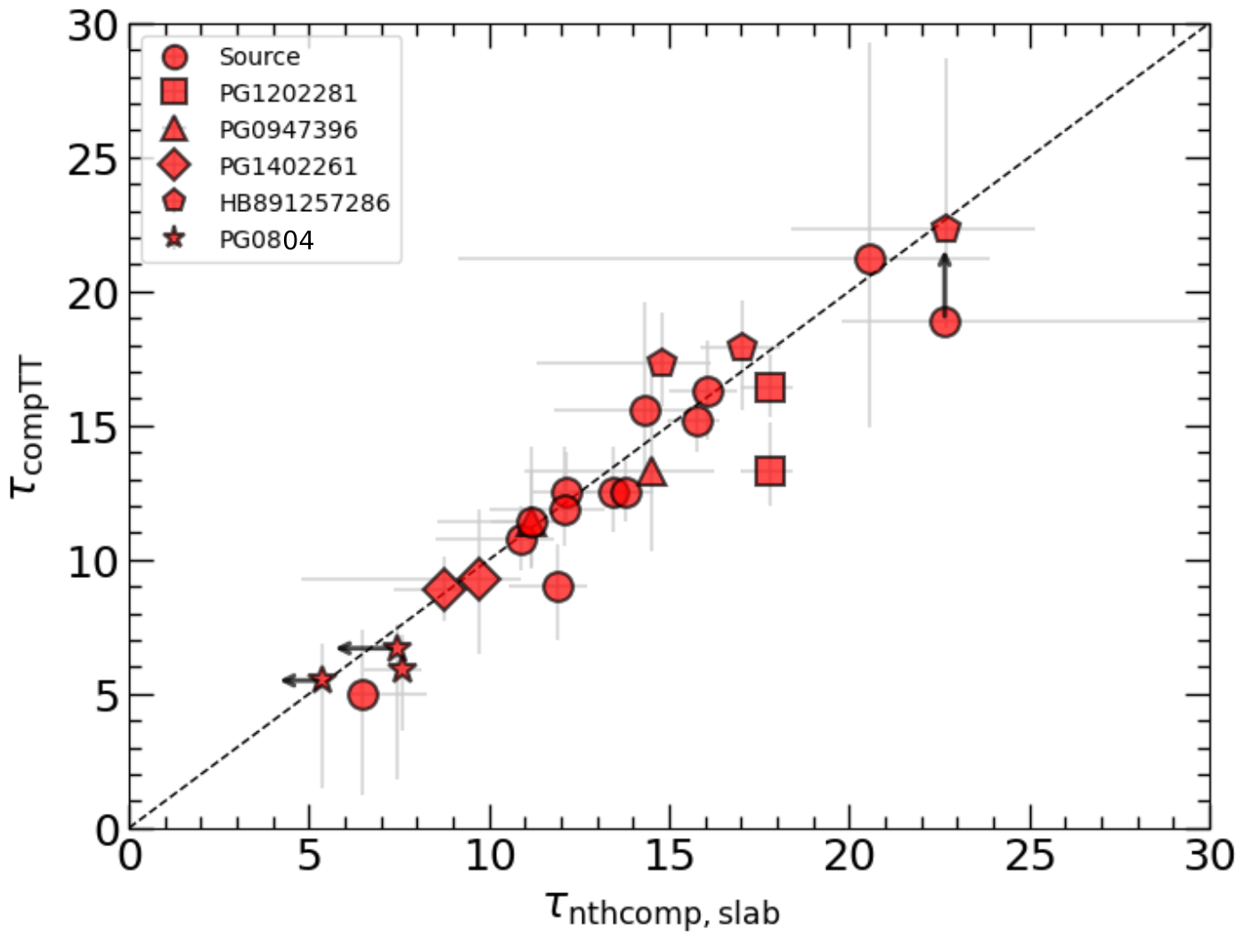}
\caption{\footnotesize \textbf{Left}: Comparison of the warm corona electron temperatures. On the x-axis, the temperatures obtained from \texttt{nthcomp} model, while on the y-axis, the ones obtained by \texttt{compTT} model, are shown.
\textbf{Right}: Comparison of the warm corona optical depths obtained by using \texttt{nthcomp} on the x-axis and \texttt{compTT} on the y-axis. For the former, $\tau$ is computed using \cref{longair_slab} (slab geometry) and considering $\Gamma$ and $kT_e^{warm}$ values from \texttt{nthcomp}.
}
\label{kte_warm}
\end{figure*}

A direct estimate of the electron temperature is obtained through Model-2.
We constrained the electron temperature $kT_e^{hot}$ in approximately one-third of the observations, 12 out of 33 ($36\%$), while the remaining observations provide lower limits in the range of 4-30 keV.
The median value and the errors at $68\%$ confidence level are $kT_e^{\text{hot}} \sim 15 \ (13-18)$ keV, marked by the dashed line in \Cref{Ecut_kt} (right), estimated by appropriately accounting for the lower limits using the parametric fit with a Log-Normal curve as described above.
For comparison, the results obtained with Asurv are 14 keV, with 9-19 keV as the 25th-75th percentiles. 

This value is lower than typical values reported in the literature ($\sim50-80$ keV, e.g. \citealp{Kamraj2022,Pal2024,Serafinelli2024}) where \texttt{xillverCp} was applied to large samples of local Seyferts.
Furthermore, this value is much smaller than what is expected from the results of Model-1, with a typical ratio E$_{cut}/kT_e\sim5$, while E$_{cut} = 2 - 3 \  kT_e$ is usually assumed  \citep{Petrucci01}.
We therefore investigate in more detail the possible origin of such a discrepancy, performing a recovery analysis in which simulated spectra with known input parameters are fitted with the same models as those used for the real data (see \Cref{simulations} for more details). In \Cref{simulation_ecut} we show the relation between the input $E_{cut}$ and $kT_e$ and the median of the recovered values, for the case $R = 1$ (results for other reflection fractions, in the range R$=0-2$ are comparable), as well as the input vs. fitted reflection fraction for completeness.
These simulations indicate that, while $E_{cut}$ can be robustly recovered up to $\sim100$ keV, \texttt{xillverCp} recovers coronal temperatures of at most $\sim15-20$ keV even for input values of up to 100 keV, highlighting the intrinsic limitations of such a model in the flux-redshift range of SUBWAYS: we speculate that it is the low redshift of the sample, coupled with the relatively faint fluxes (meaning \nustar data have constraining power only up to $\sim25$ keV) and with the sharper high-energy curvature of the Comptonization model, that prevents us from recovering higher $kT_e$ values. In contrast,  
\texttt{xillverCP} fits of low-z very bright sources, where the full \nustar band up to 70 keV can be used, have shown best-fit $kT_e$ values in the whole range $10-400$ keV \citep{Kamraj18}, while fits 
of high redshift QSOs of \xmm plus \nustar data of similar quality to the SUBWAYS sample, have shown that it is possible to constrain $kT_e$ up to $\sim50$ keV, thanks to the redshift effect \citep{Lanzuisi19,Lanzuisi26}. 
All this suggests caution when interpreting results from \texttt{xillverCP} in the SUBWAYS sample. For this reason, in the following, we will adopt the classical approach ($E_{cut}=2 kT_e$) of deriving $kT_e$ from $E_{cut}$ whenever the electron temperature is considered.

As a final test, we investigate whether different coronal properties (i.e., the high-energy cut-off) impact the presence of UFOs, as detected in \citetalias{Matzeu23}. The median E$_{cut}$ for non-UFO and UFO sources, computed with the parametric fit, points to different distributions, being the median E$_{cut}=101.0$ keV ($82.1 - 133.8$ at 68\% confidence level) for non-UFO and E$_{cut}=67.4$ keV ($56.5 - 83.1$ at 68\% confidence level) for UFO sources. However, the small UFO sample size in SUBWAYS alone prevents us from drawing firm conclusions: a log-rank test shows no significant difference between the two samples (p-value = 0.47).
A thorough investigation of any effect linking coronal temperature and the launching of nuclear winds will require the exploitation of the full sample
of sources where a systematic search for UFOs has been carried out \citep[e.g.,][]{Tombesi10,Gofford13,Igo20} which is beyond the scope of this paper and is deferred to a forthcoming paper. For the rest of this paper, we will consider the corona properties of the full SUBWAYS sample.


\subsection{Warm corona properties: \texorpdfstring{$kT_e^{warm}$,$\tau$}{kTewarm,tau}}
\label{sec:Results_warm}

The best-fit parameters for the warm corona in the SUBWAYS sample are reported in this section. We measured the warm corona electron temperature, $kT_e^{warm}$, in 23 out of 33 observations ($67\%$).
The remaining observations did not require the inclusion of a soft excess component.
Both \texttt{compTT} and \texttt{nthcomp} can constrain the $kT_e^{warm}$ values at that $90\%$ confidence level.  For consistency, we computed the median value using the previously described methods. For sources with multiple observations, the median is first computed for each source, and the resulting values are then used to compute the median for the full sample. This approach ensures that sources with multiple observations are not dominating the final median value.

The results are listed in \Cref{Model2} and \Cref{Model3}, and the comparison of the electron temperatures and optical depths derived from the two models is shown in \Cref{kte_warm}.
In both models, the median electron temperature of the warm corona is found to be $kT_e^{\text{warm}} \sim 0.40$ keV (in the range 0.1-1 keV).
The temperatures obtained with the two models are consistent with each other (left panel of \Cref{kte_warm}). 

Regarding the optical depth, the median value obtained using \texttt{compTT} is $\tau\sim12$ (in the range $5-22$), 
corresponding to an optically thick medium. This value is consistent with the values ($\tau \sim 5-15$) reported in the literature for local Seyferts \citep{Petrucci18}, and with 
the optical depth inferred indirectly from the \texttt{nthcomp} best-fit results, following \Cref{Longair} for the slab case (see right panel of \Cref{kte_warm}).
We note that if, to derive $\tau$, we instead adopt the approximations given in \citet{Beloborodov99} and widely used in the literature \citep{Tortosa18a,Petrucci18,Zhang23}, the resulting estimates remain consistent with theoretical expectations for warm coronae ($\tau\sim  5-15$), but tend to overestimate the optical depth at large $\tau$ compared to the values from \texttt{compTT} (see discussion in the \Cref{tau_comparison}). 
The discrepancies can be mainly ascribed to the assumptions in the formula \citep{Beloborodov99}, in terms of approximation and adopted geometry (see also \cref{comptmodels}). We refer to a forthcoming publication for a detailed derivation of the warm corona properties and modeling of the SUBWAYS sample, including optical/UV data (Gianolli et al. in preparation).

\begin{figure*}[!h]
    \centering \includegraphics[width=0.47\linewidth,height=0.43\linewidth]{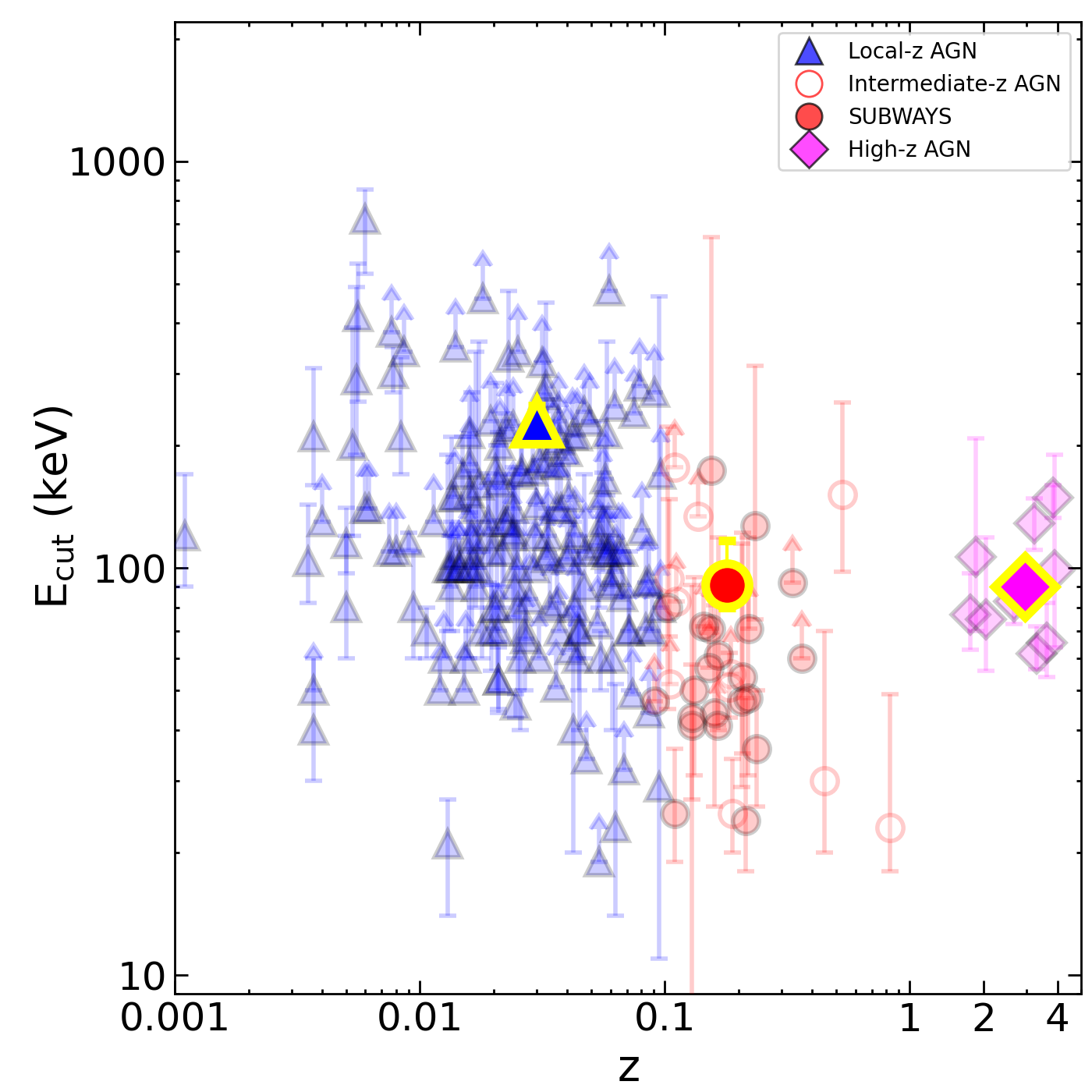}\hspace{0.5cm}    \includegraphics[width=0.47\linewidth,height=0.43\linewidth]{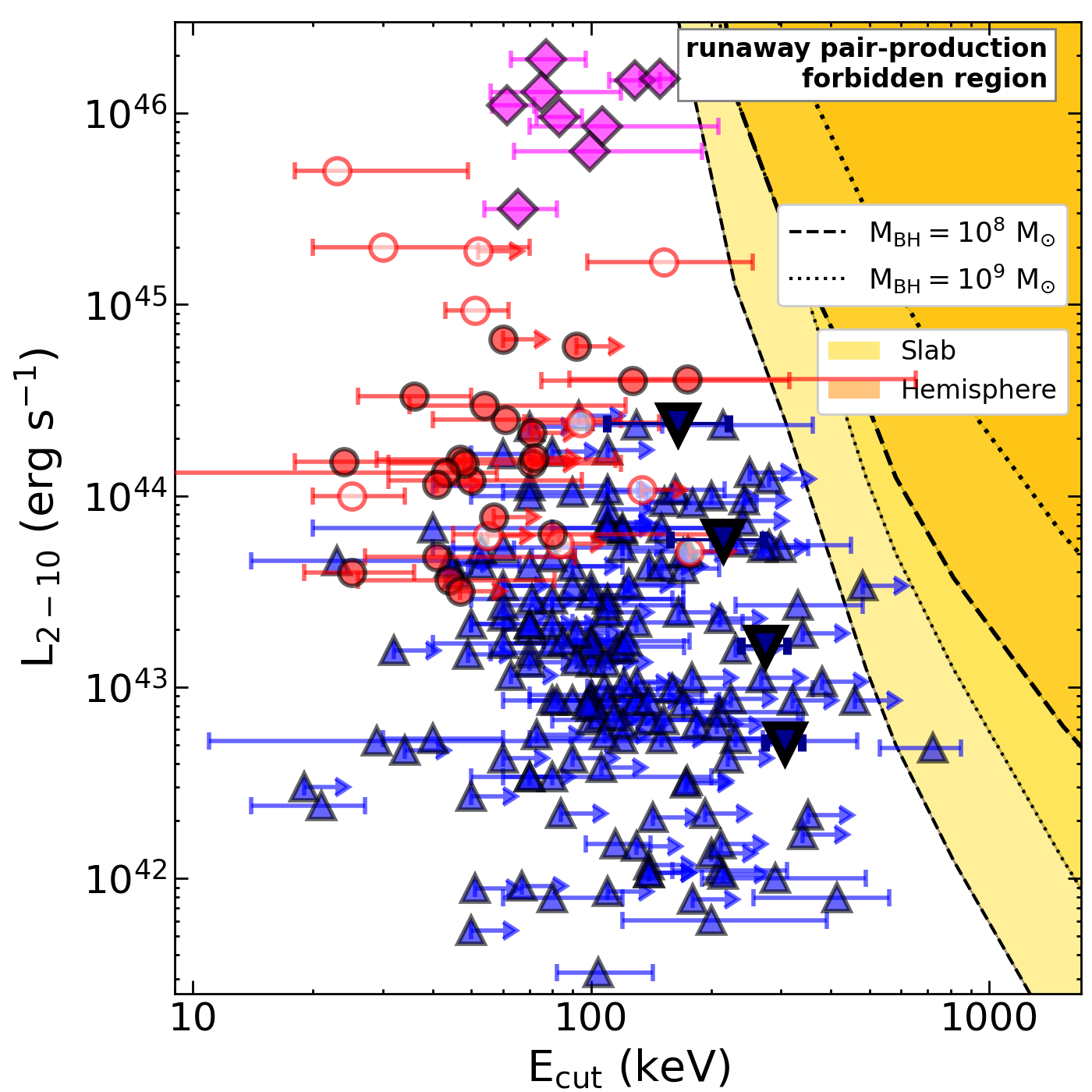}
    \caption{\footnotesize {\it Left panel}: $E_{cut}$ as a function of redshift for the low redshift sample (blue) as compiled by \cite{Bertola22}, the SUBWAYS sample  (red filled circles) the intermediate redshift AGN (in red open circles), and the high redshift sample (magenta diamonds) (Intermediate and High-z sources infos in \Cref{sample_infos}). The median point per redshift bin is overplotted with larger markers of the same colors of the redshift sample and yellow contours. 
    {\it Right panel}: The $E_{cut}$ as a function of the X-ray luminosity (i.e., the compactness-temperature ($l-\theta$) diagram, converted into observable quantities) from the same samples. Yellow regions mark the runaway pair-production limits for a BH mass of $10^8 M_\odot$ under slab (hemisphere) geometry, while dotted lines indicate the same limits for a BH mass of $10^9 M_\odot$. Dark-blue down triangles show the binned results from \citep{Ricci2017} on the BAT Sample at low-z.}
    \label{BertolaUpdated}
\end{figure*}

\section{Discussion}
\label{sec:Discussion}

The properties of the hot and warm corona have been extensively studied with \nustar for relatively large samples of low-luminosity, low-redshift Seyfert galaxies \citep[e.g.,][]{Fabian2015,Tortosa18a,Kamraj18,Balokovic_2020,Akylas21}. 
This work extends the analysis of the hot corona properties to a range of luminosities ($L_{bol}=2\times 10^{44}-2\times 10^{46}$ erg/s) and redshift ($z=0.1-0.4$) so far scarcely investigated, bridging the gap between local AGN and high-\( z \) luminous AGN. In the left panel of \Cref{BertolaUpdated}, the SUBWAYS sources (red filled points) are placed between local AGN (gray points) and the handful of high-redshift QSOs (magenta points), as collected in \citet{Bertola22}. To the latter, we added 6 sources from the ongoing WISSHFUL program (\citealp{Borrelli26} in prep.), where similar models were applied to \xmm and \nustar spectra of z=2-4 hyperluminous QSOs. Seyferts falling beyond z=0.1 are shown as empty red circles and grouped with SUBWAYS sources to populate the intermediate-redshift interval.

The median $E_{cut}$ values for each redshift sample, computed using the two methods explained above, are shown as larger symbols, with the same color used for each sample. We report the median from the parametric fit with its $1\sigma$ uncertainties. \Cref{redshift_division} shows the results for the three redshift bins and the two different methods.\\
To assess whether the distributions of high-energy cutoffs $E_{cut}$  
differ statistically between the sub-samples, we employed the non-parametric Log-Rank test \citep{Mantel1966}. This test compares the survival functions of two groups while properly accounting for censored data. For each pair, we tested the null hypothesis that the two samples are drawn from the same underlying population. The analysis was again performed using the {\sf lifelines} Python package, treating detections as observed events and lower limits as right-censored data. The second part of \Cref{redshift_division} displays the P-values for the null hypotheses obtained in this way. P-values in bold face show highly significant differences (P-value$<0.01$, i.e., significance $>99\%$).
These results clearly show that both the intermediate and high-z samples have significantly lower (at $>3\sigma$) median $E_{cut}$, and are both significantly different (at $>4\sigma$) from the low-z sample and similar to each other, according to the log-rank test. We stress that this difference is possibly even underestimated, as we are aware of cases where bright, well-studied sources with long \nustar observations do not have a reported E$_{cut}$ estimate in the literature because \nustar is not able to constrain it, being the value too high for the \nustar band (e.g., E$_{cut}\gtrsim600$ keV mentioned in \citealt{Keck15} for NGC~4151).

\begin{table}[!t]
    \caption{\footnotesize Comparison of median $E_{cut}$ results from Asurv and the fit with a Log-Normal function for the three redshift samples described in \Cref{sec:Discussion}. We have a total of 165 local AGN, 33 intermediate AGN (including the 23 SUBWAYS sources), and 9 high-z AGN.}
    \label{redshift_division}
    \centering
    \renewcommand{\arraystretch}{1.25}
    \begin{tabular}{cccc}
    \hline\hline
    \noalign{\vskip 0.5ex} 
    z   & N &  $E_{cut}$ (keV) & $E_{cut}$ (keV)\\
       & tot. (det.)   & Asurv &  Parametric fit\\  
     &   &50th (25th-75th)& Median ($1\sigma$ err.) \\
    \hline
    \noalign{\vskip 0.5ex} 
    $<0.1$ & 161 (57) &$228\ (117-428)$   & $228 \ (207-253)$\\
    $0.1-1.0$ & 35 (17) &$62\ (44-150)$   & $91 \ (80-117)$\\
    $1.0-4.0$ & 9 (9) &$80\ (68-105)$   & $90 \ (83-99)$\\ 
     \hline\hline 
     \multicolumn{3}{l}{Bin comparison}  & P-value\\
     \hline
\multicolumn{3}{l}{Low-z vs. Intermediate}  &   ${\bf 2.8\cdot 10^{-5}}$ \\
\multicolumn{3}{l}{Low-z vs. High-z}    &    ${\bf <1\cdot 10^{-5}}$ \\
\multicolumn{3}{l}{Intermediate vs. High-z}  &   $0.96$ \\
    \hline
    \end{tabular}
\end{table}

In the right panel of \Cref{BertolaUpdated}, the SUBWAYS sources are included in the $L_{x}-E_{cut}$ plane\footnote{Here the X-ray luminosities are the 2-10 keV band, absorption-corrected luminosities derived directly from the \xmm plus \nustar fits.}.
The addition of SUBWAYS sources and luminous QSOs appears to support the luminosity trend proposed by \citet{Lanzuisi19} and \citet{Bertola22}. At higher luminosities, the allowed range for $E_{cut}$ becomes increasingly limited by the pair production mechanism, resulting in a lower median value for more luminous sources.

One possible explanation is that more luminous, highly accreting sources, being more radiatively compact, experience more efficient cooling in their coronae because the pair-production mechanism acts as a thermostat, stabilizing the electron temperature, as proposed by \cite{Fabian2015,Ricci2018,Liu23}.
However, we note that a lower coronal temperature is already implied in the `standard’
Comptonization model that assume radiative equilibrium between
the disk and an optically thin corona \citep{HaardtMaraschi91}: 
the higher the accretion rate, the higher the density and hence $\tau$,
and the higher the cooling.

\subsection{Hot corona: \texorpdfstring{$E_{cut}$}{Ecut} vs. \texorpdfstring{$M_{BH}$, $L_{bol}$, $\lambda_{\rm Edd}$}{Mbh,Lbol, Lambda}   }

To better understand the dependence of $E_{cut}$ on intrinsic accretion parameters such as bolometric luminosity $L_{bol}$, $M_{BH}$ and $\lambda_{\rm Edd}$ ($L_{bol}$/$L_{Edd}$), we first obtained bolometric luminosities starting from the X-ray luminosities and applying the bolometric correction described in Eq.2 of \cite{duras20} for all the sources. Exceptions were made for luminous, X-ray selected sources, for which this correction would likely overestimate the bolometric luminosity. We used SED-based estimates for: 1) the local QSO SMSSJ11, where we adopted the results from \cite{Kammoun23}, and 2) for the WISSHFUL sample, where we used the SED-measured bolometric luminosities from \cite{Saccheo23}.
Then, we collected black hole mass estimates for the sample whenever available. We considered BH masses estimated from reverberation mapping and from virial methods applied to single-epoch measurements. We excluded those based on the M-$\sigma$ relation because less reliable on a source-by-source basis. The sample is reduced to 127 sources in total (including the 23 SUBWAYS sources), and all the relevant properties are listed in \Cref{sample_infos}.
This constitutes the largest sample of AGN with accretion properties and high-energy cut-off measurements ever analyzed in the literature, spanning five orders of magnitude in $M_{BH}$ and $\lambda_{\rm Edd}$, and six in $L_{bol}$.

\begin{figure}[t]
\includegraphics[width=0.95\linewidth,height=5.8cm]{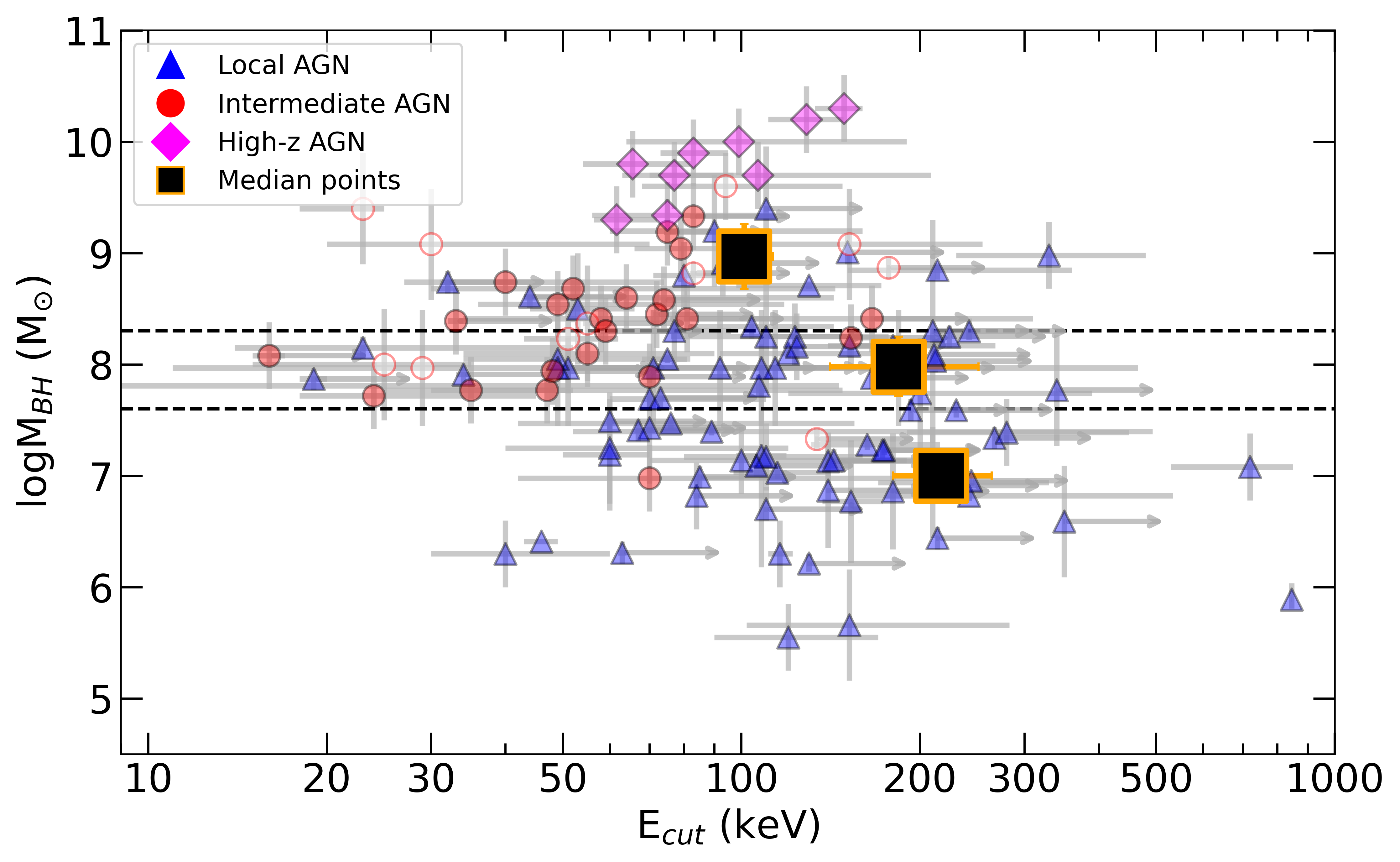}\vspace{0.5cm}
\includegraphics[width=0.95\linewidth,height=5.8cm]{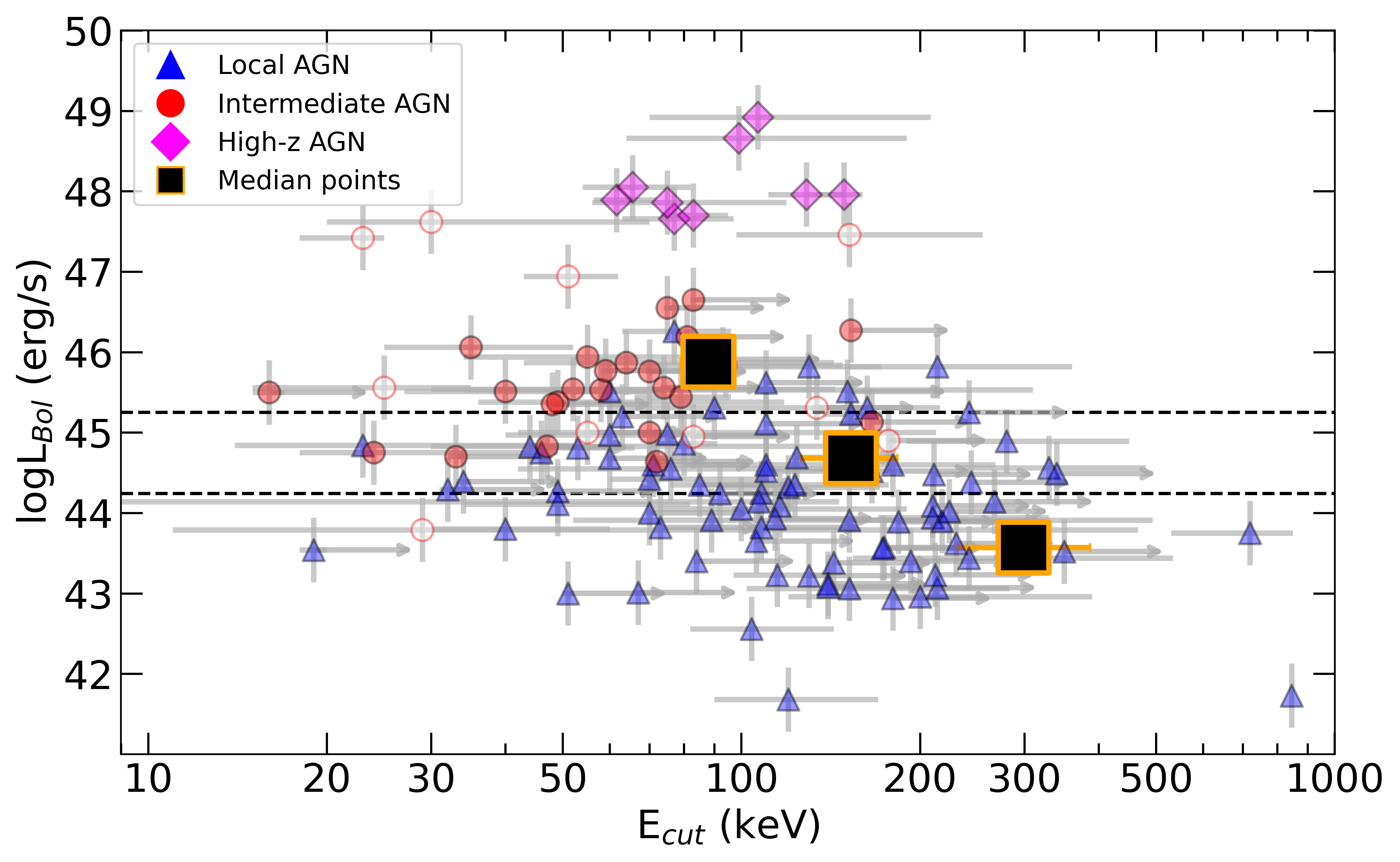}\vspace{0.5cm}
\includegraphics[width=0.95\linewidth,height=5.8cm]{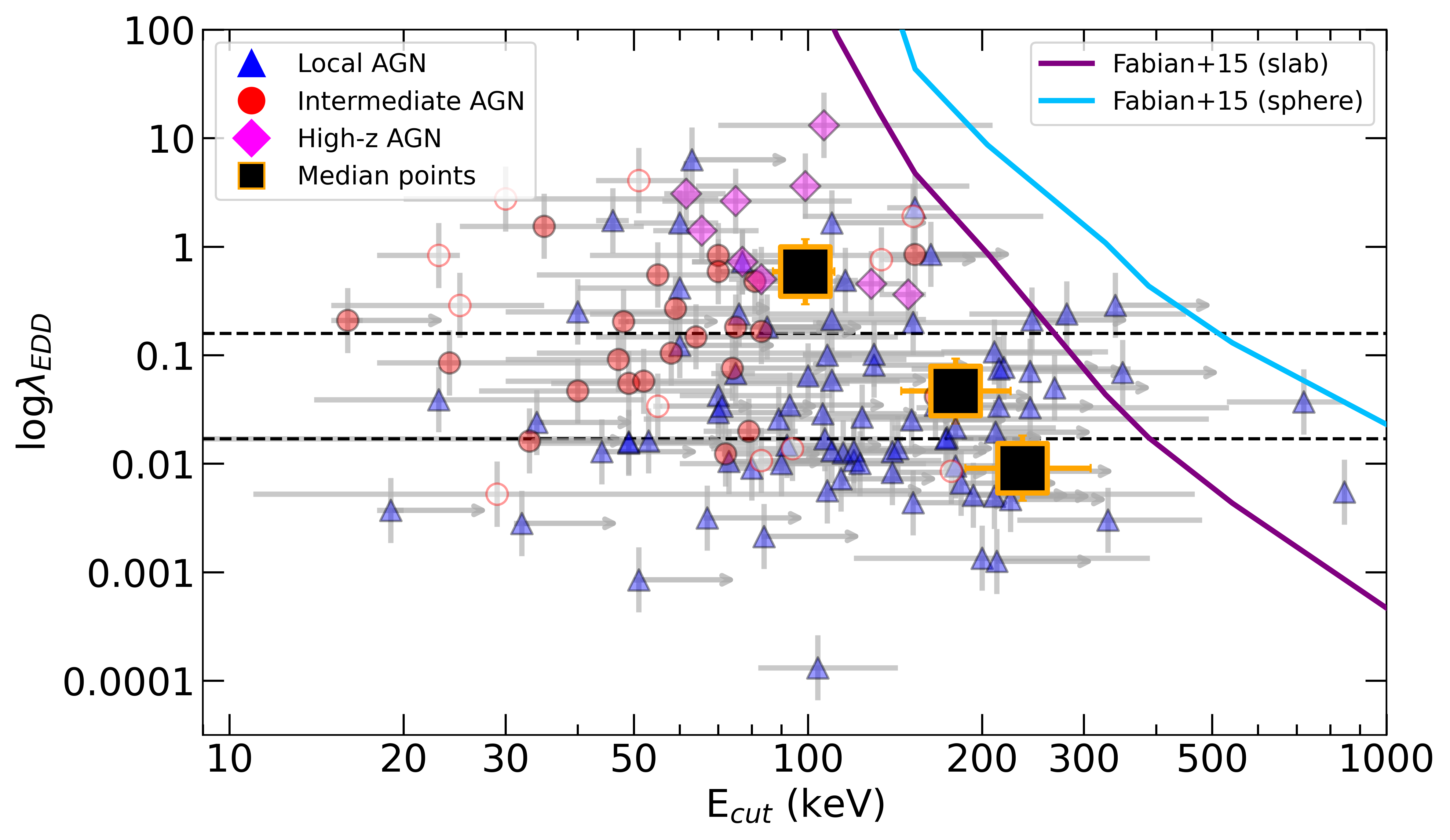}  
   \caption{\footnotesize  $M_{ BH}$, $L_{ bol}$ and $\lambda_{\rm Edd}$   vs. High-energy cutoff, for the combined sample of 127 sources at low-z (blue), intermediate-z (red empty circles and red filled circles)  and high-z AGN (pink). Symbols refer to \autoref{BertolaUpdated}. Black markers with orange contours indicate the median $E_{cut}$ values obtained by splitting the full sample into three equal-size bins (42/43 sources each) for each accretion parameter (see \Cref{Median_computation}), where the bin number (1, 2, 3) increases for increasing accretion parameter (from the bottom to the top). Dashed black lines show the bin intervals.
    }
    \label{binned}
\end{figure}

We investigate the relation between $E_{cut}$ and the accretion parameters by binning the entire sample as a function of  $L_{bol}$, $M_{BH}$ and  $\lambda_{\rm Edd}$. We sorted the sample by mass, Eddington ratio, and bolometric luminosity. For each parameter, we split the entire dataset into three bins, each containing approximately the same number of sources (42/43 per bin). We compute the median value for $E_{cut}$ by using the two methods reported in \Cref{Median_computation}. 
\Cref{binned} shows the distribution of $E_{cut}$ as a function of $M_{BH}$ (top), $L_{bol}$  (center), and $\lambda_{\rm Edd}$ (bottom) 
 for the merged sample, and the median 
$E_{cut}$ values for each bin (large black markers).
The results of the log-rank test between different bins in \Cref{binned} are reported in \Cref{logrank}, considering also the lower limits. P-values in bold face show highly significant differences (significance $>99\%$).


\begin{table}[!t]
    \centering
    \caption{\footnotesize P-value from the log-rank test for statistical comparison of the different bins in M$_{BH}$, $\lambda_{Edd}$ and L$_{bol}$.}
    \label{logrank}
   \resizebox{0.3\textwidth}{!}{
   {
  \renewcommand{\arraystretch}{1.25}
    \begin{tabular}{lc}
    \hline\hline 
     Bin comparison   & P-value\\
     \hline\hline
    M$_{BH}1$ vs. M$_{BH}2$ &  $0.48$ \\
    M$_{BH}2$ vs. M$_{BH}3$ & $0.04$ \\
    M$_{BH}1$ vs. M$_{BH}3$ & ${\bf 5.4\times10^{-3}}$ \\
    \hline
$\lambda_{Edd}$1 vs. $\lambda_{Edd}$2  & $0.29$ \\
$\lambda_{Edd}$2 vs. $\lambda_{Edd}$3 & ${\bf9.1\times 10^{-3}}$ \\
$\lambda_{Edd}$1 vs. $\lambda_{Edd}$3 &  ${\bf7.1\times 10^{-4}}$ \\
    \hline
L$_{bol}1$ vs. L$_{bol}2$    & $0.04$\\
L$_{bol}2$ vs. L$_{bol}3$    & ${\bf 6.2\times 10^{-3}}$\\
L$_{bol}1$ vs. L$_{bol}3$    & ${\bf <10^{-6}}$\\
\hline\hline
   \end{tabular}
   }
   } 
 \par\smallskip
  \noindent\begin{minipage}{0.5\textwidth}\raggedright \vspace{0.1cm}
{\footnotesize{\bf Notes:} In bold face, the bins where P-value\,$<0.01$, i.e., the difference between the two bins is significant at $>99\%$.}  
\end{minipage}
\end{table}

\begin{table}[!t]
    \centering
    \caption{\small Median $E_{cut}$ values computed in bins of $M_{BH}$, $L_{bol}$, and $\lambda_{\rm Edd}$, from the merged sample. 
    }
\label{Median_computation}  \resizebox{0.5\textwidth}{!}{
       {
\renewcommand{\arraystretch}{1.25}
    \begin{tabular}{ccc}
    \hline\hline
     Bin    & $E_{cut}$ (keV) / Asurv & $E_{cut}$(keV) / Par. fit \\
     &50th (25th-75th)& Median (C.I. $68\%$) \\
\hline\hline
logM$_{BH}1$ $    [5.55;7.59] $    & $197\ (109-454) $ &$217 \ (180-264)$  \\
logM$_{BH}2$    $ [7.59;8.30]$& $183 \ (67-261)$ &$174 \ (141-241)$ \\
logM$_{BH}3$     $[8.34;10.30]$& $128 \ (70-178)$ & $101 \ (91-112)$  \\
\hline
log$\lambda_{Edd}1~[-3.88;-1.77]$  & $206$$\ (108-342) $ & 235$ \ (185-308)$ \\
log$\lambda_{Edd}2~[-1.77;-0.78]$  & $205$$\ (76-372)$ & 180 $\ (145-224)$ \\
log$\lambda_{Edd}3~[-0.74;1.11]$ & $95$$\ (60-151)$ & 99  $\ (88-111)$ \\
\hline
logL$_{bol}1$  $    [41.68;44.14] $    &$233 \ (117-687)$ & $299 \ (231-387) $ \\
logL$_{bol}2$  $  [44.24;45.25]$& $141 \ (75-289)$ & $153 \ (126-183)$ \\
logL$_{bol}3$    $[45.31;48.92]$& $88 \ (58-142)$ & $88 \ (79-98)$  \\
\hline\hline
    \end{tabular}
    }
    }
 \par\smallskip
  \noindent\begin{minipage}{0.5\textwidth}\raggedright\vspace{0.1cm}
{\footnotesize{\bf Notes:} The first column reports the range of values used in each bin for each parameter. The median $E_{cut}$ with its 25th-75th percentiles of the distribution computed with Asurv is reported in the second column, while the median $E_{cut}$ with its 68\% confidence level interval is reported in the third.}
\end{minipage}
\end{table}

The top panel of \Cref{binned} shows little dependency of the median $E_{cut}$ with BH mass, up to $M_{BH}\sim10^8 \Msun$, with the median $E_{cut}$, around $\sim200$ keV, dominated by the low-z and some of the intermediate redshift sample. At $M_{BH}>10^8 \Msun$, the bin is dominated by intermediate and high-z QSOs and shows significantly lower median $E_{cut}$, around $\sim100$ keV. 
The middle panel shows a strong decrease of  $E_{cut}$ with increasing \lbol\ (from $\sim300$ to $\sim90$ keV from the low to high \lbol\ bin), and the last panel shows a smooth transition of the median $E_{cut}$ from $\sim250$ keV for low-accreting sources (dominated by low-z AGN), to $\sim100$ keV for highly accreting ones (dominated by intermediate/high-z QSOs).
The statistical significance of these trends is assessed with the non-parametric log-rank test \cite{Mantel1966}, which compares the $E_{cut}$ survival functions between bins while properly accounting for censored data; the results are reported in \autoref{logrank}. The strongest and most significant trend is with \lbol\ and $\lambda_{\rm Edd}$, where the highest bin differs from the lowest at greater than $3-4\sigma$ significance in both cases. The $M_{BH}$ trend is weaker, with only the highest-mass bin significantly different from the lowest ($p = 6.8\times10^{-4}$), while the first two bins are statistically indistinguishable ($p = 0.48$). 
These trends are independently confirmed by computing the generalised Kendall $\tau$ statistic for censored data using Asurv, treating the $E_{cut}$ lower limits as right-censored entries (results in \autoref{partial}:
the anticorrelation between \ecut and \ledd and \lbol are strong ($p=1.7\times10^{-5}$ an  $p=1.1\times10^{-4}$, respectively), while the one with \mbh is weak ($p = 0.019$).

However, because $M_{BH}$, \lbol, and $\lambda_{\rm Edd}$ are not independent variables (by definition, \lbol$\propto\lambda_{\rm Edd} \times M_{BH}$), 
neither the binning analysis nor the pairwise $\tau$ values can establish which parameter is the primary physical driver of the $E_{cut}$ trend. To address this, we computed partial Kendall $\tau$ values, obtained by first regressing both log $E_{cut}$ and the variable of interest against the control variable using ordinary least squares, and then computing the censored Kendall $\tau$ on the resulting residuals, following \citet{Akritas96}. This measures the residual association between $E_{cut}$ and a given parameter after the shared dependence on a third variable has been removed. Results are listed in \autoref{partial}.

\begin{table}[h]
    \centering
    \caption{Pairwise and partial correlation analysis results.}
\label{partial}  \resizebox{0.48\textwidth}{!}{
{
\renewcommand{\arraystretch}{1.25}
    \begin{tabular}{lccc}
    \hline\hline
     Pair \hspace{2.2cm}| control & $\tau$ & p-value \\
\hline\hline
log \ecut vs log \ledd & -0.196 & 
${\bf1.7\times10^{-5}}$\\
log \ecut vs log \lbol & -0.184 & ${\bf 1.1\times10^{-4}}$\\
log \ecut vs log \mbh &  -0.106 & 0.019 \\
\hline
log \ecut vs log \ledd | log \mbh & -0.169 & ${\bf 1.7\times10^{-4}}$\\
log \ecut vs log \ledd | log \lbol & -0.063 & 0.155 \\ 
log \ecut vs log \lbol | log \mbh & -0.167 & ${\bf 1.9\times10^{-4}}$ \\
log \ecut vs log \lbol | log \ledd & -0.087 & 0.052 \\
log \ecut vs log \mbh | log \ledd & -0.085 & 0.056\\
log \ecut vs log \mbh | log \lbol & +0.058 & 0.189 \\
\hline\hline
    \end{tabular}
    }
    }
 \par\smallskip
  \noindent\begin{minipage}{0.5\textwidth}\raggedright\vspace{0.1cm}
{\footnotesize{\bf Notes}: In bold face, the correlations where P-value\,$<0.01$, i.e., the difference between the two bins is significant at $>99\%$.}
\end{minipage}
\end{table}

These results establish that \ledd is the primary driver of the \ecut anticorrelation. The partial correlation of \ecut with \ledd after removing the \mbh trend, $\tau$(log \ecut, log \ledd | log \mbh) = -0.169 ($p=1.7\times10^{-4}$), is as strong as the pairwise value, while the residual partial correlation of \ecut with \mbh after removing the \ledd trend drops to $\tau$(log \ecut, log \mbh | log \ledd) = -0.085 ($p = 0.056$), consistent with zero. This means that the partial correlation between \ecut and \ledd is strong even when the dependency from the mass is removed. Similarly, $\tau$(log \ecut, log \mbh | log \lbol) = +0.058 ($p = 0.189$) confirms that \mbh carries no independent information once \lbol is accounted for. The \lbol correlation behaves as expected from its definition : $\tau$(log \ecut, log \lbol | log \mbh) = -0.168 ($p=1.9\times10^{-4}$) is strong because \lbol acts as a proxy for \ledd at fixed \mbh, while the mutual partial correlations between \ledd and \lbol ($\tau$ = -0.063 and -0.087, both $p > 0.05$) are both not significant and are consistent with the co-linearity introduced by the definition \lbol~$\propto$~\ledd$\times$~\mbh across the sample.

These results are fully consistent with the pair-production thermostat model \citet{Fabian2015}: an increase in the Eddington ratio implies a higher photon density for a given size of the emitting region, under the assumption of a fixed $R=10 r_g$, enhancing electron–positron pair production and cooling the hot corona more efficiently. In fact, in the bottom panel of \autoref{binned}, we can recast the runaway pair-production limits from \citet{Fabian2015} in this parameter space, assuming a fixed coronal radius of $R = 10 R_g$, a coronal luminosity $L_X$ in the 0.1-200 keV band, extrapolated from the 2-10 keV luminosity assuming a photon index $\Gamma=1.8$ so that $L_X = 3\times L_{2-10keV}$, and a bolometric luminosity \lbol$=10\times L_X$ (i.e., a total X-ray bolometric correction of 30) for all sources. It is easier to interpret the pair-production limits in this parameter space with respect to the $L_{2-10keV}$ vs. $E_{cut}$ one of \Cref{BertolaUpdated}, since here the curves are mass-independent. The purple curve shows the slab geometry, while the cyan curve shows the hemisphere geometry.

Despite the numerous assumptions and simplifications required to transition from theoretical curves to observable quantities, the binned data points are close to the pair production curves, specifically the one for a slab corona, and no single data point is beyond the curves in the forbidden region, confirming the limit both in the total sample (median points) and in the single sources.  However, the spread in $E_{cut}$ in each bin is large, as shown by the 25th-75th percentiles in \Cref{Median_computation}. Our results support the idea that, at least for sources near the pair-production limit, the Eddington ratio plays a key role in regulating the radiative compactness of the X-ray–emitting region. An increase in the Eddington ratio implies a higher luminosity for a given size of the emitting region, and therefore a larger photon density. Under these conditions, photon–photon interactions become more efficient, enhancing electron–positron pair production and leading to a stronger radiative cooling of the hot corona. In contrast, sources hosting cooler coronae and located far from the pair line are expected to remain only weakly affected by pair production, which plays a negligible role in shaping their coronal properties.

There are other additional effects that can influence the observed coronal temperature and can explain why the observed median $E_{cut}$ lies a factor $\sim2$ below the slab curve and $\sim4$ below the sphere curve:
i) strong general-relativistic (GR) effects expected for compact, inner coronae can redshift the observed $E_{cut}$ by a factor up to 3-4 relative to the intrinsic value in the most extreme cases, depending on coronal geometry, size, and inclination \citep{Tamborra18};
ii) a modest non-thermal tail in a hybrid electron distribution could enhance pair production and lower the equilibrium temperature, shifting the observed $E_{cut}$ to lower energies at fixed compactness \citep{Fabian_2017}. 

The large difference in E$_{cut}$ between SUBWAYS and high-z relative to local Seyferts can be explained by selection effects: higher redshift sources host more massive, more highly accreting SMBHs.
Given the strong dependence of E$_{cut}$ with $\lambda_{\rm Edd}$ and $L_{bol}$ and, to a lesser extent, with $M_{BH}$, this naturally results in systematically lower $E_{cut}$ values in the high-z and SUBWAYS subsample. Therefore, the observed redshift dependence of $E_{cut}$ does not imply an intrinsic cosmological evolution of coronal properties, but rather reflects the underlying distribution of accretion parameters in the selected population.

\subsection{Hot corona: testing thermal Comptonization models }
\label{comptmodels}

\begin{figure}[t]
    \centering
    \includegraphics[width=1\linewidth]{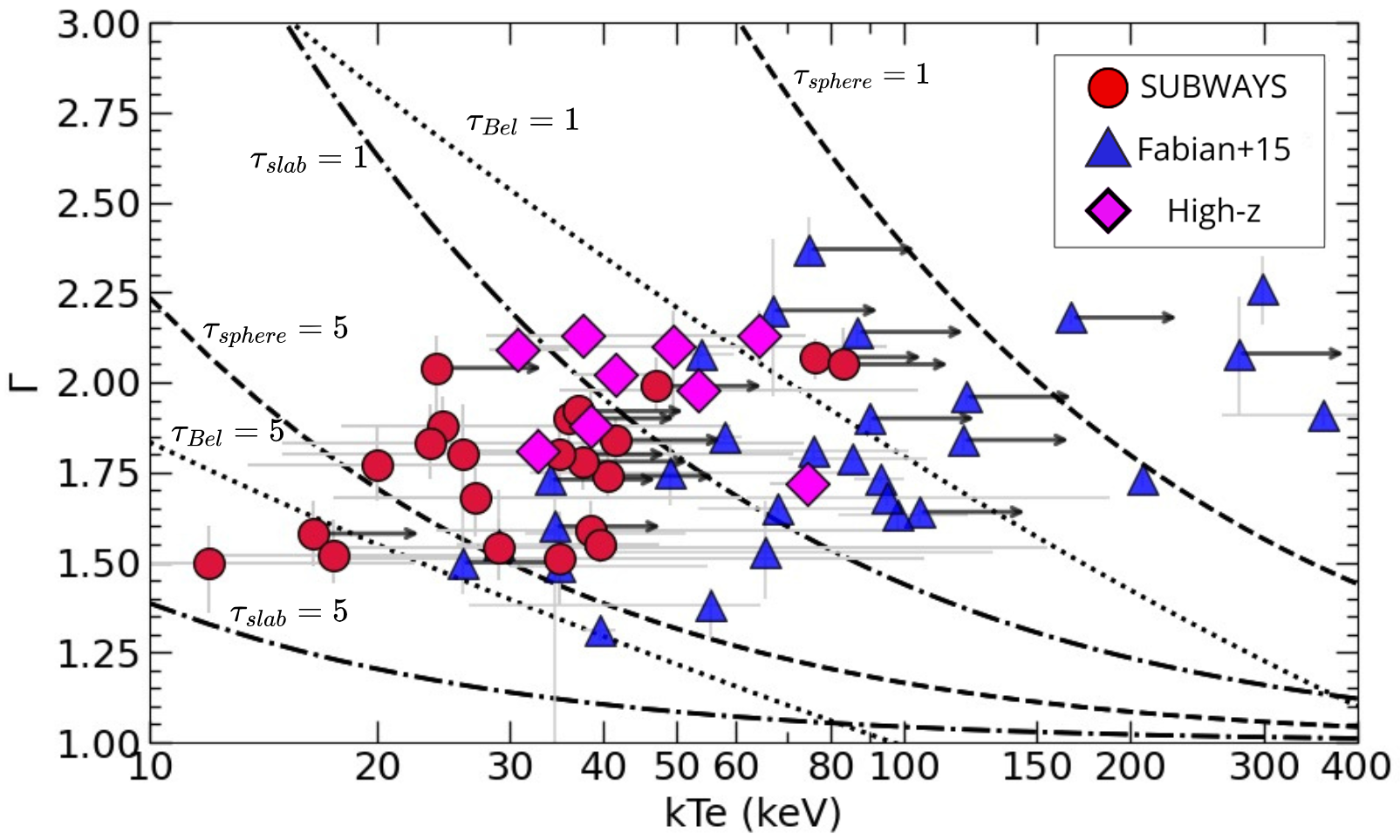}
    \caption{\footnotesize  Results of Model-1 for the hot corona in SUBWAYS sources (red), with $kT_e$ estimated from $E_{cut}$ following the empirical relation for optically thin plasma ($E_{cut}=2\ kT_e$) in order to compare them with the ones from \cite{Fabian2015} (blue). 
    For completeness, we also included the high-z QSOs in pink \citep{Lanzuisi19,Bertola22,Borrelli26}. The dashed line refers to the sphere geometry ($\tau_{sphere}$, \Cref{Longair_sphere}), the dot-dashed line to the slab geometry ($\tau_{slab}$, \Cref{longair_slab}), and the dotted line to the Beloborodov approximations ($\tau_{Bel}$, \Cref{gamma_tau}).}
    \label{Petrucci}
\end{figure}

The results of the spectral analysis of the SUBWAYS sources are compared 
in \Cref{Petrucci} with those obtained for a sample of nearby Seyferts \citep{Fabian2015} and a selection of high-redshift sources from the literature \citep{Lanzuisi19,Bertola22,Borrelli26} in order to test predictions from thermal Comptonisation models.
To convert the measured cut-offs into electron temperatures, we adopt the empirical relation $E_{cut} = 2 kT_e$ \citep{Petrucci01}. This approach enables us, starting from the results in \Cref{sec:Results}, to infer the coronal optical depth using the method described in \cite{Petrucci18}.  Furthermore, the $E_{cut} = 2 kT_e$ relation enables a direct comparison of SUBWAYS AGN with the local sample discussed in  \cite{Fabian2015}, and the few high-redshift quasars for which the temperatures were obtained using the same \citet{Petrucci01} conversion. 
Finally, for further consistency, only local sources with black hole masses $> 10^7\ M_\odot$ are considered.
In \Cref{Petrucci}, the electron temperatures are plotted against the photon index for the SUBWAYS sample, the \cite{Fabian2015} sample, and the high-z quasar sample (same of \Cref{BertolaUpdated}).  

Theoretical curves based on different analytical prescriptions are overlaid to provide approximate estimates of the plasma optical depth. The $\Gamma$ vs $kT_e$ relation from \Cref{Longair}, \Cref{Longair_sphere}, and \Cref{longair_slab} is drawn for two different values of the optical depths: $\tau=1$ and $\tau=5$, for both the slab ($\tau_{slab}$) and spherical  ($\tau_{sphere}$) geometries. We also report the relation inferred from \citet{Beloborodov99} ($\tau_{Bel}$), for the same two values of optical depth (\Cref{gamma_tau}, \Cref{Compton_parameter}).


The distribution of coronal temperatures in the SUBWAYS sources is shifted to lower values, although a non-negligible overlap is present, with respect to that of local AGN. The derived optical depths sit approximately in the range $\tau \sim1$-$5$, depending on the geometry (slab or sphere) and whether the Beloborodov or Longair prescriptions are adopted.
 
The degeneracy between the hot coronal plasma temperature and the optical depth of Comptonization models translates into a mild correlation between $kT_e$ and $\Gamma$  
for the SUBWAYS sample as well as for the combined samples: steeper (flatter) X-ray slopes correspond to hotter (colder) coronae.
This may appear counterintuitive, as it is generally assumed that higher temperatures produce flatter spectra, at fixed optical depth.
The observed trend, therefore, may be primarily driven by different optical depths: consistent with simulations reported in \citet{Middei19Moca}, a flat slope is obtained in a cold, optically thick plasma, while a hot, lower-opacity gas yields a steep slope. 
The sources analyzed in this work are well approximated by a moderately optically thick and relatively cool coronal plasma. By placing the SUBWAYS sources in a broader context, it is suggested that the coronal plasma becomes colder and thicker at higher luminosities and accretion rates, particularly at redshifts greater than 0.1.

It is interesting to note that a moderately optically thick corona may also wash out discrete emission features from the inner accretion disk. This would explain the lack of detection of broad iron lines in the SUBWAYS sample \citep{Matzeu23}.

We caution that a bias may be present that disfavors the detection of high-temperature plasma in spectra with relatively low counting statistics and limited spectral coverage at energies greater than 20-30 keV (see \Cref{simulations}). This effect may be more pronounced in the SUBWAYS sample, as it is not compensated by the redshift, which allows for probing higher energies in the rest frame for sources at $z=3-4$.

\section{Summary and Conclusions}
\label{sec:Conclusions}
Using joint \xmm plus \nustar spectral analysis, we derived with unprecedented accuracy the coronal properties of 23 AGN with bolometric luminosities $L_{bol}=2\times10^{44}-2\times10^{46}$ erg/s and redshifts $z=0.1-0.4$ as part of the SUBWAYS project. The main findings of our work are the following:

\begin{itemize}[label=\textbullet]
\item The continuum properties such as the photon index $\Gamma$ and the reflection parameter $R$, are consistent with typical values of local AGN  (median $\Gamma\sim1.8$ and $R\sim1$ with significant scatter; see \Cref{sec:Results_gamma} and \Cref{Gamma_R_comparison}).

\item The soft X-ray emission of our sources has been modeled with a warm corona, and the results of electron temperatures and optical depths are consistent with a cool and optically thick plasma with $kT^{warm}_e \sim0.4$ keV with $\tau\sim10-15$ (see (see \Cref{sec:Results_warm} and \Cref{kte_warm}), broadly in line with previous works \citep{Petrucci18} and robust across \texttt{compTT} and \texttt{nthcomp} parameterizations.

\item The median value of high energy cut-off and the electron temperature of the hot corona are respectively \ecut$\sim 87$ ($71-111$, 68\% C.I.) keV and $kT_e^{hot}\sim15$ ($13-18$, 68\% C.I.) keV, computed using the fit with a parametric log-normal function, described in \Cref{sec:hotcoronaproperties}, where the distributions of the values for both quantities are reported (\Cref{Ecut_kt}). The median high-energy cut-off is lower than the typical values for local AGN (median \ecut$\sim100-200$ keV) \citep{Fabian2015,Ricci2018}.

\end{itemize}

\noindent The sample is enlarged by comparing SUBWAYS results with literature results of low-z and low luminosities sources \citep{Fabian2015, Ricci2017} 
and high-z high luminosity sources \citep[][ see \Cref{sec:Discussion}]{Lanzuisi19,Bertola22,Borrelli26}. We assembled a sample of 207 sources with \ecut measurements from z=0 to z=4, which is reduced to 127 if we consider sources with reliable estimates of \mbh, \lbol and \ledd. This is the largest sample so far in terms of properties of the hot corona and their relation with accretion parameters. 
The results from the enlarged sample are the following:
\begin{itemize}
[label=\textbullet]
\item 
Quasars at intermediate/high redshift exhibit systematically lower high-energy cut-offs than local Seyferts; in redshift bins the median \ecut shifts from $\sim225$ keV (low-$z$) to $\sim90$ keV (intermediate/high-$z$; see \Cref{BertolaUpdated} and \Cref{redshift_division}). This can be explained by selection effects: higher redshift sources host more luminous and highly accreting SMBHs (see next point).

\item The high energy cut-off is compared with the accretion parameters in \Cref{binned}. The clearest trend is with the accretion rate and bolometric luminosity: the median \ecut decreases from $\sim250$ keV at low \ledd to $\sim100$ keV at high \ledd and from $\sim300$ at low \lbol to $\sim90$ keV at high \lbol (see \Cref{Median_computation}).
A partial correlation analysis confirms that \ledd retains a genuine independent anti-correlation with \ecut, while the \lbol trend is an indirect reflection of the \ledd dependence. The pair-production thermostat model provides a physical motivation for \ledd as the direct driver, since an increase in the Eddington ratio raises the photon density in the emitting region, enhancing electron–positron pair production and cooling the corona more efficiently.

\item The \mbh dependence is weak: \ecut remains roughly constant up to \mbh$\sim10^8$ \Msun and drops only in the highest-mass bin, which is preferentially populated by luminous, high-\ledd QSOs. The partial correlation analysis confirms this: once \lbol or \ledd are controlled for, the \mbh correlation with \ecut becomes statistically consistent with zero in both cases.

\item The SUBWAYS sample is characterized by relatively cold coronal temperatures ($kT_e \sim 20-80$ keV) and higher, moderately thick, optical depths ($\tau\sim 1-5$) with respect to a comparison sample of nearby Seyfert galaxies,(\Cref{sec:Discussion}, \Cref{Petrucci}).

\end{itemize}
 
Future X-ray observatories -- particularly the New Athena mission -- are expected to deliver the spectral resolution and effective area required to disentangle the various spectral components. Such advances will enable more precise and comprehensive investigations of coronal physics, allowing for rigorous tests of the models suggested by SUBWAYS with unprecedented accuracy.

\begin{acknowledgements} 
SP and GL acknowledge the support of the INAF Large Grant 2024  “The DEepest study of LUminous QSOs in X-ray at z=2-7 (DELUX)”.
GM acknowledges support from grants n. PID2020-115325GB-C31 and n. PID2023-147338NB-C21 funded by MICIU/AEI/10.13039/501100011033 and ERDF/EU.
VEG acknowledges funding under NASA contract 80NSSC24K1403.
POP acknowledges support from the French spatial agency CNES and from the « Action Thématique Phénomènes Extrêmes et Multimessager » of the Astronomy-Astrophysics National Programme from INSU/CNRS. B.D.M. acknowledges support via the Spanish MINECO grants PID2023-148661NB-I00, PID2022-136828NB-C44, and the AGAUR/Generalitat de Catalunya grant SGR-386/2021.
SB and FR acknowledge support from SEAWIND grant funded by the European Union - Next Generation EU, Mission 4 Component 1 CUP C53D23001330006. All the italian co-authors acknowledge support and fundings from Accordo Attuativo ASI-INAF n. 2017-14-H.0., crucial for the initial phase of the SUBWAYS project.
GP acknowledges Max-Planck-Institut für extraterrestrische Physik, Giessenbachstrasse, Garching, 85748, Germany,  Como Lake Center for Astrophysics (CLAP), DiSAT, Università degli Studi dell’Insubria, via Valleggio 11, 22100 Como, Italy, and  acknowledges financial support from the European Research Council (ERC) under the European Union’s Horizon 2020 research and innovation program HotMilk (grant agreement No. 865637) and from the Framework per l’Attrazione e il Rafforzamento delle Eccellenze (FARE) per la ricerca in Italia (R20L5S39T9).
SM (orchid: 0000-0002-4822-3559) acknowledges Astronomy Department, The Ohio State University, Columbus, OH, USA
and Center for Cosmology and Astro-Particle Physics, The Ohio State University, Columbus, OH, USA.
EB acknowledges INAF - Osservatorio Astrofisico di Arcetri, Largo E. Fermi 5, I-50125 Firenze, Italy and  the support of  the INAF GO grant ``A JWST/MIRI MIRACLE: Mid-IR Activity of Circumnuclear Line Emission'' and of the ``Ricerca Fondamentale 2024'' INAF program (mini-grant 1.05.24.07.01).
FP acknowledges "Programma di Ricerca Fondamentale INAF 2024".

\end{acknowledgements}

\bibliography{references.bib}

@string{aap = {A\&A}}

@string{apj = {ApJ}}

@string{apjl = {ApJL}}

@string{apjs = {ApJS}}

@string{araa = {Annu. Rev. Astron. Astrophys.}}

@string{mnras = {MNRAS}}

@string{nat = {Nature}}

@ARTICLE{Akritas96,
       author = {{Akritas}, M.~G. and {Siebert}, J.},
        title = "{A test for partial correlation with censored astronomical data}",
      journal = {\mnras},
         year = 1996,
        month = feb,
       volume = {278},
       number = {4},
        pages = {919-924},
          doi = {10.1093/mnras/278.4.919},
archivePrefix = {arXiv},
       eprint = {astro-ph/9508018},
 primaryClass = {astro-ph},
       adsurl = {https://ui.adsabs.harvard.edu/abs/1996MNRAS.278..919A}
}

@ARTICLE{Keck15,
       author = {{Keck}, M.~L. and {Brenneman}, L.~W. and {Ballantyne}, D.~R. and {Bauer}, F. and {Boggs}, S.~E. and {Christensen}, F.~E. and {Craig}, W.~W. and {Dauser}, T. and {Elvis}, M. and {Fabian}, A.~C. and {Fuerst}, F. and {Garc{\'\i}a}, J. and {Grefenstette}, B.~W. and {Hailey}, C.~J. and {Harrison}, F.~A. and {Madejski}, G. and {Marinucci}, A. and {Matt}, G. and {Reynolds}, C.~S. and {Stern}, D. and {Walton}, D.~J. and {Zoghbi}, A.},
        title = "{NuSTAR and Suzaku X-ray Spectroscopy of NGC 4151: Evidence for Reflection from the Inner Accretion Disk}",
      journal = {\apj},
         year = 2015,
        month = jun,
       volume = {806},
       number = {2},
          eid = {149},
        pages = {149},
          doi = {10.1088/0004-637X/806/2/149},
archivePrefix = {arXiv},
       eprint = {1504.07950},
 primaryClass = {astro-ph.HE},
       adsurl = {https://ui.adsabs.harvard.edu/abs/2015ApJ...806..149K}
}

@ARTICLE{McHardy06,
       author = {{McHardy}, I.~M. and {Koerding}, E. and {Knigge}, C. and {Uttley}, P. and {Fender}, R.~P.},
        title = "{Active galactic nuclei as scaled-up Galactic black holes}",
      journal = {\nat},
         year = 2006,
        month = dec,
       volume = {444},
       number = {7120},
        pages = {730-732},
          doi = {10.1038/nature05389},
archivePrefix = {arXiv},
       eprint = {astro-ph/0612273},
 primaryClass = {astro-ph},
       adsurl = {https://ui.adsabs.harvard.edu/abs/2006Natur.444..730M}
}

@ARTICLE{Akylas21,
       author = {{Akylas}, A. and {Georgantopoulos}, I.},
        title = "{Distribution of the coronal temperature in Seyfert 1 galaxies}",
      journal = {\aap},
         year = 2021,
        month = nov,
       volume = {655},
          eid = {A60},
        pages = {A60},
          doi = {10.1051/0004-6361/202141186},
archivePrefix = {arXiv},
       eprint = {2108.11337},
 primaryClass = {astro-ph.HE},
       adsurl = {https://ui.adsabs.harvard.edu/abs/2021A&A...655A..60A}
}

@ARTICLE{Molina2009,
       author = {{Molina}, M. and {Bassani}, L. and {Malizia}, A. and {Stephen}, J.~B. and {Bird}, A.~J. and {Dean}, A.~J. and {Panessa}, F. and {de Rosa}, A. and {Landi}, R.},
        title = "{The INTEGRAL complete sample of type 1 AGN}",
      journal = {\mnras},
         year = 2009,
        month = nov,
       volume = {399},
       number = {3},
        pages = {1293-1306},
          doi = {10.1111/j.1365-2966.2009.15257.x},
archivePrefix = {arXiv},
       eprint = {0906.2909},
 primaryClass = {astro-ph.CO},
       adsurl = {https://ui.adsabs.harvard.edu/abs/2009MNRAS.399.1293M}
}

@ARTICLE{Panessa2008,
       author = {{Panessa}, F. and {Bassani}, L. and {de Rosa}, A. and {Bird}, A.~J. and {Dean}, A.~J. and {Fiocchi}, M. and {Malizia}, A. and {Molina}, M. and {Ubertini}, P. and {Walter}, R.},
        title = "{The broad-band XMM-Newton and INTEGRAL spectra of bright type 1 Seyfert galaxies}",
      journal = {\aap},
         year = 2008,
        month = may,
       volume = {483},
       number = {1},
        pages = {151-160},
          doi = {10.1051/0004-6361:20078657},
archivePrefix = {arXiv},
       eprint = {0803.0896},
 primaryClass = {astro-ph},
       adsurl = {https://ui.adsabs.harvard.edu/abs/2008A&A...483..151P}
}

@ARTICLE{Petrucci2001,
       author = {{Petrucci}, P.~O. and {Haardt}, F. and {Maraschi}, L. and {Grandi}, P. and {Malzac}, J. and {Matt}, G. and {Nicastro}, F. and {Piro}, L. and {Perola}, G.~C. and {De Rosa}, A.},
        title = "{Testing Comptonization Models Using BeppoSAX Observations of Seyfert 1 Galaxies}",
      journal = {\apj},
         year = 2001,
        month = aug,
       volume = {556},
       number = {2},
        pages = {716-726},
          doi = {10.1086/321629},
archivePrefix = {arXiv},
       eprint = {astro-ph/0101219},
 primaryClass = {astro-ph},
       adsurl = {https://ui.adsabs.harvard.edu/abs/2001ApJ...556..716P}
}

@INPROCEEDINGS{Brusa2022,
       author = {{Brusa}, M. and {Matzeu}, G. and {Bianchi}, S. and {Piconcelli}, E. and {Dadina}, M. and {Lanzuisi}, G. and {Bongiorno}, A. and {Cappi}, M. and {Comastri}, A. and {Costanzo}, D. and {Cresci}, G. and {Duras}, F. and {Feruglio}, C. and {Gilli}, R. and {La Franca}, F. and {Luminari}, A. and {Matt}, G. and {Middei}, R. and {Nardini}, E. and {Ponti}, G. and {Tombesi}, F. and {Torresi}, E. and {Vignali}, C. and {Zaino}, A. and {Zappacosta}, L.},
        title = "{Supermassive Black Hole Winds in X-rays}",
    booktitle = {Memorie della Societa Astronomica Italiana},
         year = 2022,
       volume = {93},
        month = nov,
        pages = {48},
          doi = {10.36116/MEMSAIT_93N2_3.2022.6}
}

@article{HaardtMaraschi93,
author = {Haardt, Francesco and Maraschi, Laura},
title = {{X-ray spectra from two-phase accretion disks}},
journal = {Astrophysical Journal},
year = {1993},
volume = {413},
pages = {507--517},
month = aug
}

@ARTICLE{HaardtMaraschi91,
   author = {{Haardt}, F. and {Maraschi}, L.},
    title = "{A two-phase model for the X-ray emission from Seyfert galaxies}",
  journal = {\apjl},
     year = 1991,
    month = oct,
   volume = 380,
    pages = {L51-L54},
      doi = {10.1086/186171},
   adsurl = {http://adsabs.harvard.edu/abs/1991ApJ...380L..51H}
}

@ARTICLE{Wilms00,
   author = {{Wilms}, J. and {Allen}, A. and {McCray}, R.},
    title = "{On the Absorption of X-Rays in the Interstellar Medium}",
  journal = {\apj},
   eprint = {astro-ph/0008425},
     year = 2000,
    month = oct,
   volume = 542,
    pages = {914-924},
      doi = {10.1086/317016},
   adsurl = {http://adsabs.harvard.edu/abs/2000ApJ...542..914W}
}

@ARTICLE{Tombesi10,
   author = {{Tombesi}, F. and {Cappi}, M. and {Reeves}, J.~N. and {Palumbo}, G.~G.~C. and 
	{Yaqoob}, T. and {Braito}, V. and {Dadina}, M.},
    title = "{Evidence for ultra-fast outflows in radio-quiet AGNs. I. Detection and statistical incidence of Fe K-shell absorption lines}",
  journal = {\aap},
archivePrefix = "arXiv",
   eprint = {1006.2858},
 primaryClass = "astro-ph.HE",
     year = 2010,
    month = oct,
   volume = 521,
      eid = {A57},
    pages = {A57},
      doi = {10.1051/0004-6361/200913440},
   adsurl = {http://adsabs.harvard.edu/abs/2010A%26A...521A..57T}
}

@ARTICLE{Gofford13,
   author = {{Gofford}, J. and {Reeves}, J.~N. and {Tombesi}, F. and {Braito}, V. and 
	{Turner}, T.~J. and {Miller}, L. and {Cappi}, M.},
    title = "{The Suzaku view of highly ionized outflows in AGN - I. Statistical detection and global absorber properties}",
  journal = {\mnras},
archivePrefix = "arXiv",
   eprint = {1211.5810},
 primaryClass = "astro-ph.HE",
     year = 2013,
    month = mar,
   volume = 430,
    pages = {60-80},
      doi = {10.1093/mnras/sts481},
   adsurl = {http://adsabs.harvard.edu/abs/2013MNRAS.430...60G}
}

@ARTICLE{Balokovic20,
       author = {{Balokovi{\'c}}, M. and {Harrison}, F.~A. and {Madejski}, G. and {Comastri}, A. and {Ricci}, C. and {Annuar}, A. and {Ballantyne}, D.~R. and {Boorman}, P. and {Brandt}, W.~N. and {Brightman}, M. and {Gandhi}, P. and {Kamraj}, N. and {Koss}, M.~J. and {Marchesi}, S. and {Marinucci}, A. and {Masini}, A. and {Matt}, G. and {Stern}, D. and {Urry}, C.~M.},
        title = "{NuSTAR Survey of Obscured Swift/BAT-selected Active Galactic Nuclei. II. Median High-energy Cutoff in Seyfert II Hard X-Ray Spectra}",
      journal = {\apj},
         year = 2020,
        month = dec,
       volume = {905},
       number = {1},
          eid = {41},
        pages = {41},
          doi = {10.3847/1538-4357/abc342},
archivePrefix = {arXiv},
       eprint = {2011.06583},
 primaryClass = {astro-ph.HE},
       adsurl = {https://ui.adsabs.harvard.edu/abs/2020ApJ...905...41B}
}

@ARTICLE{Kammoun23,
       author = {{Kammoun}, E.~S. and {Igo}, Z. and {Miller}, J.~M. and {Fabian}, A.~C. and {Reynolds}, M.~T. and {Merloni}, A. and {Barret}, D. and {Nardini}, E. and {Petrucci}, P.~O. and {Piconcelli}, E. and {Barnier}, S. and {Buchner}, J. and {Dwelly}, T. and {Grotova}, I. and {Krumpe}, M. and {Liu}, T. and {Nandra}, K. and {Rau}, A. and {Salvato}, M. and {Urrutia}, T. and {Wolf}, J.},
        title = "{The first X-ray look at SMSS J114447.77-430859.3: the most luminous quasar in the last 9 Gyr}",
      journal = {\mnras},
         year = 2023,
        month = apr,
          doi = {10.1093/mnras/stad952},
       adsurl = {https://ui.adsabs.harvard.edu/abs/2023MNRAS.tmp..971K}
}

@ARTICLE{Reeves21,
       author = {{Reeves}, J.~N. and {Braito}, V. and {Porquet}, D. and {Lobban}, A.~P. and {Matzeu}, G.~A. and {Nardini}, E.},
        title = "{The flaring X-ray corona in the quasar PDS 456}",
      journal = {\mnras},
         year = 2021,
        month = jan,
       volume = {500},
       number = {2},
        pages = {1974-1991},
          doi = {10.1093/mnras/staa3377},
archivePrefix = {arXiv},
       eprint = {2010.14295},
 primaryClass = {astro-ph.HE},
       adsurl = {https://ui.adsabs.harvard.edu/abs/2021MNRAS.500.1974R}
}

@ARTICLE{SchmidtGreen83,
   author = {{Schmidt}, M. and {Green}, R.~F.},
    title = "{Quasar evolution derived from the Palomar bright quasar survey and other complete quasar surveys}",
  journal = {\apj},
     year = 1983,
    month = jun,
   volume = 269,
    pages = {352-374},
      doi = {10.1086/161048},
   adsurl = {http://adsabs.harvard.edu/abs/1983ApJ...269..352S}
}

@ARTICLE{Titarchuk94,
   author = {{Titarchuk}, L.},
    title = "{Generalized Comptonization models and application to the recent high-energy observations}",
  journal = {\apj},
     year = 1994,
    month = oct,
   volume = 434,
    pages = {570-586},
      doi = {10.1086/174760},
   adsurl = {http://adsabs.harvard.edu/abs/1994ApJ...434..570T}
}

@ARTICLE{Harrison13,
   author = {{Harrison}, F.~A. and {Craig}, W.~W. and {Christensen}, F.~E. and 
	{Hailey}, C.~J. and {Zhang}, W.~W. and {Boggs}, S.~E. and {Stern}, D. and 
	{Cook}, W.~R. and {Forster}, K. and {Giommi}, P. and {Grefenstette}, B.~W. and 
	{Kim}, Y. and {Kitaguchi}, T. and {Koglin}, J.~E. and {Madsen}, K.~K. and 
	{Mao}, P.~H. and {Miyasaka}, H. and {Mori}, K. and {Perri}, M. and 
	{Pivovaroff}, M.~J. and {Puccetti}, S. and {Rana}, V.~R. and 
	{Westergaard}, N.~J. and {Willis}, J. and {Zoglauer}, A. and 
	{An}, H. and {Bachetti}, M. and {Barri{\`e}re}, N.~M. and {Bellm}, E.~C. and 
	{Bhalerao}, V. and {Brejnholt}, N.~F. and {Fuerst}, F. and {Liebe}, C.~C. and 
	{Markwardt}, C.~B. and {Nynka}, M. and {Vogel}, J.~K. and {Walton}, D.~J. and 
	{Wik}, D.~R. and {Alexander}, D.~M. and {Cominsky}, L.~R. and 
	{Hornschemeier}, A.~E. and {Hornstrup}, A. and {Kaspi}, V.~M. and 
	{Madejski}, G.~M. and {Matt}, G. and {Molendi}, S. and {Smith}, D.~M. and 
	{Tomsick}, J.~A. and {Ajello}, M. and {Ballantyne}, D.~R. and 
	{Balokovi{\'c}}, M. and {Barret}, D. and {Bauer}, F.~E. and 
	{Blandford}, R.~D. and {Brandt}, W.~N. and {Brenneman}, L.~W. and 
	{Chiang}, J. and {Chakrabarty}, D. and {Chenevez}, J. and {Comastri}, A. and 
	{Dufour}, F. and {Elvis}, M. and {Fabian}, A.~C. and {Farrah}, D. and 
	{Fryer}, C.~L. and {Gotthelf}, E.~V. and {Grindlay}, J.~E. and 
	{Helfand}, D.~J. and {Krivonos}, R. and {Meier}, D.~L. and {Miller}, J.~M. and 
	{Natalucci}, L. and {Ogle}, P. and {Ofek}, E.~O. and {Ptak}, A. and 
	{Reynolds}, S.~P. and {Rigby}, J.~R. and {Tagliaferri}, G. and 
	{Thorsett}, S.~E. and {Treister}, E. and {Urry}, C.~M.},
    title = "{The Nuclear Spectroscopic Telescope Array (NuSTAR) High-energy X-Ray Mission}",
  journal = {\apj},
archivePrefix = "arXiv",
   eprint = {1301.7307},
 primaryClass = "astro-ph.IM",
     year = 2013,
    month = jun,
   volume = 770,
      eid = {103},
    pages = {103},
      doi = {10.1088/0004-637X/770/2/103},
   adsurl = {http://adsabs.harvard.edu/abs/2013ApJ...770..103H}
}

@ARTICLE{Kaspi00,
   author = {{Kaspi}, S. and {Brandt}, W.~N. and {Netzer}, H. and {Sambruna}, R. and 
	{Chartas}, G. and {Garmire}, G.~P. and {Nousek}, J.~A.},
    title = "{Discovery of Narrow X-Ray Absorption Lines from NGC 3783 with the Chandra High Energy Transmission Grating Spectrometer}",
  journal = {\apjl},
     year = 2000,
    month = may,
   volume = 535,
    pages = {L17-L20},
      doi = {10.1086/312697},
   adsurl = {http://adsabs.harvard.edu/abs/2000ApJ...535L..17K}
}

@ARTICLE{Salpeter64,
   author = {{Salpeter}, E.~E.},
    title = "{Accretion of Interstellar Matter by Massive Objects.}",
  journal = {\apj},
     year = 1964,
    month = aug,
   volume = 140,
    pages = {796-800},
      doi = {10.1086/147973},
   adsurl = {http://adsabs.harvard.edu/abs/1964ApJ...140..796S}
}

@ARTICLE{Shakura73,
   author = {{Shakura}, N.~I. and {Sunyaev}, R.~A.},
    title = "{Black holes in binary systems. Observational appearance.}",
  journal = {\aap},
     year = 1973,
   volume = 24,
    pages = {337-355},
   adsurl = {http://adsabs.harvard.edu/abs/1973A%26A....24..337S}
}

@ARTICLE{NandraPounds94,
   author = {{Nandra}, K. and {Pounds}, K.~A.},
    title = "{GINGA Observations of the X-Ray Spectra of Seyfert Galaxies}",
  journal = {\mnras},
     year = 1994,
    month = may,
   volume = 268,
    pages = {405},
      doi = {10.1093/mnras/268.2.405},
   adsurl = {http://adsabs.harvard.edu/abs/1994MNRAS.268..405N}
}

@ARTICLE{Singh85,
   author = {{Singh}, K.~P. and {Garmire}, G.~P. and {Nousek}, J.},
    title = "{Observation of soft X-ray spectra from a Seyfert 1 and a narrow emission-line galaxy}",
  journal = {\apj},
     year = 1985,
    month = oct,
   volume = 297,
    pages = {633-638},
      doi = {10.1086/163560},
   adsurl = {http://adsabs.harvard.edu/abs/1985ApJ...297..633S}
}

@ARTICLE{Nardini12,
   author = {{Nardini}, E. and {Fabian}, A.~C. and {Walton}, D.~J.},
    title = "{Investigating the reflection contribution to the X-ray emission of Ton S180}",
  journal = {\mnras},
archivePrefix = "arXiv",
   eprint = {1204.4451},
 primaryClass = "astro-ph.HE",
     year = 2012,
    month = jul,
   volume = 423,
    pages = {3299-3307},
      doi = {10.1111/j.1365-2966.2012.21123.x},
   adsurl = {http://adsabs.harvard.edu/abs/2012MNRAS.423.3299N}
}

@ARTICLE{Ursini15,
   author = {{Ursini}, F. and {Marinucci}, A. and {Matt}, G. and {Bianchi}, S. and 
	{Tortosa}, A. and {Stern}, D. and {Ar{\'e}valo}, P. and {Ballantyne}, D.~R. and 
	{Bauer}, F.~E. and {Fabian}, A.~C. and {Harrison}, F.~A. and 
	{Lohfink}, A.~M. and {Reynolds}, C.~S. and {Walton}, D.~J.},
    title = "{The NuSTAR X-ray spectrum of the low-luminosity active galactic nucleus in NGC 7213}",
  journal = {\mnras},
archivePrefix = "arXiv",
   eprint = {1507.01775},
 primaryClass = "astro-ph.HE",
     year = 2015,
    month = sep,
   volume = 452,
    pages = {3266-3272},
      doi = {10.1093/mnras/stv1527},
   adsurl = {http://adsabs.harvard.edu/abs/2015MNRAS.452.3266U}
}

@ARTICLE{Kara17,
       author = {{Kara}, E. and {Garc{\'\i}a}, J.~A. and {Lohfink}, A. and {Fabian}, A.~C. and {Reynolds}, C.~S. and {Tombesi}, F. and {Wilkins}, D.~R.},
        title = "{The high-Eddington NLS1 Ark 564 has the coolest corona}",
      journal = {\mnras},
         year = 2017,
        month = jul,
       volume = {468},
       number = {3},
        pages = {3489-3498},
          doi = {10.1093/mnras/stx792},
archivePrefix = {arXiv},
       eprint = {1703.09815},
 primaryClass = {astro-ph.HE},
       adsurl = {https://ui.adsabs.harvard.edu/abs/2017MNRAS.468.3489K}
}

@ARTICLE{Marinucci22,
       author = {{Marinucci}, A. and {Vietri}, G. and {Piconcelli}, E. and {Bianchi}, S. and {Guainazzi}, M. and {Lanzuisi}, G. and {Stern}, D. and {Vignali}, C.},
        title = "{Breaking the rules at z ≃ 0.45: The rebel case of RBS 1055}",
      journal = {\aap},
         year = 2022,
        month = oct,
       volume = {666},
          eid = {A169},
        pages = {A169},
          doi = {10.1051/0004-6361/202244272},
archivePrefix = {arXiv},
       eprint = {2209.01575},
 primaryClass = {astro-ph.HE},
       adsurl = {https://ui.adsabs.harvard.edu/abs/2022A&A...666A.169M}
}

@ARTICLE{Buisson18,
       author = {{Buisson}, D.~J.~K. and {Parker}, M.~L. and {Kara}, E. and {Vasudevan}, R.~V. and {Lohfink}, A.~M. and {Pinto}, C. and {Fabian}, A.~C. and {Ballantyne}, D.~R. and {Boggs}, S.~E. and {Christensen}, F.~E. and {Craig}, W.~W. and {Farrah}, D. and {Hailey}, C.~J. and {Harrison}, F.~A. and {Ricci}, C. and {Stern}, D. and {Walton}, D.~J. and {Zhang}, W.~W.},
        title = "{NuSTAR observations of Mrk 766: distinguishing reflection from absorption}",
      journal = {\mnras},
         year = 2018,
        month = nov,
       volume = {480},
       number = {3},
        pages = {3689-3701},
          doi = {10.1093/mnras/sty2081},
archivePrefix = {arXiv},
       eprint = {1808.00014},
 primaryClass = {astro-ph.HE},
       adsurl = {https://ui.adsabs.harvard.edu/abs/2018MNRAS.480.3689B}
}

@ARTICLE{Diaz20,
       author = {{Diaz}, Y. and {Ar{\'e}valo}, P. and {Hern{\'a}ndez-Garc{\'\i}a}, L. and {Bassani}, L. and {Malizia}, A. and {Gonz{\'a}lez-Mart{\'\i}n}, O. and {Ricci}, C. and {Matt}, G. and {Stern}, D. and {May}, D. and {Zezas}, A. and {Bauer}, F.~E.},
        title = "{Constraining X-ray reflection in the low-luminosity AGN NGC 3718 using NuSTAR and XMM-Newton}",
      journal = {\mnras},
         year = 2020,
        month = aug,
       volume = {496},
       number = {4},
        pages = {5399-5413},
          doi = {10.1093/mnras/staa1762},
archivePrefix = {arXiv},
       eprint = {2006.09463},
 primaryClass = {astro-ph.HE},
       adsurl = {https://ui.adsabs.harvard.edu/abs/2020MNRAS.496.5399D}
}

@ARTICLE{Lanzuisi24,
       author = {{Lanzuisi}, G. and {Matzeu}, G. and {Baldini}, P. and {Bertola}, E. and {Comastri}, A. and {Tombesi}, F. and {Luminari}, A. and {Braito}, V. and {Reeves}, J. and {Chartas}, G. and {Bianchi}, S. and {Brusa}, M. and {Cresci}, G. and {Nardini}, E. and {Piconcelli}, E. and {Zappacosta}, L. and {Serafinelli}, R. and {Gaspari}, M. and {Gilli}, R. and {Cappi}, M. and {Dadina}, M. and {Perna}, M. and {Vignali}, C. and {Veilleux}, S.},
        title = "{The XMM-Newton and NuSTAR view of IRASF11119+3257: I. Detection of multiple ultra fast outflow components and a very cold corona}",
      journal = {\aap},
         year = 2024,
        month = sep,
       volume = {689},
          eid = {A247},
        pages = {A247},
          doi = {10.1051/0004-6361/202449194},
archivePrefix = {arXiv},
       eprint = {2406.12057},
 primaryClass = {astro-ph.HE},
       adsurl = {https://ui.adsabs.harvard.edu/abs/2024A&A...689A.247L}
}

@ARTICLE{Perna17,
       author = {{Perna}, M. and {Lanzuisi}, G. and {Brusa}, M. and {Mignoli}, M. and {Cresci}, G.},
        title = "{An X-ray/SDSS sample. I. Multi-phase outflow incidence and dependence on AGN luminosity}",
      journal = {\aap},
         year = 2017,
        month = jul,
       volume = {603},
          eid = {A99},
        pages = {A99},
          doi = {10.1051/0004-6361/201630369},
archivePrefix = {arXiv},
       eprint = {1703.05335},
 primaryClass = {astro-ph.GA},
       adsurl = {https://ui.adsabs.harvard.edu/abs/2017A&A...603A..99P}
}

@ARTICLE{Saccheo23,
       author = {{Saccheo}, I. and {Bongiorno}, A. and {Piconcelli}, E. and {Testa}, V. and {Bischetti}, M. and {Bisogni}, S. and {Bruni}, G. and {Cresci}, G. and {Feruglio}, C. and {Fiore}, F. and {Grazian}, A. and {Luminari}, A. and {Lusso}, E. and {Mainieri}, V. and {Maiolino}, R. and {Marconi}, A. and {Ricci}, F. and {Tombesi}, F. and {Travascio}, A. and {Vietri}, G. and {Vignali}, C. and {Zappacosta}, L. and {La Franca}, F.},
        title = "{The WISSH quasars project. XI. The mean spectral energy distribution and bolometric corrections of the most luminous quasars}",
      journal = {\aap},
         year = 2023,
        month = mar,
       volume = {671},
          eid = {A34},
        pages = {A34},
          doi = {10.1051/0004-6361/202244296},
archivePrefix = {arXiv},
       eprint = {2211.07677},
 primaryClass = {astro-ph.GA},
       adsurl = {https://ui.adsabs.harvard.edu/abs/2023A&A...671A..34S}
}

@BOOK{Longair11,
   author = {{Longair}, M.~S.},
    title = "{High Energy Astrophysics}",
booktitle = {High Energy Astrophysics, by Malcolm S.~Longair, Cambridge, UK: Cambridge University Press, 2011},
     year = 2011,
    month = feb,
   adsurl = {http://adsabs.harvard.edu/abs/2011hea..book.....L}
}

@ARTICLE{Petrucci18,
   author = {{Petrucci}, P.-O. and {Ursini}, F. and {De Rosa}, A. and {Bianchi}, S. and 
	{Cappi}, M. and {Matt}, G. and {Dadina}, M. and {Malzac}, J.
	},
    title = "{Testing warm Comptonization models for the origin of the soft X-ray excess in AGNs}",
  journal = {\aap},
archivePrefix = "arXiv",
   eprint = {1710.04940},
 primaryClass = "astro-ph.HE",
     year = 2018,
    month = mar,
   volume = 611,
      eid = {A59},
    pages = {A59},
      doi = {10.1051/0004-6361/201731580},
   adsurl = {https://ui.adsabs.harvard.edu/abs/2018A%26A...611A..59P}
}

@ARTICLE{Garcia10,
   author = {{Garc{\'{\i}}a}, J. and {Kallman}, T.~R.},
    title = "{X-ray Reflected Spectra from Accretion Disk Models. I. Constant Density Atmospheres}",
  journal = {\apj},
archivePrefix = "arXiv",
   eprint = {1006.0485},
 primaryClass = "astro-ph.HE",
     year = 2010,
    month = aug,
   volume = 718,
    pages = {695-706},
      doi = {10.1088/0004-637X/718/2/695},
   adsurl = {https://ui.adsabs.harvard.edu/abs/2010ApJ...718..695G}
}

@ARTICLE{Ponti12,
   author = {{Ponti}, G. and {Papadakis}, I. and {Bianchi}, S. and {Guainazzi}, M. and 
	{Matt}, G. and {Uttley}, P. and {Bonilla}, N.~F.},
    title = "{CAIXA: a catalogue of AGN in the XMM-Newton archive. III. Excess variance analysis}",
  journal = {\aap},
archivePrefix = "arXiv",
   eprint = {1112.2744},
 primaryClass = "astro-ph.HE",
     year = 2012,
    month = jun,
   volume = 542,
      eid = {A83},
    pages = {A83},
      doi = {10.1051/0004-6361/201118326},
   adsurl = {https://ui.adsabs.harvard.edu/abs/2012A%26A...542A..83P}
}

@ARTICLE{HI4PI16,
   author = {{Ben Bekhti}, N. and {Fl{\"o}er}, L. and 
	{Keller}, R. and {Kerp}, J. and {Lenz}, D. and {Winkel}, B. and 
	{Bailin}, J. and {Calabretta}, M.~R. and {Dedes}, L. and {Ford}, H.~A. and 
	{Gibson}, B.~K. and {Haud}, U. and {Janowiecki}, S. and {Kalberla}, P.~M.~W. and 
	{Lockman}, F.~J. and {McClure-Griffiths}, N.~M. and {Murphy}, T. and 
	{Nakanishi}, H. and {Pisano}, D.~J. and {Staveley-Smith}, L.
	},
    title = "{HI4PI: A full-sky H I survey based on EBHIS and GASS}",
  journal = {\aap},
archivePrefix = "arXiv",
   eprint = {1610.06175},
     year = 2016,
    month = oct,
   volume = 594,
      eid = {A116},
    pages = {A116},
      doi = {10.1051/0004-6361/201629178},
   adsurl = {https://ui.adsabs.harvard.edu/abs/2016A%26A...594A.116H}
}

@ARTICLE{Porquet18,
       author = {{Porquet}, D. and {Reeves}, J.~N. and {Matt}, G. and {Marinucci}, A. and
         {Nardini}, E. and {Braito}, V. and {Lobban}, A. and
         {Ballantyne}, D.~R. and {Boggs}, S.~E. and {Christensen}, F.~E.},
        title = "{A deep X-ray view of the bare AGN Ark 120. IV. XMM-Newton and NuSTAR spectra dominated by two temperature (warm, hot) Comptonization processes}",
      journal = {\aap},
         year = "2018",
        month = "Jan",
       volume = {609},
          eid = {A42},
        pages = {A42},
          doi = {10.1051/0004-6361/201731290},
archivePrefix = {arXiv},
       eprint = {1707.08907},
 primaryClass = {astro-ph.HE},
       adsurl = {https://ui.adsabs.harvard.edu/abs/2018A&A...609A..42P}
}

@ARTICLE{Garcia14,
       author = {{Garc{\'\i}a}, J. and {Dauser}, T. and {Lohfink}, A. and
         {Kallman}, T.~R. and {Steiner}, J.~F. and {McClintock}, J.~E. and
         {Brenneman}, L. and {Wilms}, J. and {Eikmann}, W. and
         {Reynolds}, C.~S. and {Tombesi}, F.},
        title = "{Improved Reflection Models of Black Hole Accretion Disks: Treating the Angular Distribution of X-Rays}",
      journal = {\apj},
         year = "2014",
        month = "Feb",
       volume = {782},
       number = {2},
          eid = {76},
        pages = {76},
          doi = {10.1088/0004-637X/782/2/76},
archivePrefix = {arXiv},
       eprint = {1312.3231},
 primaryClass = {astro-ph.HE},
       adsurl = {https://ui.adsabs.harvard.edu/abs/2014ApJ...782...76G}
}

@ARTICLE{Garcia13,
       author = {{Garc{\'\i}a}, J. and {Dauser}, T. and {Reynolds}, C.~S. and
         {Kallman}, T.~R. and {McClintock}, J.~E. and {Wilms}, J. and
         {Eikmann}, W.},
        title = "{X-Ray Reflected Spectra from Accretion Disk Models. III. A Complete Grid of Ionized Reflection Calculations}",
      journal = {\apj},
         year = "2013",
        month = "May",
       volume = {768},
       number = {2},
          eid = {146},
        pages = {146},
          doi = {10.1088/0004-637X/768/2/146},
archivePrefix = {arXiv},
       eprint = {1303.2112},
 primaryClass = {astro-ph.HE},
       adsurl = {https://ui.adsabs.harvard.edu/abs/2013ApJ...768..146G}
}

@ARTICLE{Zdziarski96,
   author = {{Zdziarski}, A.~A. and {Johnson}, W.~N. and {Magdziarz}, P.},
    title = "{Broad-band {$\gamma$}-ray and X-ray spectra of NGC 4151 and their implications for physical processes and geometry.}",
  journal = {\mnras},
   eprint = {astro-ph/9607015},
     year = 1996,
    month = nov,
   volume = 283,
    pages = {193-206},
      doi = {10.1093/mnras/283.1.193},
   adsurl = {https://ui.adsabs.harvard.edu/abs/1996MNRAS.283..193Z}
}

@ARTICLE{Zycky99,
   author = {{{\.Z}ycki}, P.~T. and {Done}, C. and {Smith}, D.~A.},
    title = "{The 1989 May outburst of the soft X-ray transient GS 2023+338 (V404 Cyg)}",
  journal = {\mnras},
   eprint = {astro-ph/9904304},
     year = 1999,
    month = nov,
   volume = 309,
    pages = {561-575},
      doi = {10.1046/j.1365-8711.1999.02885.x},
   adsurl = {https://ui.adsabs.harvard.edu/abs/1999MNRAS.309..561Z}
}

@ARTICLE{Petrucci01,
       author = {{Petrucci}, P.~O. and {Merloni}, A. and {Fabian}, A. and {Haardt}, F. and
         {Gallo}, E.},
        title = "{The effects of a Comptonizing corona on the appearance of the reflection components in accreting black hole spectra}",
      journal = {\mnras},
         year = "2001",
        month = "Dec",
       volume = {328},
       number = {2},
        pages = {501-510},
          doi = {10.1046/j.1365-8711.2001.04897.x},
archivePrefix = {arXiv},
       eprint = {astro-ph/0108342},
 primaryClass = {astro-ph},
       adsurl = {https://ui.adsabs.harvard.edu/abs/2001MNRAS.328..501P}
}

@ARTICLE{Petrucci13,
       author = {{Petrucci}, P. -O. and {Paltani}, S. and {Malzac}, J. and
         {Kaastra}, J.~S. and {Cappi}, M. and {Ponti}, G. and {De Marco}, B. and
         {Kriss}, G.~A. and {Steenbrugge}, K.~C. and {Bianchi}, S. and {Brand
        uardi-Raymont}, G. and {Mehdipour}, M. and {Costantini}, E. and
         {Dadina}, M. and {Lubi{\'n}ski}, P.},
        title = "{Multiwavelength campaign on Mrk 509. XII. Broad band spectral analysis}",
      journal = {\aap},
         year = "2013",
        month = "Jan",
       volume = {549},
          eid = {A73},
        pages = {A73},
          doi = {10.1051/0004-6361/201219956},
archivePrefix = {arXiv},
       eprint = {1209.6438},
 primaryClass = {astro-ph.HE},
       adsurl = {https://ui.adsabs.harvard.edu/abs/2013A&A...549A..73P}
}

@ARTICLE{Middei18,
       author = {{Middei}, R. and {Bianchi}, S. and {Cappi}, M. and {Petrucci}, P. -O. and
         {Ursini}, F. and {Arav}, N. and {Behar}, E. and {Brand
        uardi-Raymont}, G. and {Costantini}, E. and {De Marco}, B. and
         {Di Gesu}, L. and {Ebrero}, J. and {Kaastra}, J. and {Kaspi}, S. and
         {Kriss}, G.~A. and {Mao}, J. and {Mehdipour}, M. and {Paltani}, S. and
         {Peretz}, U. and {Ponti}, G.},
        title = "{Multi-wavelength campaign on NCG 7469. IV. The broad-band X-ray spectrum}",
      journal = {\aap},
         year = "2018",
        month = "Jul",
       volume = {615},
          eid = {A163},
        pages = {A163},
          doi = {10.1051/0004-6361/201832726},
       adsurl = {https://ui.adsabs.harvard.edu/abs/2018A&A...615A.163M}
}

@INPROCEEDINGS{Beloborodov99,
       author = {{Beloborodov}, A.~M.},
        title = "{Accretion Disk Models}",
    booktitle = {High Energy Processes in Accreting Black Holes},
         year = "1999",
       editor = {{Poutanen}, Juri and {Svensson}, Roland},
       series = {Astronomical Society of the Pacific Conference Series},
       volume = {161},
        month = "Jan",
        pages = {295},
archivePrefix = {arXiv},
       eprint = {astro-ph/9901108},
 primaryClass = {astro-ph},
       adsurl = {https://ui.adsabs.harvard.edu/abs/1999ASPC..161..295B}
}

@ARTICLE{Tortosa18a,
       author = {{Tortosa}, A. and {Bianchi}, S. and {Marinucci}, A. and {Matt}, G. and {Petrucci}, P.~O.},
        title = "{A NuSTAR census of coronal parameters in Seyfert galaxies}",
      journal = {\aap},
         year = 2018,
        month = jun,
       volume = {614},
          eid = {A37},
        pages = {A37},
          doi = {10.1051/0004-6361/201732382},
archivePrefix = {arXiv},
       eprint = {1801.04456},
 primaryClass = {astro-ph.GA},
       adsurl = {https://ui.adsabs.harvard.edu/abs/2018A&A...614A..37T}
}

@ARTICLE{Kamraj18,
       author = {{Kamraj}, N. and {Harrison}, F.~A. and {Balokovi{\'c}}, M. and {Lohfink}, A. and {Brightman}, M.},
        title = "{Coronal Properties of Swift/BAT-selected Seyfert 1 AGNs Observed with NuSTAR}",
      journal = {\apj},
         year = 2018,
        month = oct,
       volume = {866},
       number = {2},
          eid = {124},
        pages = {124},
          doi = {10.3847/1538-4357/aadd0d},
archivePrefix = {arXiv},
       eprint = {1809.01757},
 primaryClass = {astro-ph.HE},
       adsurl = {https://ui.adsabs.harvard.edu/abs/2018ApJ...866..124K}
}

@ARTICLE{Igo20,
       author = {{Igo}, Z. and {Parker}, M.~L. and {Matzeu}, G.~A. and {Alston}, W. and
         {Alvarez Crespo}, N. and {F{\"u}rst}, F. and {Buisson}, D.~J.~K. and
         {Lobban}, A. and {Joyce}, A.~M. and {Mallick}, L. and {Schartel}, N. and
         {Santos-Lle{\'o}}, M.},
        title = "{Searching for ultra-fast outflows in AGN using variability spectra}",
      journal = {\mnras},
         year = 2020,
        month = mar,
       volume = {493},
       number = {1},
        pages = {1088-1108},
          doi = {10.1093/mnras/staa265},
archivePrefix = {arXiv},
       eprint = {2001.08208},
 primaryClass = {astro-ph.HE},
       adsurl = {https://ui.adsabs.harvard.edu/abs/2020MNRAS.493.1088I}
}

@ARTICLE{Ursini20,
       author = {{Ursini}, F. and {Petrucci}, P. -O. and {Bianchi}, S. and {Matt}, G. and {Middei}, R. and {Marcel}, G. and {Ferreira}, J. and {Cappi}, M. and {De Marco}, B. and {De Rosa}, A. and {Malzac}, J. and {Marinucci}, A. and {Ponti}, G. and {Tortosa}, A.},
        title = "{NuSTAR/XMM-Newton monitoring of the Seyfert 1 galaxy HE 1143-1810. Testing the two-corona scenario}",
      journal = {\aap},
         year = 2020,
        month = feb,
       volume = {634},
          eid = {A92},
        pages = {A92},
          doi = {10.1051/0004-6361/201936486},
archivePrefix = {arXiv},
       eprint = {1912.08720},
 primaryClass = {astro-ph.HE},
       adsurl = {https://ui.adsabs.harvard.edu/abs/2020A&A...634A..92U}
}

@ARTICLE{Piconcelli04,
       author = {{Piconcelli}, E. and {Jimenez-Bail{\'o}n}, E. and {Guainazzi}, M. and
         {Schartel}, N. and {Rodr{\'\i}guez-Pascual}, P.~M. and
         {Santos-Lle{\'o}}, M.},
        title = "{Evidence for a multizone warm absorber in the XMM-Newton spectrum of Markarian 304}",
      journal = {\mnras},
         year = 2004,
        month = jun,
       volume = {351},
       number = {1},
        pages = {161-168},
          doi = {10.1111/j.1365-2966.2004.07764.x},
archivePrefix = {arXiv},
       eprint = {astro-ph/0404263},
 primaryClass = {astro-ph},
       adsurl = {https://ui.adsabs.harvard.edu/abs/2004MNRAS.351..161P}
}

@ARTICLE{Bianchi09caixa1,
       author = {{Bianchi}, S. and {Guainazzi}, M. and {Matt}, G. and
         {Fonseca Bonilla}, N. and {Ponti}, G.},
        title = "{CAIXA: a catalogue of AGN in the XMM-Newton archive. I. Spectral analysis}",
      journal = {\aap},
         year = 2009,
        month = feb,
       volume = {495},
       number = {2},
        pages = {421-430},
          doi = {10.1051/0004-6361:200810620},
archivePrefix = {arXiv},
       eprint = {0811.1126},
 primaryClass = {astro-ph},
       adsurl = {https://ui.adsabs.harvard.edu/abs/2009A&A...495..421B}
}

@ARTICLE{KaastraBleeker16,
       author = {{Kaastra}, J.~S. and {Bleeker}, J.~A.~M.},
        title = "{Optimal binning of X-ray spectra and response matrix design}",
      journal = {\aap},
         year = 2016,
        month = mar,
       volume = {587},
          eid = {A151},
        pages = {A151},
          doi = {10.1051/0004-6361/201527395},
archivePrefix = {arXiv},
       eprint = {1601.05309},
 primaryClass = {astro-ph.IM},
       adsurl = {https://ui.adsabs.harvard.edu/abs/2016A&A...587A.151K}
}

@ARTICLE{Steenbrugge03xabs,
       author = {{Steenbrugge}, K.~C. and {Kaastra}, J.~S. and {de Vries}, C.~P. and {Edelson}, R.},
        title = "{XMM-NEWTON High resolution spectroscopy of NGC 5548}",
      journal = {\aap},
         year = 2003,
        month = may,
       volume = {402},
        pages = {477-486},
          doi = {10.1051/0004-6361:20030261},
archivePrefix = {arXiv},
       eprint = {astro-ph/0302493},
 primaryClass = {astro-ph},
       adsurl = {https://ui.adsabs.harvard.edu/abs/2003A&A...402..477S}
}

@INPROCEEDINGS{Kaastra96spex,
       author = {{Kaastra}, J.~S. and {Mewe}, R. and {Nieuwenhuijzen}, H.},
        title = "{SPEX: a new code for spectral analysis of X \& UV spectra.}",
    booktitle = {UV and X-ray Spectroscopy of Astrophysical and Laboratory Plasmas},
         year = 1996,
        month = jan,
        pages = {411-414},
       adsurl = {https://ui.adsabs.harvard.edu/abs/1996uxsa.conf..411K}
}

@ARTICLE{Middei19Moca,
       author = {{Middei}, R. and {Bianchi}, S. and {Marinucci}, A. and {Matt}, G. and {Petrucci}, P.-O. and {Tamborra}, F. and {Tortosa}, A.},
        title = "{Relations between phenomenological and physical parameters in the hot coronae of AGNs computed with the MoCA code}",
      journal = {\aap},
         year = 2019,
        month = oct,
       volume = {630},
          eid = {A131},
        pages = {A131},
          doi = {10.1051/0004-6361/201935881},
archivePrefix = {arXiv},
       eprint = {1908.10373},
 primaryClass = {astro-ph.HE},
       adsurl = {https://ui.adsabs.harvard.edu/abs/2019A&A...630A.131M}
}

@ARTICLE{Malzac01,
       author = {{Malzac}, Julien and {Beloborodov}, Andrei M. and {Poutanen}, Juri},
        title = "{X-ray spectra of accretion discs with dynamic coronae}",
      journal = {\mnras},
         year = 2001,
        month = sep,
       volume = {326},
       number = {2},
        pages = {417-427},
          doi = {10.1046/j.1365-8711.2001.04450.x},
archivePrefix = {arXiv},
       eprint = {astro-ph/0102490},
 primaryClass = {astro-ph},
       adsurl = {https://ui.adsabs.harvard.edu/abs/2001MNRAS.326..417M}
}

@ARTICLE{Akaike74,
  author={Akaike, H.},
  journal={IEEE Transactions on Automatic Control}, 
  title={A new look at the statistical model identification}, 
  year={1974},
  volume={19},
  number={6},
  pages={716-723},
  doi={10.1109/TAC.1974.1100705}}

@ARTICLE{Schwarz78,
       author = {{Schwarz}, Gideon},
        title = "{Estimating the Dimension of a Model}",
      journal = {Annals of Statistics},
         year = 1978,
        month = jul,
       volume = {6},
       number = {2},
        pages = {461-464},
       adsurl = {https://ui.adsabs.harvard.edu/abs/1978AnSta...6..461S}
}

@ARTICLE{Middei20_Mrk359,
       author = {{Middei}, R. and {Petrucci}, P.-O. and {Bianchi}, S. and {Ursini}, F. and {Cappi}, M. and {Clavel}, M. and {De Rosa}, A. and {Marinucci}, A. and {Matt}, G. and {Tortosa}, A.},
        title = "{The soft excess of the NLS1 galaxy Mrk 359 studied with an XMM-Newton-NuSTAR monitoring campaign}",
      journal = {\aap},
         year = 2020,
        month = aug,
       volume = {640},
          eid = {A99},
        pages = {A99},
          doi = {10.1051/0004-6361/202038112},
archivePrefix = {arXiv},
       eprint = {2006.09005},
 primaryClass = {astro-ph.HE},
       adsurl = {https://ui.adsabs.harvard.edu/abs/2020A&A...640A..99M}
}

@ARTICLE{Serafinelli19,
       author = {{Serafinelli}, Roberto and {Tombesi}, Francesco and {Vagnetti}, Fausto and {Piconcelli}, Enrico and {Gaspari}, Massimo and {Saturni}, Francesco G.},
        title = "{Multiphase quasar-driven outflows in PG 1114+445. I. Entrained ultra-fast outflows}",
      journal = {\aap},
         year = 2019,
        month = jul,
       volume = {627},
          eid = {A121},
        pages = {A121},
          doi = {10.1051/0004-6361/201935275},
archivePrefix = {arXiv},
       eprint = {1906.02765},
 primaryClass = {astro-ph.GA},
       adsurl = {https://ui.adsabs.harvard.edu/abs/2019A&A...627A.121S}
}

@ARTICLE{Lanzuisi19,
       author = {{Lanzuisi}, G. and {Gilli}, R. and {Cappi}, M. and {Dadina}, M. and {Bianchi}, S. and {Brusa}, M. and {Chartas}, G. and {Civano}, F. and {Comastri}, A. and {Marinucci}, A. and {Middei}, R. and {Piconcelli}, E. and {Vignali}, C. and {Brandt}, W.~N. and {Tombesi}, F. and {Gaspari}, M.},
        title = "{NuSTAR Measurement of Coronal Temperature in Two Luminous, High-redshift Quasars}",
      journal = {\apjl},
         year = 2019,
        month = apr,
       volume = {875},
       number = {2},
          eid = {L20},
        pages = {L20},
          doi = {10.3847/2041-8213/ab15dc},
archivePrefix = {arXiv},
       eprint = {1904.04784},
 primaryClass = {astro-ph.HE},
       adsurl = {https://ui.adsabs.harvard.edu/abs/2019ApJ...875L..20L}
}

@ARTICLE{Kammoun17,
       author = {{Kammoun}, E.~S. and {Risaliti}, G. and {Stern}, D. and {Jun}, H.~D. and {Graham}, M. and {Celotti}, A. and {Behar}, E. and {Elvis}, M. and {Harrison}, F.~A. and {Matt}, G. and {Walton}, D.~J.},
        title = "{Coronal properties of the luminous radio-quiet quasar QSO B2202-209}",
      journal = {\mnras},
         year = 2017,
        month = feb,
       volume = {465},
       number = {2},
        pages = {1665-1671},
          doi = {10.1093/mnras/stw2897},
archivePrefix = {arXiv},
       eprint = {1611.02306},
 primaryClass = {astro-ph.HE},
       adsurl = {https://ui.adsabs.harvard.edu/abs/2017MNRAS.465.1665K}
}

@ARTICLE{KingPounds15,
       author = {{King}, Andrew and {Pounds}, Ken},
        title = "{Powerful Outflows and Feedback from Active Galactic Nuclei}",
      journal = {\araa},
         year = 2015,
        month = aug,
       volume = {53},
        pages = {115-154},
          doi = {10.1146/annurev-astro-082214-122316},
archivePrefix = {arXiv},
       eprint = {1503.05206},
 primaryClass = {astro-ph.GA},
       adsurl = {https://ui.adsabs.harvard.edu/abs/2015ARA&A..53..115K}
}

@ARTICLE{Xie17,
       author = {{Xie}, Yanxia and {Li}, Aigen and {Hao}, Lei},
        title = "{Silicate Dust in Active Galactic Nuclei}",
      journal = {\apjs},
         year = 2017,
        month = jan,
       volume = {228},
       number = {1},
          eid = {6},
        pages = {6},
          doi = {10.3847/1538-4365/228/1/6},
archivePrefix = {arXiv},
       eprint = {1612.04293},
 primaryClass = {astro-ph.GA},
       adsurl = {https://ui.adsabs.harvard.edu/abs/2017ApJS..228....6X}
}

@ARTICLE{Bertola22,
       author = {{Bertola}, E. and {Vignali}, C. and {Lanzuisi}, G. and {Dadina}, M. and {Cappi}, M. and {Gilli}, R. and {Matzeu}, G.~A. and {Chartas}, G. and {Piconcelli}, E. and {Comastri}, A.},
        title = "{The properties of the X-ray corona in the distant (z = 3.91) quasar APM 08279+5255}",
      journal = {\aap},
         year = 2022,
        month = jun,
       volume = {662},
          eid = {A98},
        pages = {A98},
          doi = {10.1051/0004-6361/202142642},
archivePrefix = {arXiv},
       eprint = {2205.01113},
 primaryClass = {astro-ph.HE},
       adsurl = {https://ui.adsabs.harvard.edu/abs/2022A&A...662A..98B}
}

@article{Mantel1966,
  title = {Evaluation of survival data and two new rank order statistics arising in its consideration},
  author = {Mantel, Nathan},
  journal = {Cancer Chemotherapy Reports},
  volume = {50},
  number = {3},
  pages = {163--170},
  year = {1966},
  pmid = {5910392}
}

@article{DavidsonPilon2019, 
    doi = {10.21105/joss.01317}, 
    url = {https://doi.org/10.21105/joss.01317}, 
    year = {2019}, 
    publisher = {The Open Journal}, 
    volume = {4}, 
    number = {40}, 
    pages = {1317}, 
    author = {Davidson-Pilon, Cameron}, 
    title = {lifelines: survival analysis in Python}, 
    journal = {Journal of Open Source Software} }

@ARTICLE{Petrucci00,
       author = {{Petrucci}, P.~O. and {Haardt}, F. and {Maraschi}, L. and {Grandi}, P. and {Matt}, G. and {Nicastro}, F. and {Piro}, L. and {Perola}, G.~C. and {De Rosa}, A.},
        title = "{Testing Comptonizing Coronae on a Long BeppoSAX Observation of the Seyfert 1 Galaxy NGC 5548}",
      journal = {\apj},
         year = 2000,
        month = sep,
       volume = {540},
       number = {1},
        pages = {131-142},
          doi = {10.1086/309319},
archivePrefix = {arXiv},
       eprint = {astro-ph/0004118},
 primaryClass = {astro-ph},
       adsurl = {https://ui.adsabs.harvard.edu/abs/2000ApJ...540..131P}
}

@ARTICLE{Matt00,
       author = {{Matt}, G. and {Fabian}, A.~C. and {Guainazzi}, M. and {Iwasawa}, K. and {Bassani}, L. and {Malaguti}, G.},
        title = "{The X-ray spectra of Compton-thick Seyfert 2 galaxies as seen by BeppoSAX}",
      journal = {\mnras},
         year = 2000,
        month = oct,
       volume = {318},
       number = {1},
        pages = {173-179},
          doi = {10.1046/j.1365-8711.2000.03721.x},
archivePrefix = {arXiv},
       eprint = {astro-ph/0005219},
 primaryClass = {astro-ph},
       adsurl = {https://ui.adsabs.harvard.edu/abs/2000MNRAS.318..173M}
}

@ARTICLE{Madathil24,
       author = {{Madathil-Pottayil}, A. and {Walton}, D.~J. and {Garc{\'\i}a}, Javier and {Miller}, Jon and {Gallo}, Luigi C. and {Ricci}, C. and {Reynolds}, Mark T. and {Stern}, D. and {Dauser}, T. and {Jiang}, Jiachen and {Alston}, William and {Fabian}, A.~C. and {Hardcastle}, M.~J. and {Kosec}, Peter and {Nardini}, Emanuele and {Reynolds}, Christopher S.},
        title = "{Exploring the high-density reflection model for the soft excess in RBS 1124}",
      journal = {\mnras},
         year = 2024,
        month = oct,
       volume = {534},
       number = {1},
        pages = {608-620},
          doi = {10.1093/mnras/stae2104},
archivePrefix = {arXiv},
       eprint = {2409.01395},
 primaryClass = {astro-ph.HE},
       adsurl = {https://ui.adsabs.harvard.edu/abs/2024MNRAS.534..608M}
}

@ARTICLE{Tamborra18,
       author = {{Tamborra}, Francesco and {Papadakis}, Iossif and {Dov{\v{c}}iak}, Michal and {Svoboda}, Ji{\v{r}}i},
        title = "{On the high energy cut-off of accreting sources: Is general relativity relevant?}",
      journal = {\mnras},
         year = 2018,
        month = apr,
       volume = {475},
       number = {2},
        pages = {2045-2050},
          doi = {10.1093/mnras/stx2987},
archivePrefix = {arXiv},
       eprint = {1711.06183},
 primaryClass = {astro-ph.HE},
       adsurl = {https://ui.adsabs.harvard.edu/abs/2018MNRAS.475.2045T}
}

@article{Lanzuisi26,
  title={},
  author={{Lanzuisi}, Giorgio and {Borrelli}, Laura and {Piconcelli}, Enrico and {Brusa}, Marcella and {Zappacosta}, Luca and {Comastri} Andrea},
  journal = {\aap~submitted.},
  year={2026}
}

@ARTICLE{Zappacosta18,
       author = {{Zappacosta}, L. and {Comastri}, A. and {Civano}, F. and {Puccetti}, S. and {Fiore}, F. and {Aird}, J. and {Del Moro}, A. and {Lansbury}, G.~B. and {Lanzuisi}, G. and {Goulding}, A. and {Mullaney}, J.~R. and {Stern}, D. and {Ajello}, M. and {Alexander}, D.~M. and {Ballantyne}, D.~R. and {Bauer}, F.~E. and {Brandt}, W.~N. and {Chen}, C.-T.~J. and {Farrah}, D. and {Harrison}, F.~A. and {Gandhi}, P. and {Lanz}, L. and {Masini}, A. and {Marchesi}, S. and {Ricci}, C. and {Treister}, E.},
        title = "{The NuSTAR  Extragalactic Surveys: X-Ray Spectroscopic Analysis of the Bright Hard-band Selected Sample}",
      journal = {\apj},
         year = 2018,
        month = feb,
       volume = {854},
       number = {1},
          eid = {33},
        pages = {33},
          doi = {10.3847/1538-4357/aaa550},
archivePrefix = {arXiv},
       eprint = {1801.04280},
 primaryClass = {astro-ph.HE},
       adsurl = {https://ui.adsabs.harvard.edu/abs/2018ApJ...854...33Z}
}

@article{Borrelli26,
  title={},
  author={{Borrelli}, Laura and {Lanzuisi}, Giorgio and {Piconcelli}, Enrico and {Brusa}, Marcella and {Zappacosta}, Luca and {Comastri} Andrea},
  journal = {in preparation.},
  year={2026}
}

@ARTICLE{Duras20,
       author = {{Duras}, F. and {Bongiorno}, A. and {Ricci}, F. and {Piconcelli}, E. and {Shankar}, F. and {Lusso}, E. and {Bianchi}, S. and {Fiore}, F. and {Maiolino}, R. and {Marconi}, A. and {Onori}, F. and {Sani}, E. and {Schneider}, R. and {Vignali}, C. and {La Franca}, F.},
        title = "{Universal bolometric corrections for active galactic nuclei over seven luminosity decades}",
      journal = {\aap},
         year = 2020,
        month = apr,
       volume = {636},
          eid = {A73},
        pages = {A73},
          doi = {10.1051/0004-6361/201936817},
archivePrefix = {arXiv},
       eprint = {2001.09984},
 primaryClass = {astro-ph.GA},
       adsurl = {https://ui.adsabs.harvard.edu/abs/2020A&A...636A..73D}
}

@ARTICLE{Matzeu23,
       author = {{Matzeu}, G.~A. and {Brusa}, M. and {Lanzuisi}, G. and {Dadina}, M. and {Bianchi}, S. and {Kriss}, G. and {Mehdipour}, M. and {Nardini}, E. and {Chartas}, G. and {Middei}, R. and {Piconcelli}, E. and {Gianolli}, V. and {Comastri}, A. and {Longinotti}, A.~L. and {Krongold}, Y. and {Ricci}, F. and {Petrucci}, P.~O. and {Tombesi}, F. and {Luminari}, A. and {Zappacosta}, L. and {Miniutti}, G. and {Gaspari}, M. and {Behar}, E. and {Bischetti}, M. and {Mathur}, S. and {Perna}, M. and {Giustini}, M. and {Grandi}, P. and {Torresi}, E. and {Vignali}, C. and {Bruni}, G. and {Cappi}, M. and {Costantini}, E. and {Cresci}, G. and {De Marco}, B. and {De Rosa}, A. and {Gilli}, R. and {Guainazzi}, M. and {Kaastra}, J. and {Kraemer}, S. and {La Franca}, F. and {Marconi}, A. and {Panessa}, F. and {Ponti}, G. and {Proga}, D. and {Ursini}, F. and {Baldini}, P. and {Fiore}, F. and {King}, A.~R. and {Maiolino}, R. and {Matt}, G. and {Merloni}, A.},
        title = "{Supermassive Black Hole Winds in X-rays: SUBWAYS. I. Ultra-fast outflows in quasars beyond the local Universe}",
      journal = {\aap},
         year = 2023,
        month = feb,
       volume = {670},
          eid = {A182},
        pages = {A182},
          doi = {10.1051/0004-6361/202245036},
archivePrefix = {arXiv},
       eprint = {2212.02960},
 primaryClass = {astro-ph.HE},
       adsurl = {https://ui.adsabs.harvard.edu/abs/2023A&A...670A.182M}
}

@ARTICLE{Kompaneets57,
       author = {{Kompaneets}, A.~S.},
        title = "{The Establishment of Thermal Equilibrium between Quanta and Electrons}",
      journal = {Soviet Journal of Experimental and Theoretical Physics},
         year = 1957,
        month = may,
       volume = {4},
       number = {5},
        pages = {730-737},
       adsurl = {https://ui.adsabs.harvard.edu/abs/1957JETP....4..730K}
}

@article{Ricci2018,
   title={BAT AGN Spectroscopic Survey – XII. The relation between coronal properties of active galactic nuclei and the Eddington ratio},
   volume={480},
   ISSN={1365-2966},
   url={http://dx.doi.org/10.1093/mnras/sty1879},
   DOI={10.1093/mnras/sty1879},
   number={2},
   journal={MNRAS},
   publisher={Oxford University Press (OUP)},
   author={Ricci, C and Ho, L C and Fabian, A C and Trakhtenbrot, B and Koss, M J and Ueda, Y and Lohfink, A and Shimizu, T and Bauer, F E and Mushotzky, R and Schawinski, K and Paltani, S and Lamperti, I and Treister, E and Oh, K},
   year={2018},
   month=jul, pages={1819–1830} }

@article{Kaplan1958,
author = {E. L. Kaplan and Paul Meier},
title = {Nonparametric Estimation from Incomplete Observations},
journal = {Journal of the American Statistical Association},
volume = {53},
number = {282},
pages = {457--481},
year = {1958},
publisher = {ASA Website},
doi = {10.1080/01621459.1958.10501452},
}

@article{Ricci2017,
   title={BAT AGN Spectroscopic Survey. V. X-Ray Properties of the Swift/BAT 70-month AGN Catalog},
   volume={233},
   ISSN={1538-4365},
   url={http://dx.doi.org/10.3847/1538-4365/aa96ad},
   DOI={10.3847/1538-4365/aa96ad},
   number={2},
   journal={The Astrophysical Journal Supplement Series},
   publisher={American Astronomical Society},
   author={Ricci, C. and Trakhtenbrot, B. and Koss, M. J. and Ueda, Y. and Del Vecchio, I. and Treister, E. and Schawinski, K. and Paltani, S. and Oh, K. and Lamperti, I. and Berney, S. and Gandhi, P. and Ichikawa, K. and Bauer, F. E. and Ho, L. C. and Asmus, D. and Beckmann, V. and Soldi, S. and Baloković, M. and Gehrels, N. and Markwardt, C. B.},
   year={2017},
   month=dec, pages={17} }

@article{Balokovic_2020,
   title={NuSTAR Survey of Obscured Swift/BAT-selected Active Galactic Nuclei. II. Median High-energy Cutoff in Seyfert II Hard X-Ray Spectra},
   volume={905},
   ISSN={1538-4357},
   url={http://dx.doi.org/10.3847/1538-4357/abc342},
   DOI={10.3847/1538-4357/abc342},
   number={1},
   journal={ApJ},
   publisher={American Astronomical Society},
   author={Baloković, M. and Harrison, F. A. and Madejski, G. and Comastri, A. and Ricci, C. and Annuar, A. and Ballantyne, D. R. and Boorman, P. and Brandt, W. N. and Brightman, M. and Gandhi, P. and Kamraj, N. and Koss, M. J. and Marchesi, S. and Marinucci, A. and Masini, A. and Matt, G. and Stern, D. and Urry, C. M.},
   year={2020},
   month=dec, pages={41} }

@INPROCEEDINGS{Asurv1990,
       author = {{Isobel}, T. and {Feigelson}, E.~D.},
        title = "{ASURV. The Pennsylvania State University. Report for the period Sep 1987 - Jan 1990.}",
    booktitle = {Bulletin of the American Astronomical Society},
         year = 1990,
       volume = {22},
        month = mar,
        pages = {917-918},
       adsurl = {https://ui.adsabs.harvard.edu/abs/1990BAAS...22..917I}
}

@article{Kamraj2022,
   title={X-Ray Coronal Properties of Swift/BAT-selected Seyfert 1 Active Galactic Nuclei},
   volume={927},
   ISSN={1538-4357},
   url={http://dx.doi.org/10.3847/1538-4357/ac45f6},
   DOI={10.3847/1538-4357/ac45f6},
   number={1},
   journal={ApJ},
   publisher={American Astronomical Society},
   author={Kamraj, Nikita and Brightman, Murray and Harrison, Fiona A. and Stern, Daniel and García, Javier A. and Baloković, Mislav and Ricci, Claudio and Koss, Michael J. and Mejía-Restrepo, Julian E. and Oh, Kyuseok and Powell, Meredith C. and Urry, C. Megan},
   year={2022},
   month=mar, pages={42} }

@ARTICLE{Fabian2015,
       author = {{Fabian}, A.~C. and {Lohfink}, A. and {Kara}, E. and {Parker}, M.~L. and {Vasudevan}, R. and {Reynolds}, C.~S.},
        title = "{Properties of AGN coronae in the NuSTAR era}",
      journal = {\mnras},
         year = 2015,
        month = aug,
       volume = {451},
       number = {4},
        pages = {4375-4383},
          doi = {10.1093/mnras/stv1218},
archivePrefix = {arXiv},
       eprint = {1505.07603},
 primaryClass = {astro-ph.HE},
       adsurl = {https://ui.adsabs.harvard.edu/abs/2015MNRAS.451.4375F}
}

@ARTICLE{Mehdipour23,
       author = {{Mehdipour}, M. and {Kriss}, G.~A. and {Brusa}, M. and {Matzeu}, G.~A. and {Gaspari}, M. and {Kraemer}, S.~B. and {Mathur}, S. and {Behar}, E. and {Bianchi}, S. and {Cappi}, M. and {Chartas}, G. and {Costantini}, E. and {Cresci}, G. and {Dadina}, M. and {De Marco}, B. and {De Rosa}, A. and {Dunn}, J.~P. and {Gianolli}, V.~E. and {Giustini}, M. and {Kaastra}, J.~S. and {King}, A.~R. and {Krongold}, Y. and {La Franca}, F. and {Lanzuisi}, G. and {Longinotti}, A.~L. and {Luminari}, A. and {Middei}, R. and {Miniutti}, G. and {Nardini}, E. and {Perna}, M. and {Petrucci}, P. -O. and {Piconcelli}, E. and {Ponti}, G. and {Ricci}, F. and {Tombesi}, F. and {Ursini}, F. and {Vignali}, C. and {Zappacosta}, L.},
        title = "{Supermassive Black Hole Winds in X-rays: SUBWAYS. II. HST UV spectroscopy of winds at intermediate redshifts}",
      journal = {\aap},
         year = 2023,
        month = feb,
       volume = {670},
          eid = {A183},
        pages = {A183},
          doi = {10.1051/0004-6361/202245047},
archivePrefix = {arXiv},
       eprint = {2212.02961},
 primaryClass = {astro-ph.HE},
       adsurl = {https://ui.adsabs.harvard.edu/abs/2023A&A...670A.183M}
}

@article{Fabian_2017,
   title={Properties of AGN coronae in the NuSTAR era – II. Hybrid plasma},
   volume={467},
   ISSN={1365-2966},
   url={http://dx.doi.org/10.1093/mnras/stx221},
   DOI={10.1093/mnras/stx221},
   number={3},
   journal={MNRAS},
   publisher={Oxford University Press (OUP)},
   author={Fabian, A. C. and Lohfink, A. and Belmont, R. and Malzac, J. and Coppi, P.},
   year={2017},
   month=feb, pages={2566–2570} }

@ARTICLE{Wik2014,
       author = {{Wik}, Daniel R. and {Hornstrup}, A. and {Molendi}, S. and {Madejski}, G. and {Harrison}, F.~A. and {Zoglauer}, A. and {Grefenstette}, B.~W. and {Gastaldello}, F. and {Madsen}, K.~K. and {Westergaard}, N.~J. and {Ferreira}, D.~D.~M. and {Kitaguchi}, T. and {Pedersen}, K. and {Boggs}, S.~E. and {Christensen}, F.~E. and {Craig}, W.~W. and {Hailey}, C.~J. and {Stern}, D. and {Zhang}, W.~W.},
        title = "{NuSTAR Observations of the Bullet Cluster: Constraints on Inverse Compton Emission}",
      journal = {\apj},
         year = 2014,
        month = sep,
       volume = {792},
       number = {1},
          eid = {48},
        pages = {48},
          doi = {10.1088/0004-637X/792/1/48},
archivePrefix = {arXiv},
       eprint = {1403.2722},
 primaryClass = {astro-ph.HE},
       adsurl = {https://ui.adsabs.harvard.edu/abs/2014ApJ...792...48W}
}

@ARTICLE{Sunyaev1980,
       author = {{Sunyaev}, R.~A. and {Titarchuk}, L.~G.},
        title = "{Comptonization of X-Rays in Plasma Clouds - Typical Radiation Spectra}",
      journal = {\aap},
         year = 1980,
        month = jun,
       volume = {86},
        pages = {121},
       adsurl = {https://ui.adsabs.harvard.edu/abs/1980A&A....86..121S}
}

@ARTICLE{Holt1980,
       author = {{Holt}, S.~S. and {Mushotzky}, R.~F. and {Becker}, R.~H. and {Boldt}, E.~A. and {Serlemitsos}, P.~J. and {Szymkowiak}, A.~E. and {White}, N.~E.},
        title = "{X-ray spectral constraints on the broad-line cloud geometry of NGC 4151.}",
      journal = {\apjl},
         year = 1980,
        month = oct,
       volume = {241},
        pages = {L13-L17},
          doi = {10.1086/183350},
       adsurl = {https://ui.adsabs.harvard.edu/abs/1980ApJ...241L..13H}
}

@ARTICLE{Pravdo81,
       author = {{Pravdo}, S.~H. and {Nugent}, J.~J. and {Nousek}, J.~A. and {Jensen}, K. and {Wilson}, A.~S. and {Becker}, R.~H.},
        title = "{Discovery of a Seyfert 1 galaxy with an unusually soft X-ray spectrum}",
      journal = {\apj},
         year = 1981,
        month = dec,
       volume = {251},
        pages = {501-507},
          doi = {10.1086/159489},
       adsurl = {https://ui.adsabs.harvard.edu/abs/1981ApJ...251..501P}
}

@article{Arnaud1996,
       author = {{Arnaud}, K.A.},
        title = "{XSPEC: The First Ten Years}",
    booktitle = {Astronomical Data Analysis Software and Systems V},
         year = 1996,
       editor = {{Jacoby}, George H. and {Barnes}, Jeannette},
       series = {Astronomical Society of the Pacific Conference Series},
       volume = {101},
        month = jan,
        pages = {17},
journal={Astronomical Society of the Pacific Conference Series},
       adsurl = {https://ui.adsabs.harvard.edu/abs/1996ASPC..101...17A}
}

@misc{Tagliacozzo2023,
      title={The geometry of the hot corona in MCG-05-23-16 constrained by X-ray polarimetry}, 
      author={D. Tagliacozzo and A. Marinucci and F. Ursini and G. Matt and S. Bianchi and L. Baldini and T. Barnouin and N. Cavero Rodriguez and A. De Rosa and L. Di Gesu and M. Dovciak and D. Harper and A. Ingram and V. Karas and D. E. Kim and H. Krawczynski and G. Madejski and F. Marin and R. Middei and H. L. Marshall and F. Muleri and C. Panagiotou and P. O. Petrucci and J. Podgorny and J. Poutanen and S. Puccetti and P. Soffitta and F. Tombesi and A. Veledina and W. Zhang and I. Agudo and L. A. Antonelli and M. Bachetti and W. H. Baumgartner and R. Bellazzini and S. D. Bongiorno and R. Bonino and A. Brez and N. Bucciantini and F. Capitanio and S. Castellano and E. Cavazzuti and C. T. Chen and S. Ciprini and E. Costa and E. Del Monte and N. Di Lalla and A. Di Marco and I. Donnarumma and V. Doroshenko and S. R. Ehlert and T. Enoto and Y. Evangelista and S. Fabiani and R. Ferrazzoli and J. A. Garcia and S. Gunji and J. Heyl and W. Iwakiri and S. G. Jorstad and P. Kaaret and F. Kislat and T. Kitaguchi and J. J. Kolodziejczak and F. La Monaca and L. Latronico and I. Liodakis and S. Maldera and A. Manfreda and A. P. Marscher and F. Massaro and I. Mitsuishi and T. Mizuno and M. Negro and C. Y. Ng and S. L. O'Dell and N. Omodei and C. Oppedisano and A. Papitto and G. G. Pavlov and A. L. Peirson and M. Perri and M. Pesce Rollins and M. Pilia and A. Possenti and B. D. Ramsey and J. Rankin and A. Ratheesh and O. J. Roberts and R. W. Romani and C. Sgrò and P. Slane and G. Spandre and D. A. Swartz and T. Tamagawa and F. Tavecchio and R. Taverna and Y. Tawara and A. F. Tennant and N. E. Thomas and A. Trois and S. S. Tsygankov and R. Turolla and J. Vink and M. C. Weisskopf and K. Wu and F. Xie and S. Zane},
      year={2023},
      eprint={2305.10213},
      archivePrefix={arXiv},
      primaryClass={astro-ph.HE},
      url={https://arxiv.org/abs/2305.10213}, 
}

@ARTICLE{Cavaliere1980,
       author = {{Cavaliere}, A. and {Morrison}, P.},
        title = "{Extreme nonthermal radiation from active galactic nuclei}",
      journal = {\apjl},
         year = 1980,
        month = jun,
       volume = {238},
        pages = {L63-L66},
          doi = {10.1086/183259},
       adsurl = {https://ui.adsabs.harvard.edu/abs/1980ApJ...238L..63C}
}

@article{Wardzinski2000,
   title={Thermal synchrotron radiation and its Comptonization in compact X-ray sources},
   volume={314},
   ISSN={1365-2966},
   url={http://dx.doi.org/10.1046/j.1365-8711.2000.03297.x},
   DOI={10.1046/j.1365-8711.2000.03297.x},
   number={1},
   journal={MNRAS},
   publisher={Oxford University Press (OUP)},
   author={Wardzinski, G. and Zdziarski, A. A.},
   year={2000},
   month=may, pages={183–198} }

@ARTICLE{Pal2024,
       author = {{Pal}, Indrani and {Anju}, A. and {Sreehari}, H. and {Rameshan}, Gitika and {Stalin}, C.~S. and {Ricci}, Claudio and {Marchesi}, S.},
        title = "{On the Properties of X-Ray Corona in Seyfert 1 Galaxies}",
      journal = {\apj},
         year = 2024,
        month = nov,
       volume = {976},
       number = {1},
          eid = {145},
        pages = {145},
          doi = {10.3847/1538-4357/ad8088},
archivePrefix = {arXiv},
       eprint = {2310.18196},
 primaryClass = {astro-ph.HE},
       adsurl = {https://ui.adsabs.harvard.edu/abs/2024ApJ...976..145P}
}

@article{Serafinelli2024,
   title={Investigating the interplay between the coronal properties and the hard X-ray variability of active galactic nuclei with NuSTAR},
   volume={690},
   ISSN={1432-0746},
   url={http://dx.doi.org/10.1051/0004-6361/202450777},
   DOI={10.1051/0004-6361/202450777},
   journal={A \& A},
   publisher={EDP Sciences},
   author={Serafinelli, Roberto and De Rosa, Alessandra and Tortosa, Alessia and Stella, Luigi and Vagnetti, Fausto and Bianchi, Stefano and Ricci, Claudio and Kammoun, Elias and Petrucci, Pierre-Olivier and Middei, Riccardo and Lanzuisi, Giorgio and Marinucci, Andrea and Ursini, Francesco and Matt, Giorgio},
   year={2024},
   month=oct, pages={A145} }

@ARTICLE{Zhang23,
       author = {{Zhang}, Zuobin and {Jiang}, Jiachen and {Liu}, Honghui and {Bambi}, Cosimo and {Reynolds}, Christopher S. and {Fabian}, Andrew C. and {Dauser}, Thomas and {Madsen}, Kristin and {Young}, Andrew and {Gallo}, Luigi and {Yu}, Zhibo and {Tomsick}, John},
        title = "{The Low-temperature Corona in ESO 511-G030 Revealed by NuSTAR and XMM-Newton}",
      journal = {\apj},
         year = 2023,
        month = may,
       volume = {949},
       number = {1},
          eid = {4},
        pages = {4},
          doi = {10.3847/1538-4357/acc38f},
archivePrefix = {arXiv},
       eprint = {2208.01452},
 primaryClass = {astro-ph.HE},
       adsurl = {https://ui.adsabs.harvard.edu/abs/2023ApJ...949....4Z}
}

@article{Liu23,
    author = {Liu, Jie-Ying and Mao, Jirong and Liu, B F},
    title = {Magnetic-reconnection-heated corona model: implication of hybrid electrons for hard X-ray emission of luminous active galactic nuclei},
    journal = {Monthly Notices of the Royal Astronomical Society},
    volume = {527},
    number = {3},
    pages = {5627-5637},
    year = {2023},
    month = {11},
    issn = {0035-8711},
    doi = {10.1093/mnras/stad3615},
    url = {https://doi.org/10.1093/mnras/stad3615},
    eprint = {https://academic.oup.com/mnras/article-pdf/527/3/5627/53971189/stad3615.pdf},
}

@inproceedings{Fabian2005,
    author = "Fabian, Andrew C. and Miniutti, Giovanni",
    title = "{The X-ray spectra of accreting Kerr black holes}",
    booktitle = "{Kerr Fest: Black Holes in Astrophysics, General Relativity and Quantum Gravity}",
    eprint = "astro-ph/0507409",
    archivePrefix = "arXiv",
    month = "7",
    year = "2005"
}

@ARTICLE{Gianolli23,
       author = {{Gianolli}, V.~E. and {Kim}, D.~E. and {Bianchi}, S. and {Ag{\'\i}s-Gonz{\'a}lez}, B. and {Madejski}, G. and {Marin}, F. and {Marinucci}, A. and {Matt}, G. and {Middei}, R. and {Petrucci}, P.-O. and {Soffitta}, P. and {Tagliacozzo}, D. and {Tombesi}, F. and {Ursini}, F. and {Barnouin}, T. and {De Rosa}, A. and {Di Gesu}, L. and {Ingram}, A. and {Loktev}, V. and {Panagiotou}, C. and {Podgorny}, J. and {Poutanen}, J. and {Puccetti}, S. and {Ratheesh}, A. and {Veledina}, A. and {Zhang}, W. and {Agudo}, I. and {Antonelli}, L.~A. and {Bachetti}, M. and {Baldini}, L. and {Baumgartner}, W.~H. and {Bellazzini}, R. and {Bongiorno}, S.~D. and {Bonino}, R. and {Brez}, A. and {Bucciantini}, N. and {Capitanio}, F. and {Castellano}, S. and {Cavazzuti}, E. and {Chen}, C.-T. and {Ciprini}, S. and {Costa}, E. and {Del Monte}, E. and {Di Lalla}, N. and {Di Marco}, A. and {Donnarumma}, I. and {Doroshenko}, V. and {Dov{\v{c}}iak}, M. and {Ehlert}, S.~R. and {Enoto}, T. and {Evangelista}, Y. and {Fabiani}, S. and {Ferrazzoli}, R. and {Garc{\'\i}a}, J.~A. and {Gunji}, S. and {Heyl}, J. and {Iwakiri}, W. and {Jorstad}, S.~G. and {Kaaret}, P. and {Karas}, V. and {Kislat}, F. and {Kitaguchi}, T. and {Kolodziejczak}, J.~J. and {Krawczynski}, H. and {La Monaca}, F. and {Latronico}, L. and {Liodakis}, I. and {Maldera}, S. and {Manfreda}, A. and {Marscher}, A.~P. and {Marshall}, H.~L. and {Massaro}, F. and {Mitsuishi}, I. and {Mizuno}, T. and {Muleri}, F. and {Negro}, M. and {Ng}, C.-Y. and {O'Dell}, S.~L. and {Omodei}, N. and {Oppedisano}, C. and {Papitto}, A. and {Pavlov}, G.~G. and {Peirson}, A.~L. and {Perri}, M. and {Pesce-Rollins}, M. and {Pilia}, M. and {Possenti}, A. and {Ramsey}, B.~D. and {Rankin}, J. and {Roberts}, O.~J. and {Romani}, R.~W. and {Sgr{\`o}}, C. and {Slane}, P. and {Spandre}, G. and {Swartz}, D.~A. and {Tamagawa}, T. and {Tavecchio}, F. and {Taverna}, R. and {Tawara}, Y. and {Tennant}, A.~F. and {Thomas}, N.~E. and {Trois}, A. and {Tsygankov}, S.~S. and {Turolla}, R. and {Vink}, J. and {Weisskopf}, M.~C. and {Wu}, K. and {Xie}, F. and {Zane}, S.},
        title = "{Uncovering the geometry of the hot X-ray corona in the Seyfert galaxy NGC 4151 with IXPE}",
      journal = {\mnras},
         year = 2023,
        month = aug,
       volume = {523},
       number = {3},
        pages = {4468-4476},
          doi = {10.1093/mnras/stad1697},
archivePrefix = {arXiv},
       eprint = {2303.12541},
 primaryClass = {astro-ph.GA},
       adsurl = {https://ui.adsabs.harvard.edu/abs/2023MNRAS.523.4468G}
}

@ARTICLE{Gianolli24,
       author = {{Gianolli}, V.~E. and {Bianchi}, S. and {Kammoun}, E. and {Gnarini}, A. and {Marinucci}, A. and {Ursini}, F. and {Parra}, M. and {Tortosa}, A. and {De Rosa}, A. and {Kim}, D.~E. and {Marin}, F. and {Matt}, G. and {Serafinelli}, R. and {Soffitta}, P. and {Tagliacozzo}, D. and {Di Gesu}, L. and {Done}, C. and {Marshall}, H.~L. and {Middei}, R. and {Mikusincova}, R. and {Petrucci}, P.-O. and {Ravi}, S. and {Svoboda}, J. and {Tombesi}, F.},
        title = "{A second view on the X-ray polarization of NGC 4151 with IXPE}",
      journal = {\aap},
         year = 2024,
        month = oct,
       volume = {691},
          eid = {A29},
        pages = {A29},
          doi = {10.1051/0004-6361/202451645},
archivePrefix = {arXiv},
       eprint = {2407.17243},
 primaryClass = {astro-ph.HE},
       adsurl = {https://ui.adsabs.harvard.edu/abs/2024A&A...691A..29G}
}

@INPROCEEDINGS{2024Gianolli,
       author = {{Gianolli}, Vittoria Elvezia and {Bianchi}, Stefano and {Petrucci}, Pierre-Olivier and {Marinucci}, Andrea and {Ingram}, Adam and {Tagliacozzo}, Daniele and {Kim}, Dawoon E. and {Marin}, Frederic and {Matt}, Giorgio and {Soffitta}, Paolo and {Tombesi}, Francesco},
        title = "{Unobscured radio-quiet active galactic nuclei under the eyes of IXPE}",
    booktitle = {Memorie della Societa Astronomica Italiana},
         year = 2024,
       volume = {95},
        month = jan,
        pages = {27-31},
          doi = {10.36116/MEMSAIT_95N1.2024.27},
archivePrefix = {arXiv},
       eprint = {2406.19519},
 primaryClass = {astro-ph.GA},
       adsurl = {https://ui.adsabs.harvard.edu/abs/2024MmSAI..95a..27G}
}

@ARTICLE{Friederic2024,
       author = {{Marin}, Fr{\'e}d{\'e}ric and {Gianolli}, Vittoria E. and {Ingram}, Adam and {Kim}, Dawoon E. and {Marinucci}, Andrea and {Tagliacozzo}, Daniele and {Ursini}, Francesco},
        title = "{An Examination of the Very First Polarimetric X-ray Observations of Radio-Quiet Active Galactic Nuclei}",
      journal = {Galaxies},
         year = 2024,
        month = jul,
       volume = {12},
       number = {4},
          eid = {35},
        pages = {35},
          doi = {10.3390/galaxies12040035},
       adsurl = {https://ui.adsabs.harvard.edu/abs/2024Galax..12...35M}
}

@ARTICLE{Gilfanov2014,
       author = {{Merloni}, A. and {Bongiorno}, A. and {Brusa}, M. and {Iwasawa}, K. and {Mainieri}, V. and {Magnelli}, B. and {Salvato}, M. and {Berta}, S. and {Cappelluti}, N. and {Comastri}, A. and {Fiore}, F. and {Gilli}, R. and {Koekemoer}, A. and {Le Floc'h}, E. and {Lusso}, E. and {Lutz}, D. and {Miyaji}, T. and {Pozzi}, F. and {Riguccini}, L. and {Rosario}, D.~J. and {Silverman}, J. and {Symeonidis}, M. and {Treister}, E. and {Vignali}, C. and {Zamorani}, G.},
        title = "{The incidence of obscuration in active galactic nuclei}",
      journal = {\mnras},
         year = 2014,
        month = feb,
       volume = {437},
       number = {4},
        pages = {3550-3567},
          doi = {10.1093/mnras/stt2149},
archivePrefix = {arXiv},
       eprint = {1311.1305},
 primaryClass = {astro-ph.CO},
       adsurl = {https://ui.adsabs.harvard.edu/abs/2014MNRAS.437.3550M}
}

@ARTICLE{Ballantyne2001,
       author = {{Ballantyne}, D.~R. and {Iwasawa}, K. and {Fabian}, A.~C.},
        title = "{Evidence for ionized accretion discs in five narrow-line Seyfert 1 galaxies}",
      journal = {\mnras},
         year = 2001,
        month = may,
       volume = {323},
       number = {2},
        pages = {506-516},
          doi = {10.1046/j.1365-8711.2001.04234.x},
archivePrefix = {arXiv},
       eprint = {astro-ph/0011360},
 primaryClass = {astro-ph},
       adsurl = {https://ui.adsabs.harvard.edu/abs/2001MNRAS.323..506B}
}

@ARTICLE{Costa2020,
       author = {{Costa}, Tiago and {Pakmor}, R{\"u}diger and {Springel}, Volker},
        title = "{Powering galactic superwinds with small-scale AGN winds}",
      journal = {\mnras},
         year = 2020,
        month = oct,
       volume = {497},
       number = {4},
        pages = {5229-5255},
          doi = {10.1093/mnras/staa2321},
archivePrefix = {arXiv},
       eprint = {2006.05997},
 primaryClass = {astro-ph.GA},
       adsurl = {https://ui.adsabs.harvard.edu/abs/2020MNRAS.497.5229C}
}

@misc{WARD2024,
      title={AGN-driven outflows in clumpy media: multiphase structure and scaling relations}, 
      author={Samuel Ruthven Ward and Tiago Costa and Chris M. Harrison and Vincenzo Mainieri},
      year={2024},
      eprint={2407.17593},
      archivePrefix={arXiv},
      primaryClass={astro-ph.GA},
      url={https://arxiv.org/abs/2407.17593}, 
}

@article{Molina2019,
    author = {Molina, M and Malizia, A and Bassani, L and Ursini, F and Bazzano, A and Ubertini, P},
    title = {Swift/XRT–NuSTAR spectra of type 1 AGN: confirming INTEGRAL results on the high-energy cut-off},
    journal = {Monthly Notices of the Royal Astronomical Society},
    volume = {484},
    number = {2},
    pages = {2735-2746},
    year = {2019},
    month = {01},
    issn = {0035-8711},
    doi = {10.1093/mnras/stz156},
    url = {https://doi.org/10.1093/mnras/stz156},
    eprint = {https://academic.oup.com/mnras/article-pdf/484/2/2735/27662562/stz156.pdf},
}

@article{Younes2019,
doi = {10.3847/1538-4357/aaf38b},
url = {https://doi.org/10.3847/1538-4357/aaf38b},
year = {2019},
month = {jan},
publisher = {The American Astronomical Society},
volume = {870},
number = {2},
pages = {73},
author = {Younes, George and Ptak, Andrew and Ho, Luis C. and Xie, Fu-Guo and Terasima, Yuichi and Yuan, Feng and Huppenkothen, Daniela and Yukita, Mihoko},
title = {NuStarHard X-Ray View of Low-luminosity Active Galactic Nuclei: High-energy Cutoff and Truncated Thin Disk},
journal = {The Astrophysical Journal},
}

@INPROCEEDINGS{Hongjun2014,
       author = {{An}, Hongjun and {Madsen}, Kristin K. and {Westergaard}, Niels J. and {Boggs}, Steven E. and {Christensen}, Finn E. and {Craig}, William W. and {Hailey}, Charles J. and {Harrison}, Fiona A. and {Stern}, Daniel K. and {Zhang}, William W.},
        title = "{In-flight PSF calibration of the NuSTAR hard X-ray optics}",
    booktitle = {Space Telescopes and Instrumentation 2014: Ultraviolet to Gamma Ray},
         year = 2014,
       editor = {{Takahashi}, Tadayuki and {den Herder}, Jan-Willem A. and {Bautz}, Mark},
       series = {Society of Photo-Optical Instrumentation Engineers (SPIE) Conference Series},
       volume = {9144},
        month = jul,
          eid = {91441Q},
        pages = {91441Q},
          doi = {10.1117/12.2055481},
archivePrefix = {arXiv},
       eprint = {1406.7419},
 primaryClass = {astro-ph.IM},
       adsurl = {https://ui.adsabs.harvard.edu/abs/2014SPIE.9144E..1QA}
}
\clearpage
\onecolumn

\begin{appendix}
\label{Appendix}

\section{SUBWAYS sample}
\Cref{SUBWAYS_ID} and \Cref{SUBWAYS_info_obs} report the log of the \xmm and \nustar observations used in this work and the main properties of the SUBWAYS sources, respectively.

\begin{table}[h]
    \centering
      \resizebox{0.9\textwidth}{!}{
    \begin{tabular}{|l|c|c|c|c|c|c|}
    \hline
      Name  &XMM ObsID & NuSTAR ObsID&Obs date (XMM)&Obs date (NuSTAR)&$T_{exp}(\rm XMM)$&$T_{exp}(\rm NuSTAR)$\\
         & & & & & ks & ks \\
         \hline
PG0052+251&0841480101&60661001002&15.07.2019&21.08.2020&34&24\\
PG0953+414&0841480201&60663001002&14.04.2020&02.11.2021&45&68\\
... & ...&... & ...& ...&... &...\\ 
    \end{tabular}
    }
    \caption{ \footnotesize \xmm and \nustar observations of the SUBWAYS sample utilized in this work. From left to right, the table includes the source name, observation ID, date and exposure time (net from pn in the case of \xmm). Sources with multiple observations are identified by the last three digits of their Obs IDs, except for PG1402+261, where observations are distinguished by the first two Obs. ID digits. 
    This table is available in its entirety in machine-readable form at the CDS \url{http...}. A portion is shown here for guidance regarding its form and content.
    }
    \label{SUBWAYS_ID}
\end{table}

\begin{table*}[h!]
    \centering
      \resizebox{0.9\textwidth}{!}{
    \begin{tabular}{|l|c|c|c|c|c|c|c|}
    \hline
      Name  & z&N$_{H}^{Gal^1}$&logL$_x  ^2$&logL$_{Bol}^3$& logM$_{BH}^4$&$log \lambda_{EDD}$&logF$_{2-10 keV}$\\
      
         && $(10^{20} \ cm^{-2})$&(erg/s)&(erg/s)&$(M_{\odot}$)&&(erg $cm^{-2} s^{-1}$)\\
         \hline
PG0052+251&0.154&3.96&44.60&46.19&8.4$^{(a)}$&-0.32&-11.19\\
PG0953+414&0.234&1.09&44.63&46.27&8.2$^{(a)}$&-0.07&-11.55\\
PG1626+554&0.132&1.55&44.08&45.38&8.5$^{(a)}$&-1.26&-11.55\\
PG1202+281&0.165&1.74&44.41&45.87&8.6$^{(a)}$&-0.83&-11.44\\
PG1435-067&0.129&4.84&43.67&44.83&7.8$^{(b)}$&-1.04&-11.93\\
SDSSJ144414+0633&0.208&2.57&44.45&45.94&8.1$^{(c)}$&-0.26&-11.59\\
2MASSJ165315+2349&0.103&4.15&43.79&45.00&7.0$^{(d)}$&-0.08&-11.52\\
PG1216+069&0.33&1.51&44.78&46.55&9.2$^{(a)}$&-0.74&-11.73\\
PG0947+396(001)&0.205&1.91&44.19&45.54&8.7$^{(a)}$&-1.24&-11.83\\
PG0947+396(301)&&&44.16&45.49&&-1.29&-11.87\\
WISEJ0537560245&0.110&15.0&43.60&44.75&7.7$^{(d)}$&-1.07&-11.86\\
HB891529+050&0.219&3.93&44.17&45.51&8.7$^{(c)}$&-1.33&-11.92\\
PG1307+085&0.154&2.10&44.33&45.76&7.9$^{(a)}$&-0.23&-11.44\\
PG1425+267&0.364&1.57&44.83&46.65&9.2$^{(c)}$&-0.78&-11.77\\
PG1352+183&0.151&2.10&43.89&45.13&8.4$^{(a)}$&-1.38&-11.85\\
2MASXJ105144+353&0.159&2.20&43.56&44.7&8.4$^{(c)}$&-1.79&-12.20\\
2MASSJ02200728&0.213&2.42&44.18&45.53&8.4$^{(d)}$&-0.98&-11.89\\
LBQS13380038&0.237&1.68&44.51&46.06&7.8$^{(c)}$&0.19&-11.67\\
PG0804+761(401)&0.100&3.34& 44.50 & 45.84 & 8.3$^{(e)}$ & -0.36 &  -10.95 \\
PG0804+761(101)&&& 44.38 & 45.70 &  & -0.56 &  -11.02 \\
PG0804+761(201)&&& 44.26 & 45.70 &  & -0.75 &  -11.14 \\

PG1416-129&0.129&3.34&44.12&45.44&9.0$^{(a)}$&-1.70&-11.48\\
PG1402+261(04)&0.164&1.22&44.06&45.35&7.9$^{(a)}$&-0.70&-11.77\\
PG1402+261(08)&&&44.05&45.34&7.0$^{(a)}$&-0.69&-11.76\\
PG1427+480&0.221&1.61&44.17&45.50&8.1$^{(d)}$&-0.68&-11.94\\
HB891257+286(101)&0.091&1.04&43.57&44.72&7.5$^{(c)}$&-1.33&-11.84\\
HB891257+286(201)&&&43.50&44.64&&-1.40&-11.91\\
HB891257+286(301)&&&43.48&44.62&&-1.93&-11.70\\
PG1114+445(301)&0.144&1.93&44.16&45.50&8.6$^{(a)}$&-1.18&-11.60\\
PG1114+445(401)&&&44.36&45.81&&-0.87&-11.67\\
PG1114+445(501)&&&44.12&45.44&&-1.24&-11.57\\
PG1114+445(701)&&&44.21&45.56&&-1.12&-11.60\\
PG1114+445(801)&&&44.13&45.46&&-1.22&-11.59\\
PG1114+445(901)&&&44.24&45.62&&-1.06&-11.65\\
\hline   
    \end{tabular}
    }
    \caption{ \footnotesize SUBWAYS sample. From left to right, the table includes the source name, redshift, Galactic column density, absorption-corrected X-ray luminosity in the 2-10 keV band, bolometric luminosity, black hole mass, Eddington ratio, and flux in the 2-10 keV band. 
    Notes: 1. Galactic absorption measured by \cite{HI4PI16}. 2. Instantaneous luminosity extracted from the spectral best-fit. 3. Bolometric luminosity derived from L$_X$ with bolometric corrections from \cite{duras20}. 4. Log of black hole mass from the measurements carried out in \cite{Bianchi09caixa1}$^{(a)}$,  \cite{Xie17}$^{(b)}$,  \cite{Perna17}$^{(c)}$,  \cite{Matzeu23}$^{(d)}$, \cite{Kaspi00}$^{(e)}$. 5. The 2–10 keV fluxes from the spectral best-fit are measured with \texttt{cflux} in \texttt{XSPEC}.}
    \label{SUBWAYS_info_obs}
\end{table*}

\begin{table*}
 \centering
\begin{tabular}{|l|c|c|c|c|c|c|c|}
\hline
Name & z & logM$_{BH}$ & logL$_x$ & logL$_{Bol}$ & $\rm log\lambda_{EDD}$ & $\rm E_{cut}$ & References \\
     &   & ($M_{\odot}$) & (erg/s) & (erg/s) & (erg $cm^{-2} s^{-1}$) & (keV) & \\
\hline
\multicolumn{8}{|l|}{\bf Local AGN}\\
\hline 
NGC~4395&0.0011&$5.5^{+0.3}_{-0.3}$&40.63&$41.7^{+0.3}_{-0.3}$&$-2.0^{+0.4}_{-0.4}$&120$^{+50}_{-30}$&(a)\\
NGC~4051&0.002&$5.89^{+0.08}_{-0.14}$&40.68&$41.7^{+0.3}_{-0.3}$&$-2.3^{+0.3}_{-0.3}$&$>$846&(b)\\
... & ...& ...&... & ...& ...&... & ...\\
\hline
\multicolumn{8}{|l|}{\bf Intermediate-z AGN}\\
\hline
4C74.24 &0.104&$9.6^{+0.3}_{-0.3}$&44.38&$45.84^{+0.3}_{-0.3}$&$-1.9^{+0.3}_{-0.3}$&$94^{+54}_{-26}$&(e)\\
Mrk813&0.110&$8.87^{+0.1}_{-0.1}$&43.71&$44.9^{+0.3}_{-0.3}$&$-2.1^{+0.3}_{-0.3}$&$>187$&(f)\\
... & ...& ...&... & ...& ...&... & ...\\
\hline
\multicolumn{8}{|l|}{\bf High-z AGN}\\
\hline
SDSSJ0950+4329  &1.77&$9.7^{+0.3}_{-0.3}$&46.28&$47.7^{+0.3}_{-0.3}$&$-0.14^{+0.4}_{-0.4}$&$77^{+20}_{-14}$&(p)\\
2MASSJ16&1.86&$9.74^{+0.3}_{-0.3}$&45.93&$48.9^{+0.3}_{-0.3}$&$1.1^{+0.4}_{-0.4}$&$106^{+102}_{-37}$&(q)\\
... & ...& ...&... & ...& ...&... & ...\\
\label{sample_infos}
\end{tabular}
\caption{ \footnotesize AGN Sample from the literature presented in \Cref{binned}. This table is available in its entirety in machine-readable form at the CDS \url{http...}. A portion is shown here for guidance regarding its form and content.
References: (a) \cite{Balokovic20}, (b) \cite{Akylas21}, (c) \cite{Younes2019}, (d) \cite{Fabian2015},  (e)  \cite{Molina2019}, (f) \cite{Kamraj18}, (g) \cite{Buisson18}, (h)  \cite{Kara17}, (i) \cite{Tortosa18a}, (j)  \cite{Ursini20}, (k) \cite{Reeves21}, (l) \cite{Lanzuisi24}, (m) \cite{Kammoun17}, (n) \cite{Marinucci22}, (o) \cite{Kammoun23}, (p) \cite{Borrelli26}, (q) \cite{Lanzuisi19}, (r) \cite{Bertola22}.}
\end{table*}

\clearpage

\section{Results}
\subsection{Best fit values}
In this Section,  we report the best-fit values for Model-1 and Model-2.
Sources WISEJ053756-0245,
HB891529+050, 2MASSJ165315+2349, 2MASSJ10514428+3539, 2MASXJ02201453-0728, and PG1114+445 do not require an additional component for the soft excess.
For PG1114+445, we included three WAs using three \texttt{XABS} tables, following the analysis in \cite{Serafinelli19}. 
For PG0804+761, one extra blackbody component in excess to the warm corona model is needed, following \citep{Matzeu23}.

\begin{table}[h!]
\begin{center}
\label{Model2}
\begin{tabular}{lcccccc}
    \hline
         Source& $\tau$ & kT$_e^{warm}$ (kev) & $ \Gamma$&$E_{cut}$  (keV)&R&C$_{TOT}/\nu$ \\
          \hline
        PG0052+251&$12.5^{+1.5}_{-1.0}$ & $0.35^{+0.05}_{-0.05}$&$1.74^{+0.05}_{-0.06}$& $>81$& $0.30^{+0.16}_{-0.15}$&626.8/536\\

          PG0953+414&$16.3^{+1.9}_{-1.8}$ & $0.23^{+0.03}_{-0.02}$&$2.07^{+0.05}_{-0.06}$&$>153$&$1.2^{+0.3}_{-0.3}$&695.1/543\\
          
          PG1626+554&$15.2^{+1.4}_{-1.2}$ & $0.26^{+0.33}_{-0.30}$&$1.88^{+0.08}_{-0.06}$&$49^{+69}_{-13}$&$1.09^{+0.33}_{-0.30}$&519.9/503\\
        
          PG1202+281(02)&$16.4^{+1.3}_{-1.1}$ & $0.31^{+0.20}_{-0.17}$&$1.63^{+0.08}_{-0.07}$&$77^{+116}_{-29}$&$0.37^{+0.20}_{-0.17}$&515.1/534\\
        
          PG1202+281(04)&$13.3^{+1.8}_{-1.3}$ & $0.36^{+0.06}_{-0.06}$&$1.55^{+0.08}_{-0.07}$&$50^{+42}_{-13}$&$0.44^{+0.20}_{-0.19}$&532.8/534\\
       
          PG1435-067&$>18.9$ &$0.16^{+0.03}_{-0.01}$ &$1.83^{+0.11}_{-0.10}$&$47^{+101}_{-17}$&$>0.93$&545.9/496\\
          
          SDSSJ144414+0633&$10.8^{+1.2}_{-1.2}$ &$0.43^{+0.10}_{-0.07}$&$1.68^{+0.08}_{-0.11}$&$55^{+55}_{-21}$&$0.87^{+0.27}_{-0.24}$&579.5/503\\
           
           2MASSJ165315+2349&-&-&$1.51^{+0.13}_{-0.13}$&$70^{+143}_{-28}$&$0.79^{+0.22}_{-0.20}$&534.9/477\\
          PG1216+069&$12.5^{+1.7}_{-1.5}$  &$0.33^{+0.06}_{-0.05}$&$1.78^{+0.06}_{-0.08}$&$>75$&$0.58^{+0.26}_{-0.23}$&568.3/499\\
        
          PG0947+396(01)&$13.3^{+4.0}_{-3.0}$ &$0.33^{+0.16}_{-0.09}$&$1.85^{+0.09}_{-0.18}$&$59^{+102}_{-28}$&$>1.09$&498.8/466\\
          
        PG0947+396(02)&$11.4^{+2.8}_{-1.7}$ &$0.37^{+0.11}_{-0.10}$ &$1.76^{+0.19}_{-0.13}$&$45^{+81}_{-15}$&$>0.98$&484.1/464\\
          
          WISEJ053756&-&-&$1.50^{+0.10}_{-0.14}$&$24^{+21}_{-6}$&$0.84^{+0.38}_{-0.34}$&517.5/474\\
        
          HB891528+050&$15.6^{+4.0}_{-2.3}$&$0.32^{+0.06}_{-0.06}$&$1.77^{+0.11}_{-0.10}$&$40^{+34}_{-13}$&$1.38^{+0.49}_{-0.40}$&473.4/445\\
                 
           PG1307+085&$11.4^{+2.0}_{-1.7}$ &$0.35^{+0.08}_{-0.05}$&$1.80^{+0.09}_{-0.08}$&$>70$&$0.39^{+0.73}_{-0.27}$&593.4/526\\
         
          PG1425+267&$9.0^{+1.6}_{-2.0}$ &$0.54^{+0.11}_{-0.26}$&$1.84^{+0.06}_{-0.11}$&$>83$&$1.04^{+0.29}_{-0.25}$&576.1/499\\
        
  PG1352+183&$11.9^{+2.3}_{-1.4}$ &$0.32^{+0.07}_{-0.08}$&$2.05^{+0.10}_{-0.04}$&$>166$&$>1.29$&596.0/487\\
        2MASSJ10514428+3539&-&-&$1.58^{+0.09}_{-0.09}$&$>33$&$>1.19$&449.6/403\\
        
          2MASSJ0220-0728&-&-&$1.54^{+0.08}_{-0.08}$&$58^{+252}_{-24}$&$0.51^{+0.28}_{-0.25}$&474.8/448\\
       
LBQS1338 &$12.5^{+1.2}_{-1.1}$&$0.40^{+0.06}_{-0.05}$& $1.52^{+0.1}_{-0.08}$&$35^{+17}_{-10}$&$0.25^{+0.18}_{-0.16}$&497.5/469\\
        
PG0804+761(401) &$6.7_{-4.9}^{+0.9}$ &$0.74_{-0.12}^{+0.68}$ &$1.94_{-0.34}^{+0.10}$&$>59$ & $0.76_{-0.41}^{+0.30}$ &483.0/422\\

PG0804+761(101) &$5.5_{-4.0}^{+1.4}$ &$0.94_{-0.04}^{+...}$ &$1.95_{-0.08}^{+0.04}$&$>117$ & $0.81_{-0.26}^{+0.24}$ &505.5/458\\

PG0804+761(201) &$5.9_{-2.3}^{+1.3}$ &$0.81_{-0.19}^{+...}$ &$2.00_{-0.14}^{+0.06}$&$>88$ & $1.18_{-0.26}^{+0.29}$ &535.5/456\\

  PG1416-129&$21.2^{+8.1}_{-6.3}$ &$0.23^{+0.01}_{-0.05}$&$1.55^{+0.05}_{-0.05}$&$79^{+17}_{-13}$&$0.41^{+0.21}_{-0.20}$&514.7/482\\
 
     PG1402+261(02)&$9.3^{+2.6}_{-2.82}$ &$0.40^{+0.23}_{-0.12}$&$2.04^{+0.09}_{-0.15}$&$>48$&$1.30^{+0.60}_{-0.56}$&491.3/491\\
   
     PG1402+261(04)&$8.9^{+1.2}_{-1.2}$ &$0.47^{+0.12}_{-0.08}$&$2.09^{+0.07}_{-0.07}$&$>171$&$0.45^{+0.33}_{-0.28}$&565.6/518\\

  PG1427+480&$5.21^{+2.4}_{-3.8}$ &$0.64^{+0.15}_{-0.42}$&$1.99^{+0.08}_{-0.07}$&$>94$&$1.29_{-0.57}^{+0.59}$&458.6/440\\
  
 HB891257+286(101) &$17.9^{+1.8}_{-2.3}$ &$0.24^{+0.03}_{-0.01}$&$2.00^{+0.04}_{-0.08}$&$>97$&$1.25^{+0.36}_{-0.40}$&546.5/480\\

    HB891257+286(201) &$22.3^{+6.4}_{-3.7}$ &$0.20^{+0.03}_{-0.03}$&$1.90^{+0.05}_{-0.06}$&$>74$&$1.10^{+0.38}_{-0.35}$&468.8/469\\

  HB891257+286(301) &$17.3^{+1.9}_{-1.6}$ &$0.25^{+0.03}_{-0.02}$ &$1.90^{+0.07}_{-0.06}$&$>72$&$0.98^{+0.39}_{-0.35}$&565.0/478\\

 PG1114+445(301)&-&-&$1.92^{+0.03}_{-0.04}$&$>74$&$0.79^{+0.20}_{-0.19}$&514.6/467\\
    PG1114+445(401)&-&-&$2.11^{+0.02}_{-0.02}$&$>363$&$0.97^{+0.20}_{-0.28}$&597.7/478\\
    PG1114+445(501)&-&-&$2.08^{+0.02}_{-0.02}$&$>229$&$1.34^{+0.36}_{-0.33}$&587.8/474\\
    PG1114+445(701)&-&-&$2.04^{+0.02}_{-0.02}$&$>307$&$1.17^{+0.31}_{-0.29}$&559.1/477\\
    PG1114+445(801)&-&-&$2.00^{+0.02}_{-0.02}$&$>160$&$1.18^{+0.36}_{-0.30}$&507.1/474\\
    PG1114+445(901)&-&-&$1.97^{+0.02}_{-0.02}$&$>121$&$0.75^{+0.35}_{-0.24}$&504.9/476\\
\hline
\end{tabular}
\caption{Model-1 best fit values.  From left to right: optical depth and electron temperature of the warm corona, given by \texttt{compTT}; photon index, high-energy cut-off, and reflection scaling factor provided by \texttt{xillver}. The last column reports the C-statistic over the degrees of freedom $\nu$.}
\end{center}
\end{table}

\begin{table}[h]
\begin{center}
\label{Model3}
\begin{tabular}{lcccccc}
\hline
Source&$\Gamma^{asymptotic}$& kT$_e^{warm}$ (kev)&$ \Gamma$&kT$_e^{hot}$ (keV)&R&C$_{TOT}/\nu$\\
\hline
PG0052+251&$2.58^{+0.07}_{-0.06}$ & $0.35^{+0.05}_{-0.04}$&$1.80^{+0.03}_{-0.03}$&$>12$&$0.31^{+0.17}_{-0.15}$&622.7/536\\
PG0953+414&$2.44^{+0.11}_{-0.13}$ & $0.23^{+0.03}_{-0.02}$&$2.08^{+0.04}_{-0.04}$&$>21$&$1.5^{+0.3}_{-0.4}$&682.1/542\\
PG1626+554&$2.41^{+0.08}_{-0.09}$ & $0.25^{+0.02}_{-0.02}$&$1.80^{+0.04}_{-0.04}$&$>15$&$1.03^{+0.31}_{-0.28}$&513.5/503\\
PG1202+281(02)&$2.06^{+0.06}_{-0.07}$ & $0.29^{+0.02}_{-0.02}$&$1.76^{+0.03}_{-0.03}$&$12^{+7}_{-3}$&$0.39^{+0.20}_{-0.18}$&514.9/534\\
PG1202+281(04)&$2.06^{+0.06}_{-0.06}$ & $0.28^{+0.02}_{-0.02}$&$1.75^{+0.04}_{-0.03}$&$12^{+6}_{-3}$&$0.39^{+0.20}_{-0.18}$&514.9/534\\
PG1435-067&$<2.39$ &$0.13^{+0.03}_{-0.01}$&$1.99^{+0.02}_{-0.02}$&$13^{+25}_{-5}$&$>1.68$&556.4/471\\
SDSSJ144414+0633&$2.57^{+0.08}_{-0.09}$ &$0.44^{+0.14}_{-0.08}$&$1.80^{+0.05}_{-0.06}$&$8^{+2}_{-1}$&$>1.1$&571.2/503\\
2MASSJ165315+2349&-&-&$1.70^{+0.05}_{-0.04}$&$14^{+8}_{-4}$&$0.81^{+0.22}_{-0.20}$&532.6/477\\
PG1216+069&$2.52^{+0.11}_{-0.12}$  &$0.31^{+0.05}_{-0.04}$&$1.82^{+0.04}_{-0.04}$&$>14$&$0.58^{+0.27}_{-0.23}$&569.6/499\\
PG0947+396(01)&$2.40^{+0.28}_{-0.23}$ &$0.29^{+0.10}_{-0.07}$&$1.92^{+0.06}_{-0.08}$&$13^{+31}_{-5}$&$>0.98$&498.2/463\\
PG0947+396(02)&$2.66^{+0.13}_{-0.12}$ &$0.39^{+0.13}_{-0.09}$&$1.88^{+0.07}_{-0.07}$&$8^{+2}_{-1}$&$>1.41$&483.7/464\\
WISEJ053756&-&-&$1.78^{+0.01}_{-0.01}$&$9^{+2.5}_{-2}$& $0.77^{+0.30}_{-0.24}$&557.0/477\\
HB891528+050&$2.19^{+0.15}_{-0.20}$&$0.37^{+0.10}_{-0.07}$&$1.86^{+0.05}_{-0.06}$&$7^{+1}_{-1}$&$>1.53$&467.6/445\\
PG1307+085&$2.74^{+0.12}_{-0.13}$ &$0.36^{+0.06}_{-0.04}$&$1.87^{+0.03}_{-0.03}$&$>14$&$0.59^{+0.20}_{-0.18}$&589.63/526\\
PG1425+267&$2.36^{+0.13}_{-0.17}$ &$0.45^{+0.08}_{-0.06}$ &$1.86^{+0.05}_{-0.05}$&$>23$&$1.05^{+0.27}_{-0.12}$&560.6/498\\
PG1352+183&$2.71^{+0.08}_{-0.08}$ &$0.32^{+0.08}_{-0.06}$ &$2.05^{+0.03}_{-0.05}$&$>16$&$>1.4$&590.1/487\\
2MASSJ105144+3539&-&-&$1.66^{+0.05}_{-0.06}$&$>8$&$>0.94$&487.9/440\\
2MASSJ0220-0728&-&-&$1.70^{+0.05}_{-0.04}$&$>8$&$0.46^{+0.26}_{-0.24}$&463.4/434\\
LBQS1338 &$2.32^{+0.08}_{-0.07}$ &$0.35^{+0.05}_{-0.04}$&$1.76^{+0.03}_{-0.03}$&$7^{+1.5}_{-1.1}$&$0.47^{+0.32}_{-0.26}$&505.5/496\\
PG0804+761(401) &$2.95_{-0.37}^{+0.39}$ &$0.66_{-0.16}^{+...}$ &$1.92_{-0.18}^{+0.10}$&$>16$ & $0.70_{-0.40}^{+0.44}$ &482.0/422\\
PG0804+761(101) &$3.23_{-0.05}^{+0.06}$ &$0.98_{-0.32}^{+...}$ &$1.94_{-0.10}^{+0.05}$&$>18$ & $0.81_{-0.24}^{+0.25}$ &501.8/458\\
PG0804+761(201) &$3.05_{-0.04}^{+0.04}$ &$0.59_{-0.08}^{+0.13}$ &$2.00_{-0.06}^{+0.04}$&$>27$ & $1.09_{-0.25}^{+0.25}$ &530.4/456\\
PG1416-129 &$1.97^{+0.30}_{-0.50}$ &$0.24^{+0.10}_{-0.06}$&$1.68^{+0.03}_{-0.03}$&$9.0^{+6.7}_{-1.6}$&$0.87^{+0.68}_{-0.42}$&512.2/484\\
PG1402+261(02)&$3.07^{+0.14}_{-0.16}$ &$0.37^{+0.20}_{-0.10}$&$2.11^{+0.06}_{-0.06}$&$>13$&$1.38^{+0.57}_{-0.53}$&489.4/490\\
PG1402+261(04)&$2.94^{+0.07}_{-0.06}$ &$0.49^{+0.12}_{-0.09}$&$2.10^{+0.06}_{-0.07}$&$>33$&$0.50^{+0.85}_{-0.32}$&565.5/518\\
PG1427+480 &$3.25^{+0.03}_{-0.09}$ &$0.69^{+0.06}_{-0.54}$&$1.89^{+0.01}_{-0.02}$&$23.4_{-3.3}^{+2.3   }$&$1.1_{-0.48}^{+0.57}$&466.5/440\\
HB891257+286(101)&$2.26^{+0.12}_{-0.12}$ &$0.25^{+0.03}_{-0.02}$&$1.99^{+0.05}_{-0.04}$&$>13$&$1.24^{+0.40}_{-0.33}$&543.7/480\\
HB891257+286(201) &$1.99^{+0.19}_{-0.24}$ &$0.19^{+0.03}_{-0.02}$&$1.93^{+0.03}_{-0.03}$&$>9$&$1.17^{+0.37}_{-0.34}$&467.0/469\\
HB891257+286(301) &$2.37^{+0.20}_{-0.11}$ &$0.29^{+0.10}_{-0.04}$&$1.90^{+0.05}_{-0.05}$&$7.4$&$>1.19$&563.3/477\\
PG1114+445(301)&-&-&$1.99^{+0.02}_{-0.02}$&$>19$&$0.89^{+0.34}_{-0.32}$&522.7/466\\
PG1114+445(401)&-&-&$2.14^{+0.02}_{-0.02}$&$>49$&$1.29^{+0.32}_{-0.31}$&600.3/478\\
PG1114+445(501)&-&-&$2.07^{+0.02}_{-0.02}$&$>27$&$1.4^{+0.37}_{-0.34}$&586.6/474\\
PG1114+445(701)&-&-&$2.04^{+0.02}_{-0.02}$&$>22$&$1.26^{+0.33}_{-0.30}$&558.6/477\\
PG1114+445(801)&-&-&$1.92^{+0.02}_{-0.02}$&$>18$&$0.73^{+0.29}_{-0.27}$&506.5/473\\
PG1114+445(901)&-&-&$2.04^{+0.02}_{-0.02}$&$>16$&$1.38^{+0.32}_{-0.30}$&510.0/476\\
\hline
\end{tabular}
\caption{Model-2 best fit values. From left to right, asymptotic photon index and electron temperature of the warm corona, obtained from \texttt{nthcomp}; photon index, electron temperature, and reflection scaling factor for the hot corona,  provided by \texttt{xillverCp}. The last column reports the C-statistic over the degrees of freedom $\nu$.}
\end{center}
\vspace{-0.3cm}
\end{table}

\clearpage

\section{Comparison of high energy cut-off and electron temperature and recovery analysis}
\label{simulations}

In \Cref{kratio} we plot the comparison between  $kT_e^{hot}$ from Model-2 and E$_{cut}$ for Model-1, source-by-source: the data points align along the relation E$_{cut} = 5\times kT_e$. This would require either i) peculiar coronal optical depth and geometry (Monte Carlo grids show that the ratio increases with optical depth, and the increase is stronger in slab geometry \citealp{Middei19Moca} or ii) introducing more complex/exotic corona models, such as hybrid (thermal plus non-thermal) electron distributions \citep{Fabian_2017}, or a dynamic, outflowing corona with substantial outflow velocity that would blueshift in the observed E$_{cut}$ \citep{Malzac01}, or patchy/multizone coronae. In all these cases, however, it is difficult to explain why this effect appears in the SUBWAYS sample but not in others. 

\begin{wrapfigure}{r}{0.5\columnwidth}
 \centering
    \includegraphics[width=0.9\linewidth]{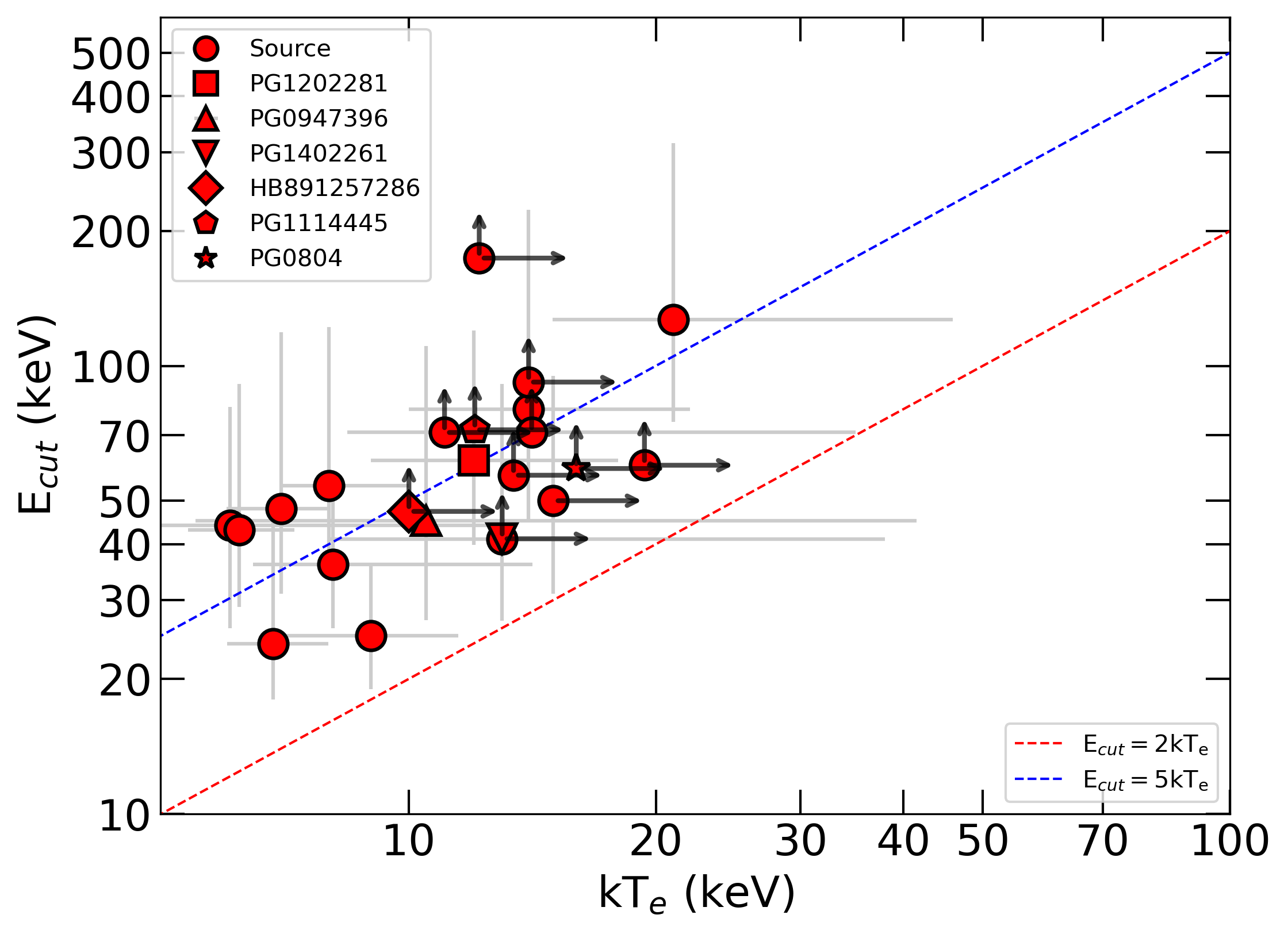}
   \caption{ \footnotesize High-energy cut-off vs. electron temperature for the SUBWAYS sample (red points). Dashed lines represent empirical relations with scaling factors of 2 (red) and 5 (blue). The median value is reported for sources with multiple observations.}
    \label{kratio}
\end{wrapfigure} 

We decided to investigate this peculiar behavior in the results for SUBWAYS sources, by performing a recovery analysis on both Model-1 and Model-2. The goal is to test the reliability of our measurements in high-energy cut-off $E_{cut}$ and coronal temperature $kT_e$ in the redshift and flux regimes of SUBWAYS. We selected the \xmm plus \nustar spectrum of PG~1416-129 as our starting point (a second test was performed using LBQS~1338, with similar results). This source exhibits the largest discrepancy between the electron temperature derived using \texttt{xillverCP} ($kT \sim 9$ keV) and the high-energy cut-off obtained with \texttt{xillver} ($E_{cut} \sim 80$ keV). PG~1416-129 is located at an intermediate redshift ($z = 0.129$) and has a moderate flux (Log$F_{2-10} = -11.48$ erg cm$^{-2}$ s$^{-1}$), making it a reasonably representative source. Its spectrum is particularly flat ($ \Gamma \sim  1.5-1.7$, depending on the model), which, in principle, should help constrain the hard X-ray component due to the relatively higher flux in the hard band.

Starting from the best-fit models with both \texttt{xillver} and \texttt{xillverCp}, we performed a series of simulations by varying the high-energy cut-off ($E_{cut}=10, 20, 40, 80, 100, 200$ keV) and coronal temperature ($kT_e = 5, 10, 18, 30, 50, 100 $ keV)  for reflection fractions $R = $0.1, 0.5, 1, 1.5, and 2. For each parameter combination, we simulated 30 spectra, refitted them, recorded the best-fit values, and computed the median, accounting for lower limits as explained in \Cref{sec:Results}. Figure \ref{simulation_ecut} shows the results of these simulations, for the $R=1$ case.

\begin{figure*}[h!]
    \centering
    \includegraphics[width=0.45\linewidth]{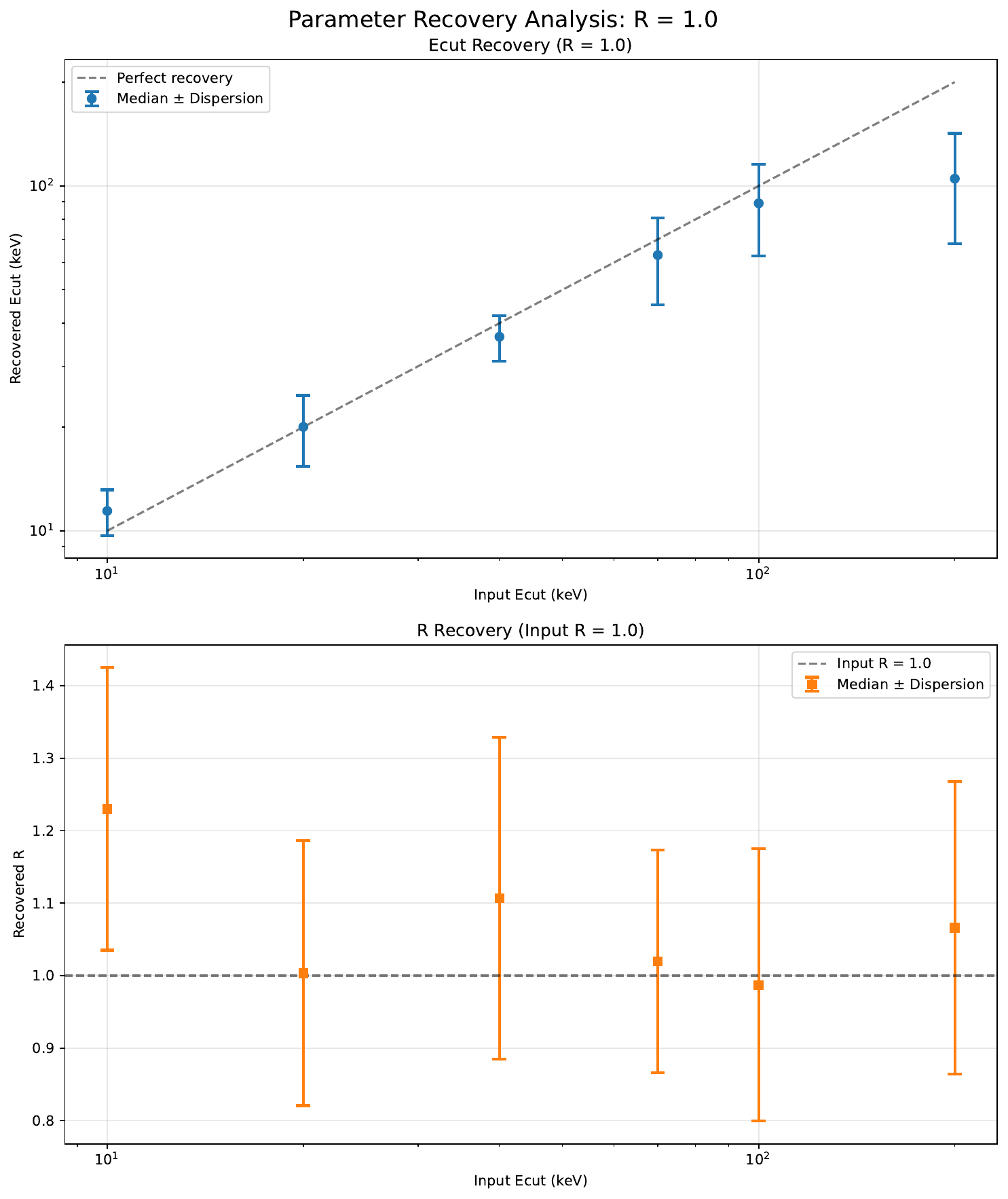}
    \hspace{0.5cm}
    \includegraphics[width=0.45\linewidth]{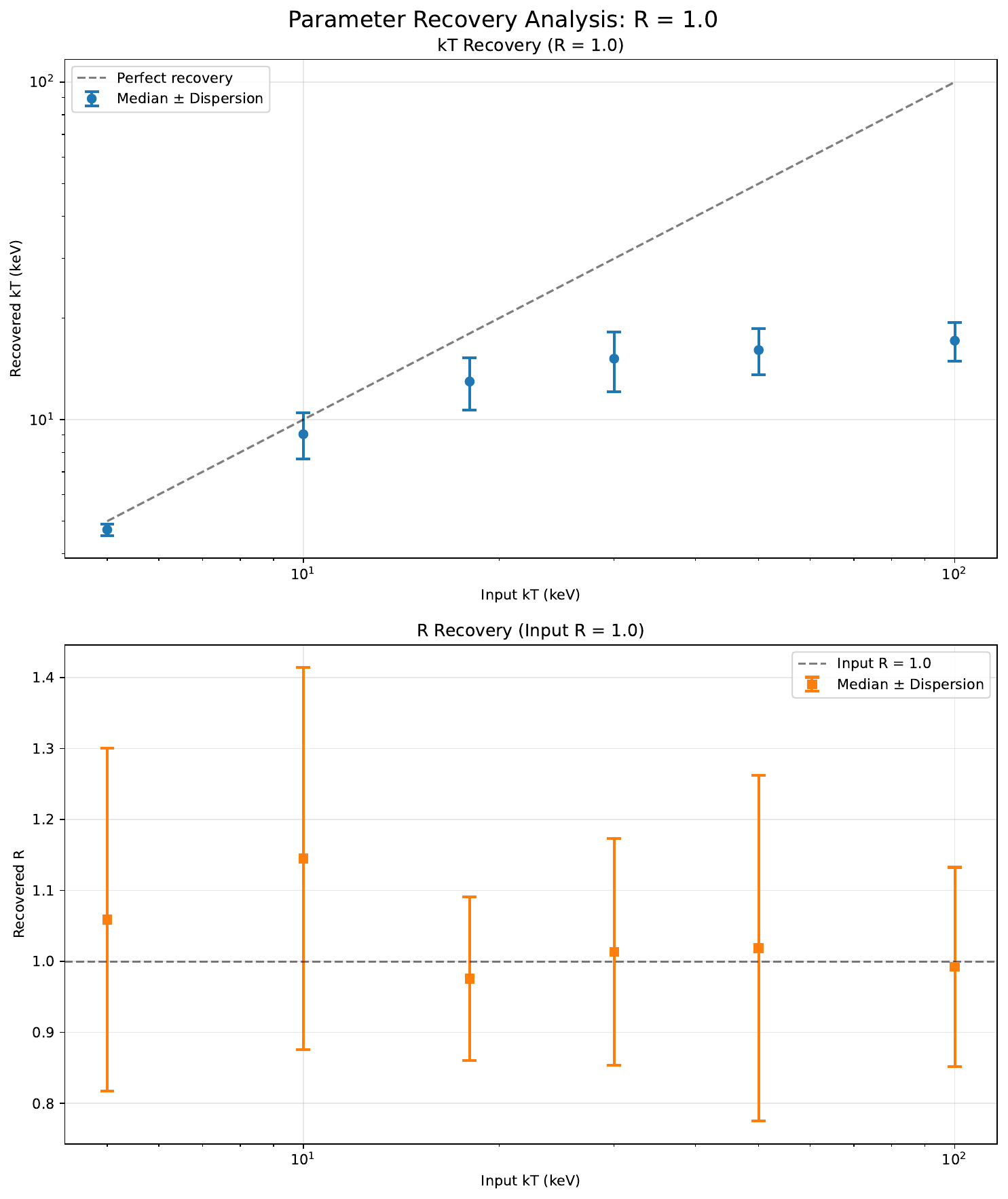}
    \caption{\footnotesize    
    Recovery simulations of \texttt{xillver} (left) and \texttt{xillverCp} (right) on the spectrum PG~1416-129. The median $E_{cut}$ obtained from fitting the simulated spectra is in agreement with the input value up to $E_{cut}\sim100$ keV, while the median $kT_e$ is in agreement with the input value only up to $kT_e\sim10$ keV. At higher input $kT_e$ values, the median fit results stall at $\sim15-20$ keV.}
    \label{simulation_ecut}
\end{figure*}

In the case of \texttt{xillver}, the median $E_{cut}$ obtained from fitting the simulated spectra is in accordance with the input value up to $E_{cut}\sim100$ keV, while in the case of \texttt{xillverCp}. The median $kT_e$ is in agreement with the input value only up to $kT_e\sim10$ keV, while at higher input $kT_e$ values, the median fit results stall at $\sim15-20$ keV, implying that we cannot recover any $kT_e$ value larger than that. We stress that R is always correctly recovered, within the $1\sigma$ dispersion shown in the plots, at all input $E_{cut}$ and $kT_e$ values, as well as in the other R cases, not shown here. Therefore, there is no clear bias in the recovery of R correlated with the bias observed in the recovered $kT_e$.

\section{Comparison of  optical depths estimates}
\label{tau_comparison}
In \Cref{sec:Discussion}, we recovered the optical depth values for the warm corona using the relation contained in \cite{Longair11}.
However, in the literature, the relations described in \citet{Beloborodov99} are the most used one: 
 \begin{equation}
\Gamma \sim \frac{9}{4} y^{-\frac{2}{9}}
\label{gamma_tau}
\end{equation}
where the Compton $y$-parameter is given by:
\begin{equation}
y = 4(\theta + 4\theta^2)\tau(\tau + 1)
\label{Compton_parameter}
\end{equation}
The optical depth obtained by the \texttt{compTT} model is compared with the one computed from the equations above, using the \texttt{nthcomp} results (\Cref{Comptt_Longair}).
\begin{figure}[b]
    \centering
    \includegraphics[width=0.45\linewidth]{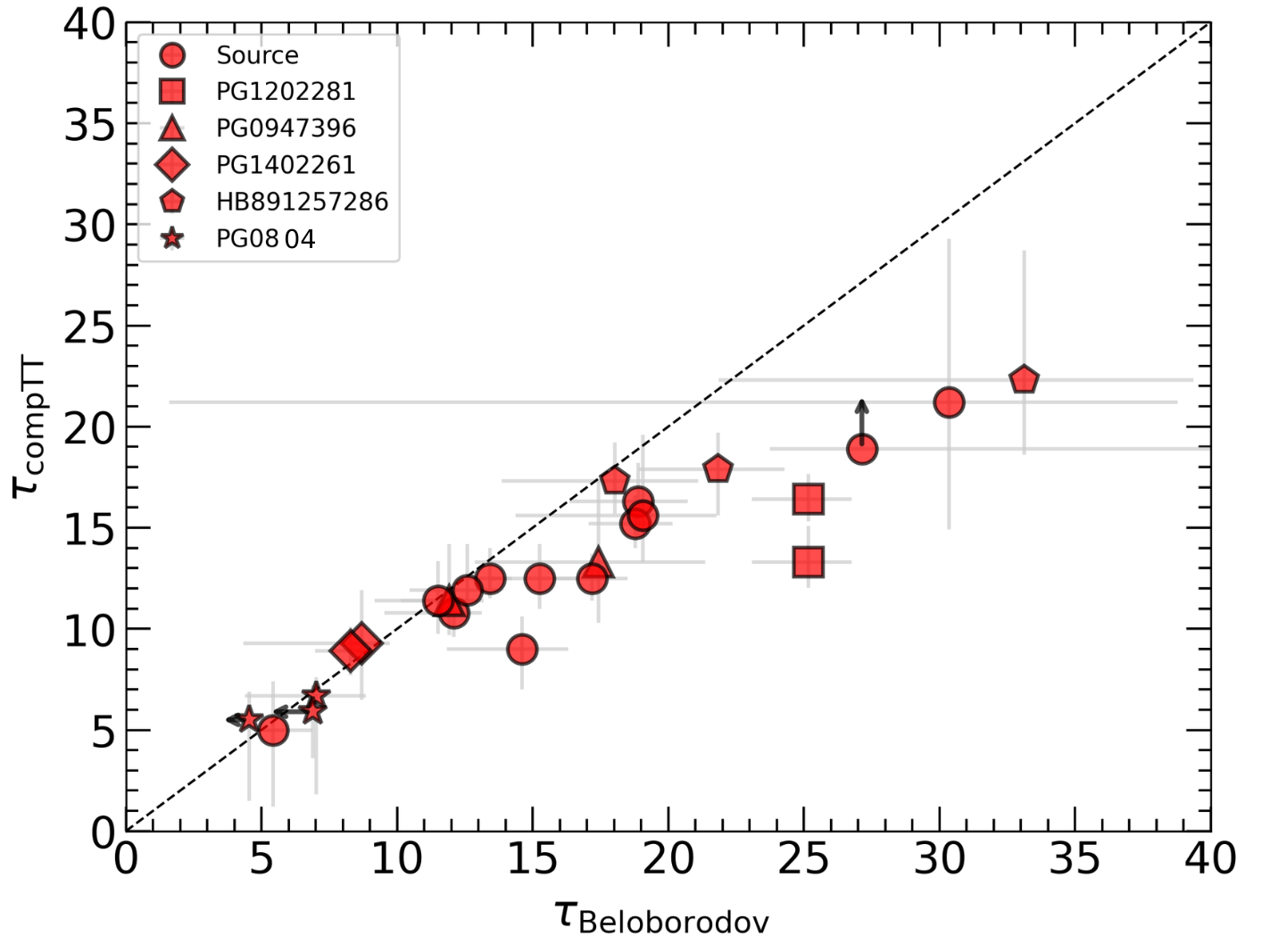}
   \caption{\footnotesize Comparison of optical depth estimates. On the y-axis, the one obtained directly from the \texttt{compTT} model, while on the x-axis, the values are inferred using \Cref{gamma_tau} and \Cref{Compton_parameter} from Beloborodov approximations \citep{Beloborodov99}.}
    \label{Comptt_Longair}
\end{figure}
The Beloborodov approximation ($\tau_B$) overestimates the optical depth for values higher than $\sim 10$, while the formulation reported in \cite{Longair11} show a better agreement with the \texttt{compTT} model findings (\Cref{kte_warm}).
\end{appendix}

\end{document}